%% file: main.tex
\documentclass[twocolumn]{aastex701}
\usepackage{overpic}
\usepackage[outline]{contour}
\usepackage{xcolor}
\usepackage[printonlyused]{acronym}
\makeatletter
\newcommand{\accite}[2]{\AC@placelabel{#1}\acl{#1} (\acs{#1};~\citealp{#2})\acused{#1}}
\makeatother
\usepackage[utf8]{inputenc}
\usepackage[T1]{fontenc}
\usepackage{comets}

\newcommand{\labelcolor}{green}

\contourlength{1.2pt}
\newcommand{\labelpicAWithTrim}[6]{\begin{overpic}[width=#4\linewidth,trim=#5,clip]{#1}\put (5,7) {\Large\color{\labelcolor}\textbf{\contour{black}{#2}}}\put (55,8) {\large\color{\labelcolor}\textbf{\contour{black}{#3}}}\if\relax\detokenize{#6}\relax
  \else
    \put (2,62) {\includegraphics[width=0.07\linewidth,trim=0pt 0pt 0pt 0pt]{#6}}\fi
\end{overpic}}
\newcommand{\labelpicA}[5]{\labelpicAWithTrim{#1}{#2}{#3}{#4}{0pt 0pt 0pt 0pt}{#5}}

\begin{document}

\title{Tails, Trails, and Transients: Comets, NEOs, and Supernovae in Rubin Early Data Preview 2}

\newcommand{\diracuw}{Department of Astronomy \& the DiRAC Institute, University of Washington, Seattle, USA}
\newcommand{\nau}{Department of Astronomy and Planetary Science, Northern Arizona University, Flagstaff, USA}
\newcommand{\planetaryscienceinst}{Planetary Science Institute, 1700 East Fort Lowell Rd., Suite 106, Tucson, AZ 85719, USA}
\newcommand{\marylandastro}{Department of Astronomy, University of Maryland
College Park, MD 20742, USA}
\newcommand{\qubuk}{Astrophysics Research Centre, School of Mathematics and Physics, Queen's University Belfast, BT7 1NN, UK}
\newcommand{\usp}{Departamento de Astronomia, Instituto de Astronomia, Geof\'{\i}sica e Ci\^{e}ncias Atmosf\'{e}ricas, Universidade de S\~{a}o Paulo, 05508-090, S\~{a}o Paulo, SP, Brazil}

\author[0000-0001-7335-1715]{Colin Orion Chandler}
\email{coc123@uw.edu}
\affiliation{\diracuw}
\affiliation{\nau}

\author[0000-0001-9505-1131]{Joseph Murtagh}\altaffiliation{DiRAC Postdoctoral Fellow}
\email{murtagh@uw.edu}
\affiliation{\diracuw}

\author[0009-0005-9428-9590]{Ian Chow}
\email{chowian@uw.edu}
\affiliation{\diracuw}

\author[0009-0003-3171-3118]{Maxine West}
\email{maxwest@uw.edu}
\affiliation{\diracuw}

\author[0000-0003-4094-9408]{A. Fraser Gillan}
\email{andrew.gillan@ncbj.gov.pl}\affiliation{Astrophysics Division, National Centre for Nuclear Research, Pasteura 7, 02-093 Warsaw, Poland}

\author[0000-0001-6957-1627]{Peter S. Ferguson}
\email{pferguso@uw.edu}
\affiliation{\diracuw}

\author[0000-0002-6702-7676]{Michael S. P. Kelley}
\email{msk@astro.umd.edu}
\affiliation{\marylandastro}

\author[0000-0002-9154-3136]{Melissa L. Graham}
\email{mlg3k@uw.edu}
\affiliation{\diracuw}

\author[0009-0005-5452-0671]{Jacob A. Kurlander}
\email{jkurla@uw.edu}
\affiliation{\diracuw}

\author[0000-0001-7225-9271]{Henry H. Hsieh}
\affiliation{\planetaryscienceinst}
\email{hhsieh@psi.edu}  

\author[0000-0003-0743-9422]{Pedro H. Bernardinelli}
\affiliation{\usp}
\email{pedro.bernardinelli@usp.br}

\author[0009-0003-1791-8707]{Wilson Beebe}
\email{wbeebe@uw.edu}
\affiliation{\diracuw}

\author[0009-0009-2281-7031]{Jeremy Kubica}
\affiliation{McWilliams Center for Cosmology and Astrophysics, Department of Physics, Carnegie Mellon University, Pittsburgh, PA 15213, USA}
\email{jkubica@andrew.cmu.edu}

\author[0000-0001-5576-8189]{Andrew Connolly}
\email{ajc26@uw.edu}
\affiliation{\diracuw}

\author[0000-0002-0637-835X]{James R. A. Davenport}
\affiliation{\diracuw}
\email{jrad@uw.edu}

\begin{abstract}
We identify effective single-visit imaging hidden within the deep coadded images of the NSF-DOE Vera C. Rubin Observatory Early Data Preview 2 (EDP2). We develop a workflow to locate known small Solar System bodies in these data, using our Ponder orchestrator with the Sorcha survey simulator to predict their positions and the Rubin Science Platform to identify shallow coadds and extract image cutouts. Because cometary activity and near-Earth object (NEO) trails produce extended morphologies, both remain identifiable in shallow coadds. From approximately 700 comet cutouts, we identify 59 activity detections spanning 32 objects. We discover activity in 2025 NC$_6$, previously unknown to be active, and independently rediscover two comets through their activity. We recover 336 NEO cutouts, including 29 with predicted trails over 25~pixels. We demonstrate how the same provenance-driven methods can recover stationary transients in Rubin coadds, including ones predating the alert stream. Of 47 publicly-reported transients coinciding with EDP2, 36 occupy shallow coadds. We show Rubin photometry agrees with published lightcurves to within 0.2~mag. The imaging provides additional bands, host-galaxy context, and deeper supernova pre-discovery limits than discovery surveys could reach. We showcase seven transients, including SN~2025nbh, which we identify as lying in the outskirts of AGC~114905, an ultra-diffuse galaxy central to the dark-matter-free debate; the supernova offers an independent galaxy distance measurement. Shallow coadds are a relatively unexplored facet of Rubin data with applications beyond the examples presented here; regions with varying numbers of contributing visits will persist throughout the Legacy Survey of Space and Time (LSST).
\end{abstract}

\acresetall{}

\keywords{
Comets (280) ---
Near-Earth objects (1092) ---
Small Solar System bodies (1469) ---
Astronomical techniques (1684) ---
Astronomical data analysis (1858) ---
Sky surveys (1464)
}

\section{Introduction} \label{sec:intro}

The 8.4~m flagship NSF-DOE Vera C. Rubin Observatory (hereafter, Rubin) began acquiring data with its 3.2 gigapixel LSSTCam in 2025 April, and the ensuing \accite{LSST}{ivezicLSSTScienceDrivers2019} is expected to revolutionize Solar System science with discoveries of vast numbers of comets and minor planets \citep{solontoi_detecting_2010,kurlanderPredictionsLSSTSolar2025,dorsey_visibility_2025,murtaghPredictionsLSSTSolar2025,murtaghPredictionsLSSTSolar2026,chow_predictions_2026}. 
The commissioning/Science Validation data that followed, through roughly 2026 January, are to be fully included in the \accite{DP2}{verac.rubinobservatoryteamRTN115VeraRubin2026,nsf-doeverac.rubinobservatoryLegacySurveySpace2026} dataset, to be nominally released in 2026 December. 

A preview data product, \ac{EDP2}, was released on 2026 July 27.  This pre-\ac{LSST} \ac{DP2} product contains only $\sim$25,000 visits (camera exposures); \added{the \ac{DP2}} documentation lists 28,698 visits on 131 nights between 2025 April and 2026 January \citep{ObservationsDP2} -- though the number of visits available on the \ac{RSP} was slightly higher at 29,756 (Section \ref{sec:transientmethods}) -- so some areas of the sky, in some bands, were imaged only once or twice, resulting in ``shallow'' areas in the otherwise deep image data. This anticipated scenario offers a special opportunity to carry out science enabled by single exposures, such as transient phenomena including small Solar System bodies (e.g., asteroids, comets), before single-visit image products are available. The \ac{EDP2} catalogs, including the \texttt{Object} catalog and associated visit-level measurements (e.g., photometry), provide complementary information and support additional Solar System investigations (e.g., Vavilov et al., in preparation).

With rare exceptions such as \comet{3I} and \comet{433P} \citep{chandlerNSFDOEVeraRubin2026,hsieh2026ReactivationMainbelt2026}, few images of comets from LSSTCam have been seen by the community, but the scientific utility of examining images of comets and other active bodies (e.g., for morphology analyses) is high \citep[e.g.,][]{farnhamComaMorphologyJupiterfamily2009,jewittActiveAsteroids2015a}. Similarly, some \ac{NEO} images may also be found in the Rubin Alert Stream \citep{rubinObservatoryPromptAlerts2026}, but nominal single-image access is not yet available to data-rights holders. Both of these types of objects are unique among the small-body population in that they can sometimes be confidently identified in single exposures due to their extended nature that clearly sets them apart from stars and other background \added{point} sources: comets display activity, while \acp{NEO} appear trailed if their on-sky motion during one exposure is greater than the telescope's resolution.

Before we can access images of these objects, we must first know where they are \added{expected to be}. Several services exist to help identify which small solar system bodies may be in the \ac{FOV} of an astronomical image, such as \texttt{SkyBot} \citep{berthierSkyBoTNewVO2006}, operated by the \ac{IMCCE}, \texttt{mpchecker}\footnote{\url{https://www.minorplanetcenter.net/cgi-bin/checkmp.cgi}}, and the JPL Small-Body Identification Tool\footnote{\url{https://ssd.jpl.nasa.gov/tools/sb_ident.html}}. Unsurprisingly, however, these tools have significant limitations at Rubin scale. A more sustainable approach is to make use of a survey simulator, such as \texttt{Kete} \citep{dahlenKetePredictingKnown2025} or \texttt{Sorcha} \citep{merrittSorchaSolarSystem2025a,2025AJ....170..100M,holmanSorchaOptimizedSolar2025}, that simulates the solar system and provides ephemerides (i.e., the object's \added{predicted} position at a given time) for all known solar system objects. This approach has several advantages, but the most important is that it relies on local resources rather than an external service.

Studies that rely on understanding \added{cometary} activity benefit from knowledge of onset, duration, astrometry, photometry, and morphological evolution, all potentially derivable from archival precovery; comparison-sample studies can begin with this catalog. \added{The value of the archival searches we describe also extends more broadly beyond Solar System science to other time-domain phenomena that benefit from image-derived information beyond what catalogs alone can provide.} Deep coadds are useful for discovery and visualization products, though the dataset presents its own challenges due to the ``patchwork quilt'' effects (Section \ref{subsec:provenance}) needed to produce large, contiguous deep coadd images and catalogs. We describe working within those parameters in this work.

\added{Rubin's value to time-domain astronomy is often framed in terms of its alert stream, but the images themselves carry information that the detection catalogs do not preserve. Moreover, not all transient events make it to the alert stream; for example, when template coverage is insufficient to produce a difference image. Furthermore, the alert stream itself did not begin until 2026 February, well after the timespan covered by \ac{DP2}. For example, any survey-reported transient candidate can be confirmed (or rejected) by inspecting the pixels at its reported position in Rubin data. This can determine whether the source is point-like, visually associated with a host galaxy, or in empty sky, and whether anything was there before. Forced photometry at a known position can recover flux below the single-epoch detection threshold and turn pre-supernova explosion non-detections into limits on the explosion time and precursor emission \citep[e.g.,][]{gaglianoFindingFuseProspects2025}. Pre-explosion images can constrain or identify the progenitor \citep{smarttProgenitorsCoreCollapseSupernovae2009}, and Rubin can extend light curve coverage, including into bands not observed by the discovering survey; Rubin's $u$, $z$, and $y$ coverage and depth can complement other surveys, e.g., \accite{ZTF}{masci_zwicky_2019}, \accite{ATLAS}{tonry_atlas_2018}. Host identification and offset also inform classification and redshift priors \citep{gaglianoGHOSTUsingOnly2021}. These science opportunities are available for any transient falling within Rubin's footprint at the relevant epochs, whether or not Rubin itself issued an alert. \ac{EDP2} provides an early opportunity to explore this capability; its coadds include visits that overlap the lifetimes of hundreds of publicly reported transients, and the same tools developed here to recover comets and \acp{NEO} from shallow coadds can recover those transients as well.}

\section{Methods} \label{sec:methods}

We carried out our procedures with \added{four} datasets, in this order: Rubin 3I/ATLAS observations \citep{chandlerNSFDOEVeraRubin2026}, comets, \acp{NEO}, \added{and transients}. These techniques apply to other populations, such as the \acp{TNO} of \cite{bernardinelliTransNeptunianObjectsFound2020}, but we restrict this work to tailed, trailed, \added{and transient} sources.

\subsection{3I/ATLAS}\label{subec:3I}

\cometL{3I} data from Rubin were previously made available to data-rights holders and included in the \textit{Rubin Comet Catchers} citizen science project \citep{chandlerRubinCometCatchers2025}, so we already knew where to find them (i.e., visit + detector, sky coordinates) in the Rubin data. Our first goal was to identify the images in \ac{EDP2} that corresponded to the locations of \comet{3I}, proving the cutout system functioned, and to create ``comparison images'' that depicted the same area on the sky \added{at a different time}, ideally with deeper \added{limiting magnitude}. This allows us to understand, for example, more about the background sources blended with the object. To this end, we produced a table \added{of observation dates and times} from \cite{chandlerNSFDOEVeraRubin2026}, and we queried JPL Horizons \citep{giorginiHorizonsJPLsOnLine1996} at the \added{corresponding} dates/times to retrieve \cometL{3I}'s RA, Dec, rates, apparent magnitude, and heliocentric distance. 

\begin{figure}\centering
    \setlength{\tabcolsep}{3pt}
    \begin{tabular}{@{}c@{\hspace{3pt}}c@{\hspace{3pt}}c@{}}
        & \textbf{PVI} & \textbf{\texttt{deep\_coadd}} \\[2pt]
        \rotatebox{90}{\small first Rubin image} &
        \includegraphics[width=0.44\linewidth,trim=262.5bp 262.5bp 262.5bp 262.5bp,clip]{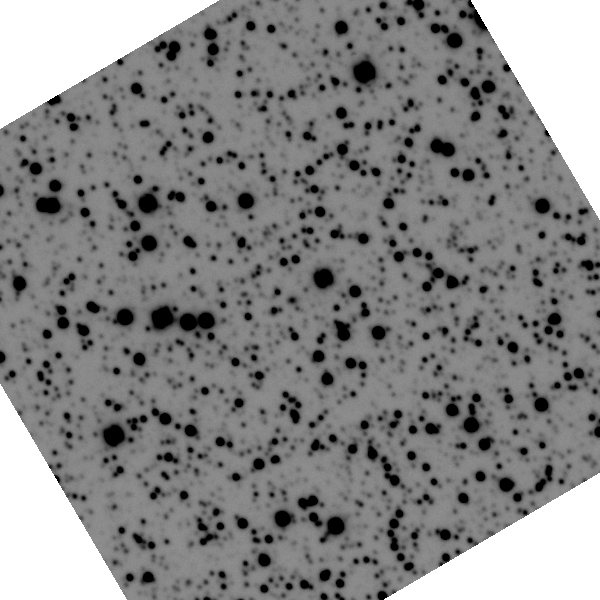} &
        \includegraphics[width=0.44\linewidth,trim=262.5bp 262.5bp 262.5bp 262.5bp,clip]{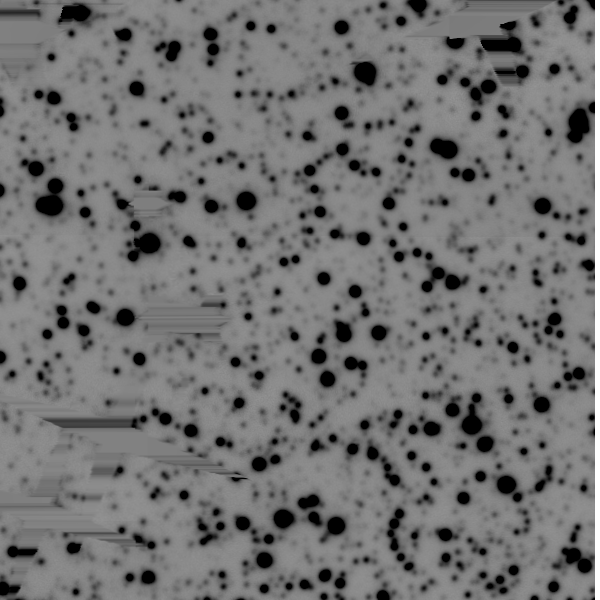} \\[3pt]
        \rotatebox{90}{\small contamination identified} &
        \includegraphics[width=0.44\linewidth,trim=262.5bp 262.5bp 262.5bp 262.5bp,clip]{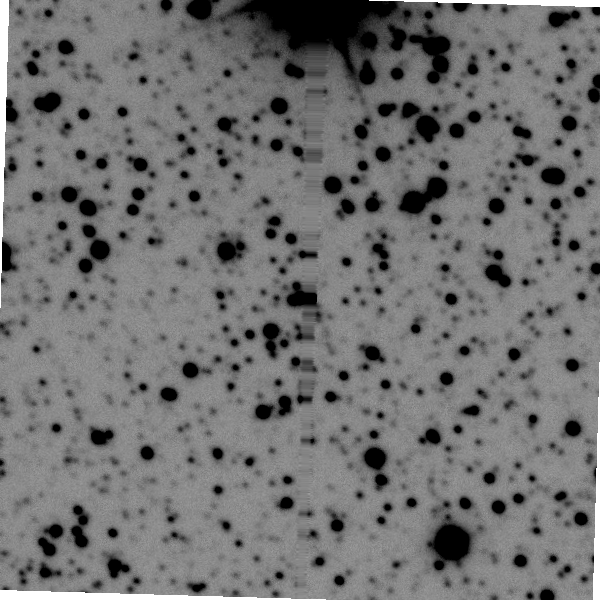} &
        \includegraphics[width=0.44\linewidth,trim=262.5bp 262.5bp 262.5bp 262.5bp,clip]{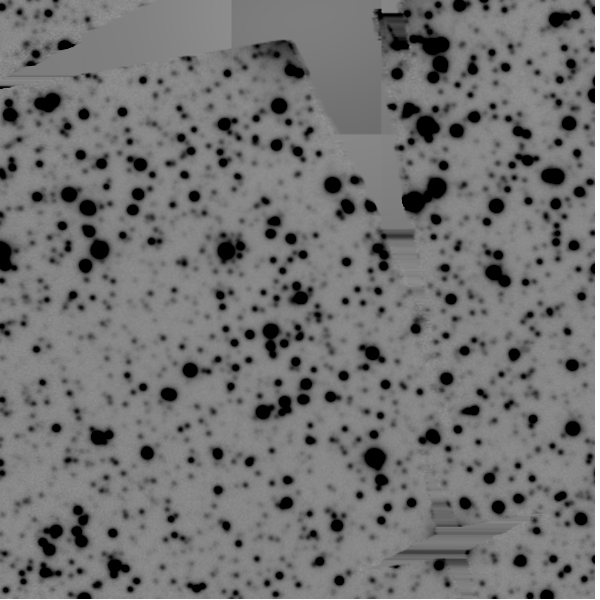} \\
    \end{tabular}
    \caption{Centered eightfold zooms ($75\times75$ pixels, $15\arcsec\times15\arcsec$) of two 3I/ATLAS comparison stages in Rubin \ac{EDP2} imaging. The left column shows the \ac{PVI} products and the right column the corresponding \texttt{deep\_coadd} products. The upper row is the first Rubin image of 3I/ATLAS (2025-06-21 08:11:32), and the lower row represents a stage demonstrating contamination identification (2025-06-22 02:32:47). The timestamps are predicted source-visit epochs in TAI (not UTC), the standard time system for Rubin; they are selection/provenance times and do not timestamp the \texttt{deep\_coadd} itself. 
    \label{fig:3i-gallery}}
\end{figure}

For each \cometL{3I} (RA, Dec) coordinate and corresponding \textit{ugrizy} band, we queried the \ac{RSP} \ac{SIA} service to find overlapping \texttt{lsst.deep\_coadd} image products (\texttt{calib\_level=3}), and tried to select the specified band. If the band was not available, we tried (in order) \textit{rigzyu} -- the expected \added{order for} brightest-to-faintest flux yield for a typical small Solar System body (Kurlander et al., in preparation). We applied this cascading alternate selection only for \cometL{3I}, as a feature of our comparison image tool (Figure \ref{fig:3i-gallery}). The \ac{SIA} query identified the available tract/patch datasets \added{(tracts are $\sim$1.66\degr\ per side sky tiles; each is divided into 100 slightly overlapping patches, and those are the size of the  \texttt{deep\_coadd} images that are produced)}, and a single (RA, Dec) position could return multiple overlapping patches.

\begin{figure*}[ht!]
\centering
\setlength{\tabcolsep}{0pt}\renewcommand{\arraystretch}{0}\begin{tabular*}{\linewidth}{@{\extracolsep{\fill}}ccccc}
\labelpicA{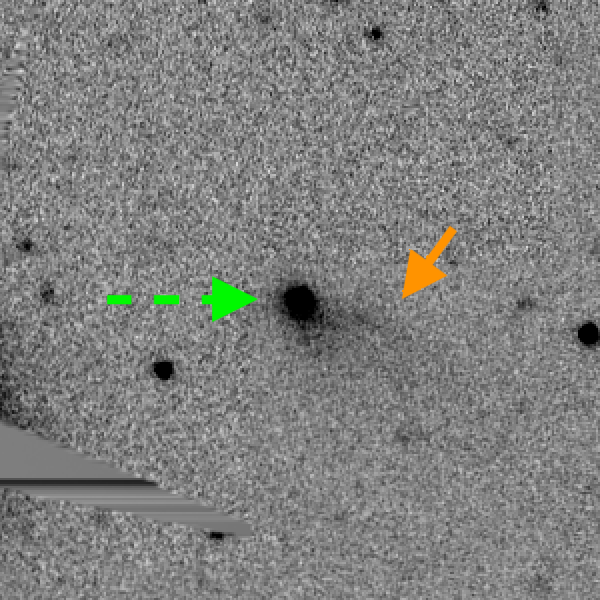}{(a)}{{\scriptsize 2008 QZ44}}{0.19}{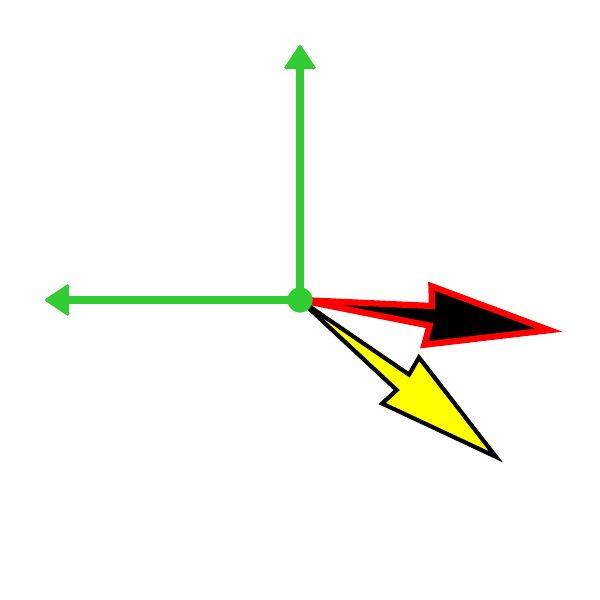}
&
\labelpicA{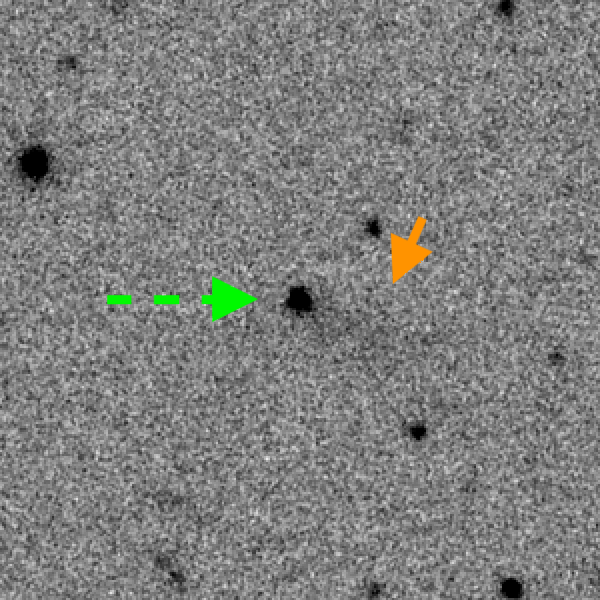}{(b)}{{\scriptsize 2025 NC$_6$}}{0.19}{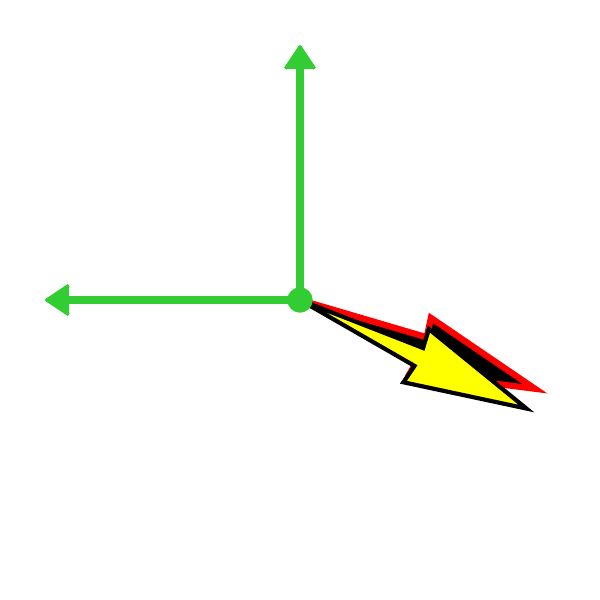}
&
\labelpicA{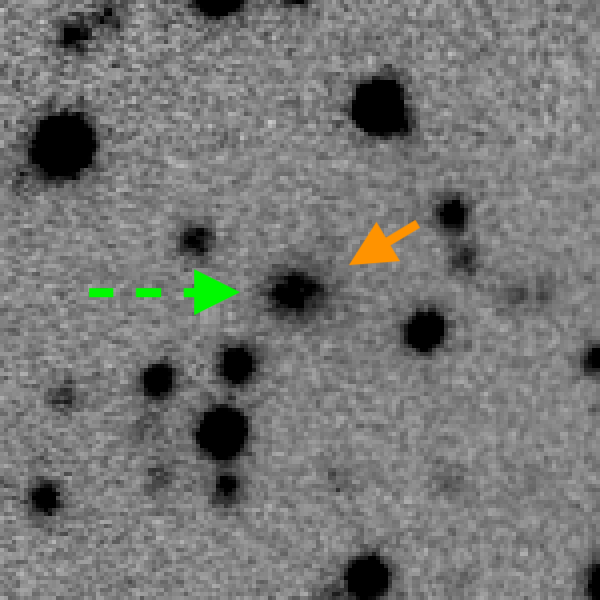}{(c)}{3I}{0.19}{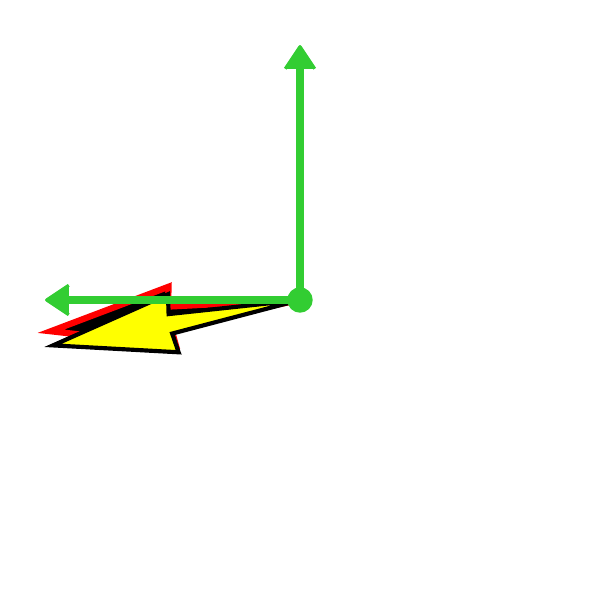}
&
\labelpicA{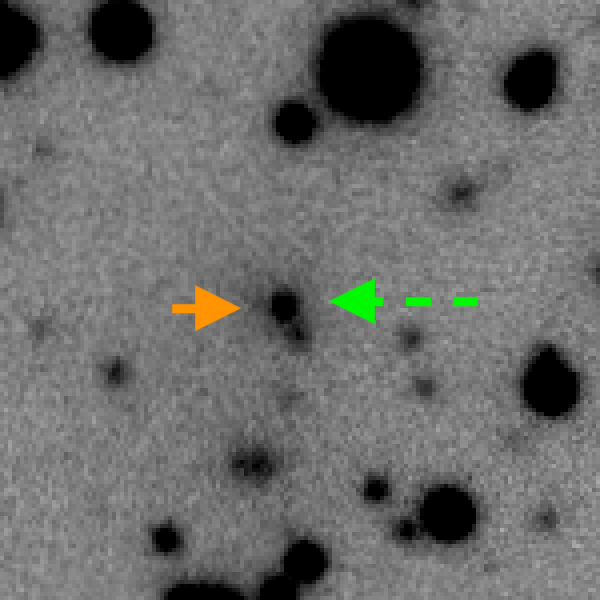}{(d)}{12P}{0.19}{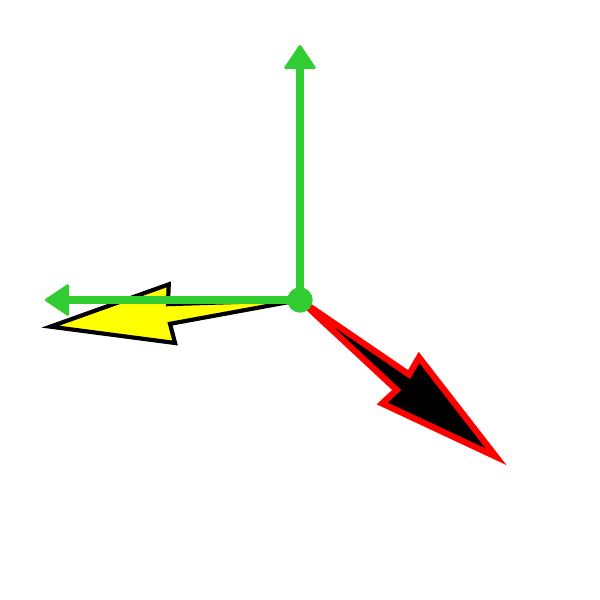}
&
\labelpicA{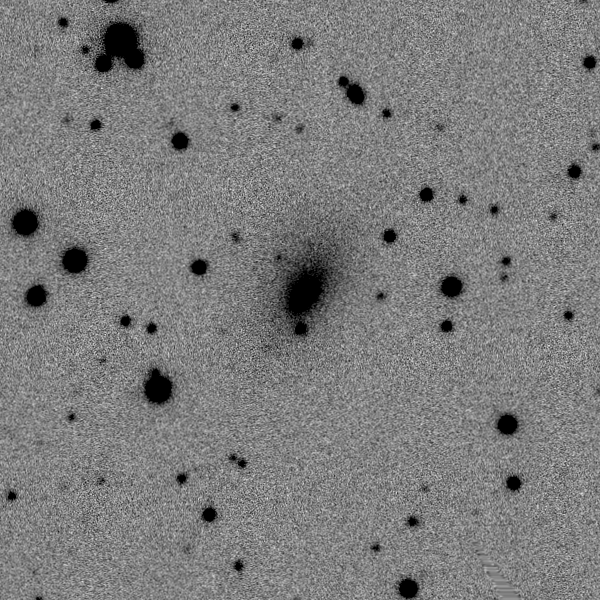}{(e)}{13P}{0.19}{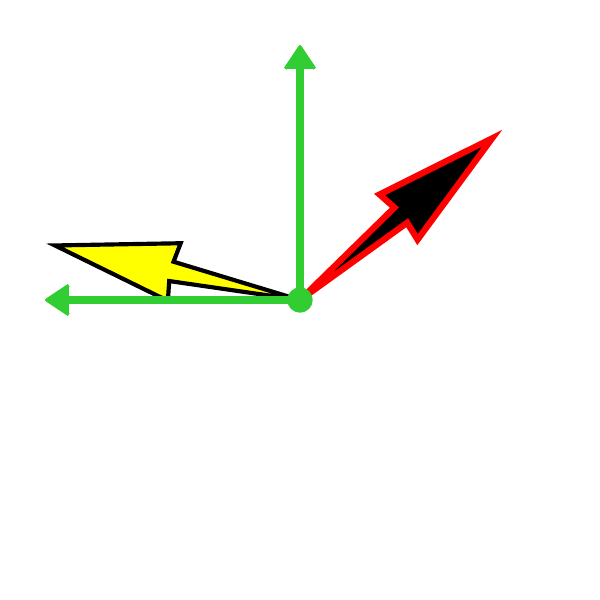} \\[0.0144\linewidth]
\labelpicA{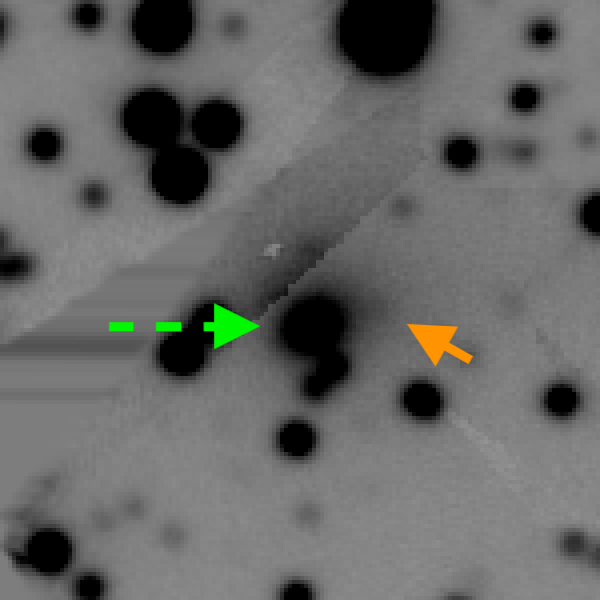}{(f)}{65P}{0.19}{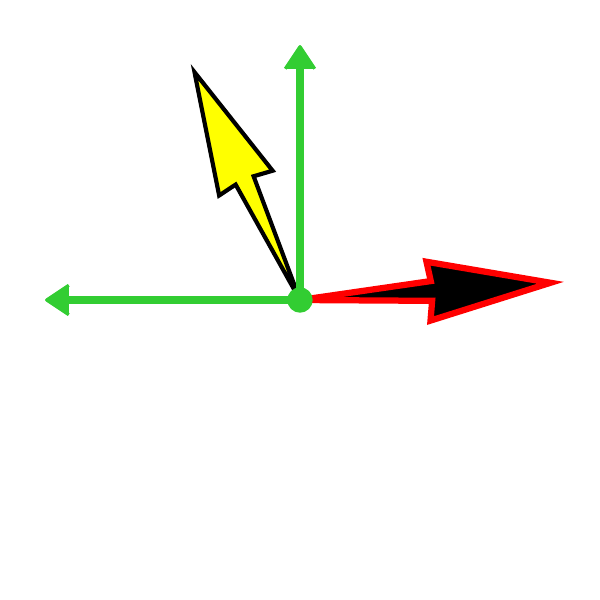}
&
\labelpicA{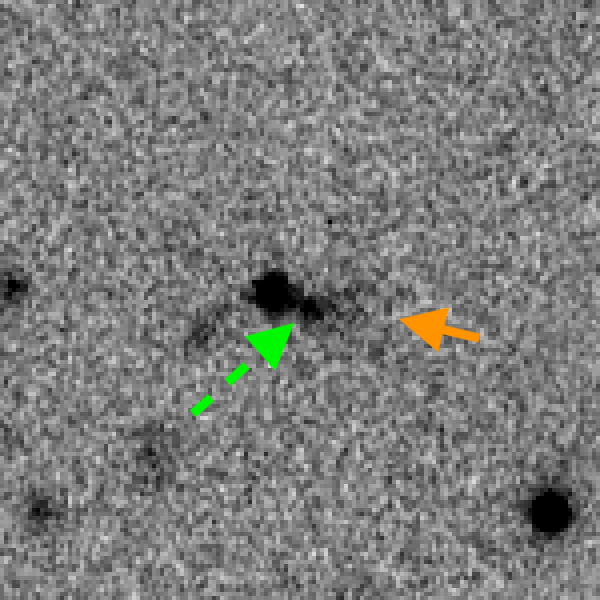}{(g)}{77P}{0.19}{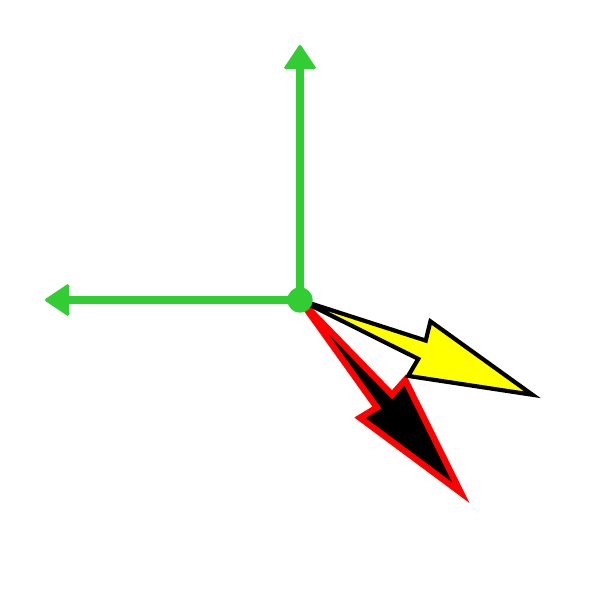}
&
\labelpicA{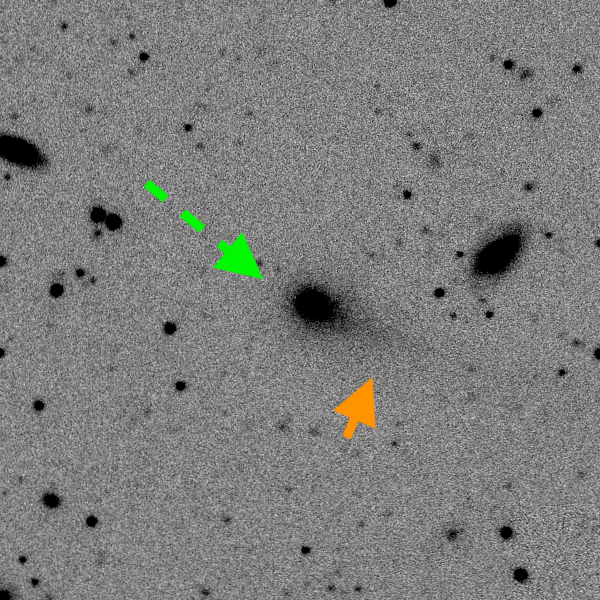}{(h)}{78P}{0.19}{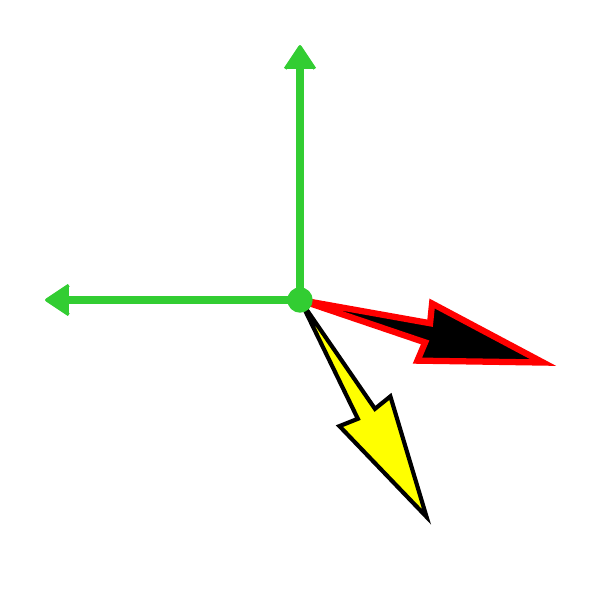}
&
\labelpicA{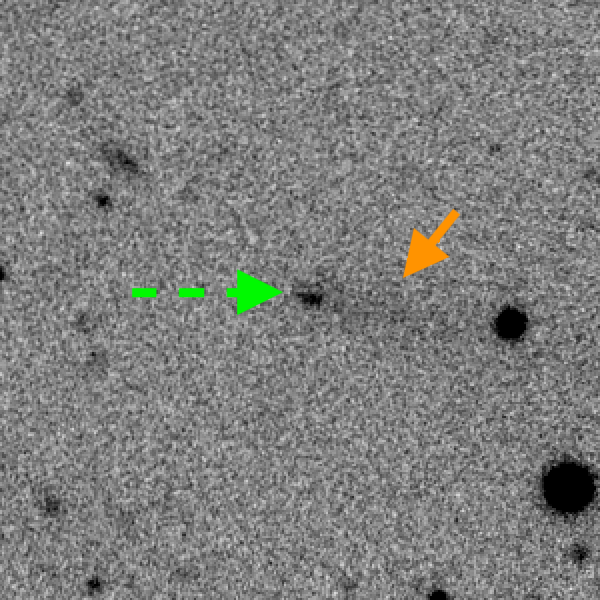}{(i)}{99P}{0.19}{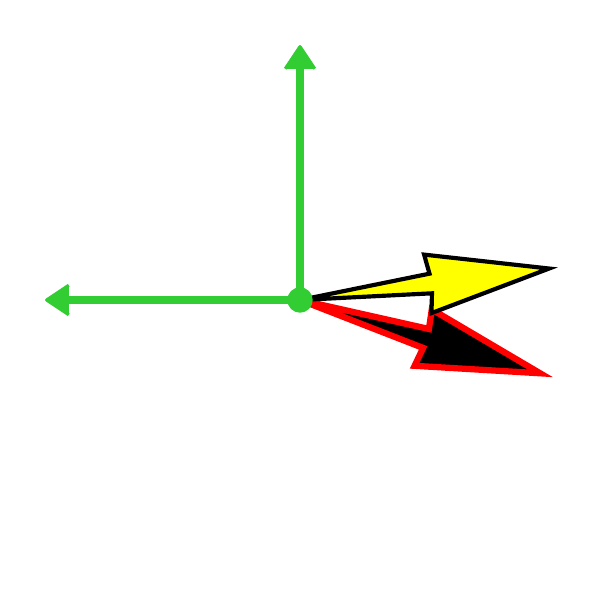}
&
\labelpicA{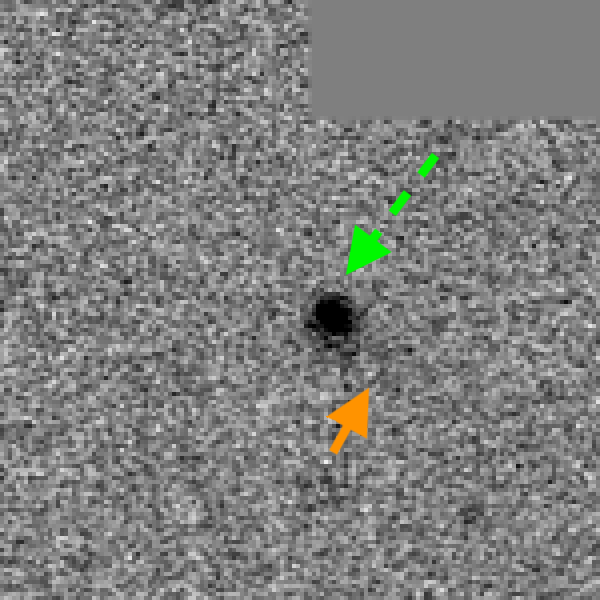}{(j)}{131P}{0.19}{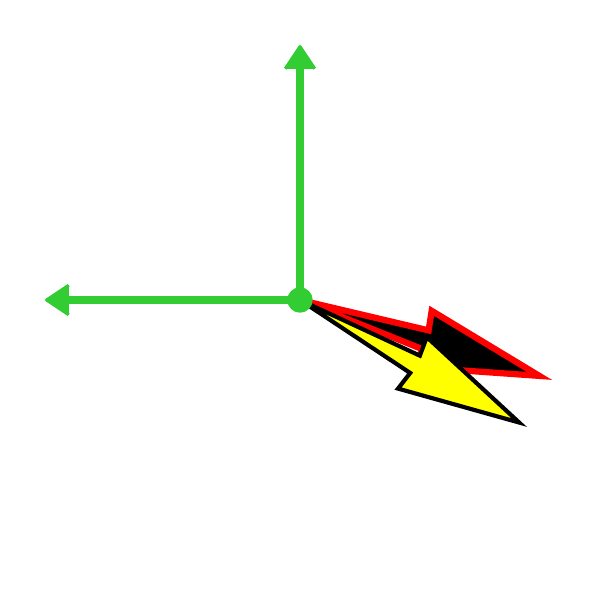} \\[0.0144\linewidth]
\labelpicA{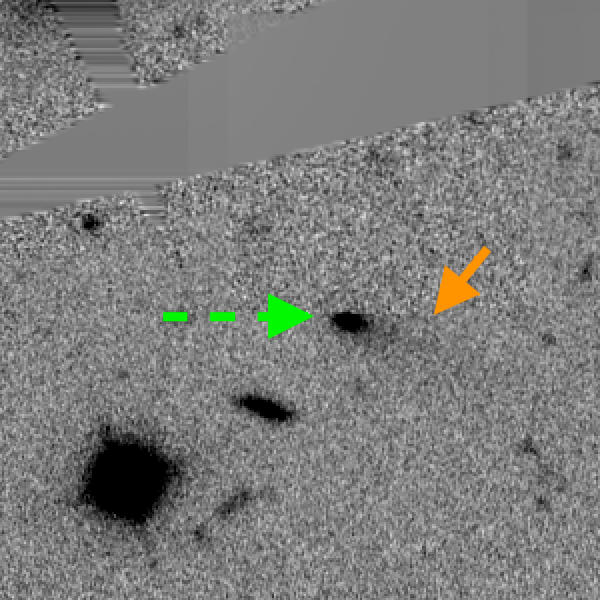}{(k)}{188P}{0.19}{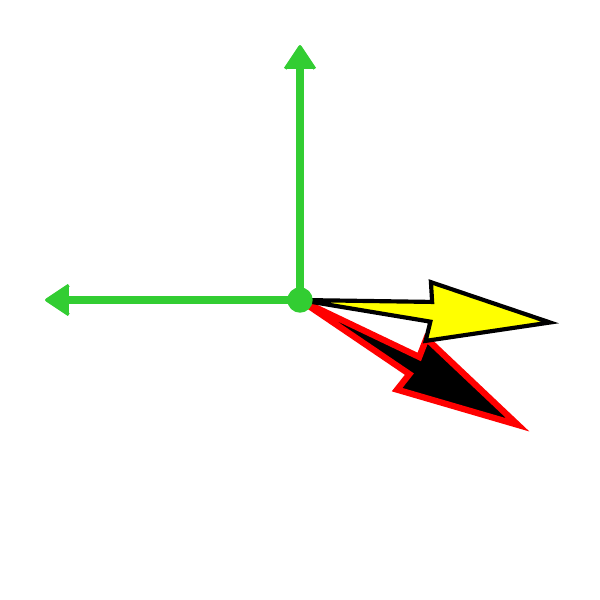}
&
\labelpicA{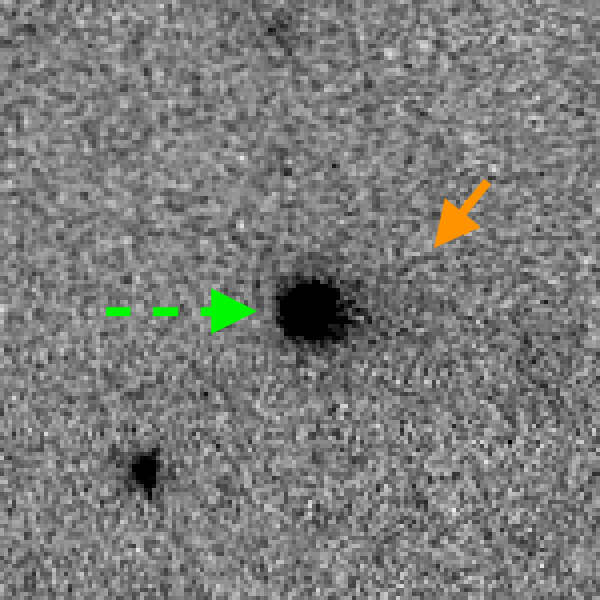}{(l)}{243P}{0.19}{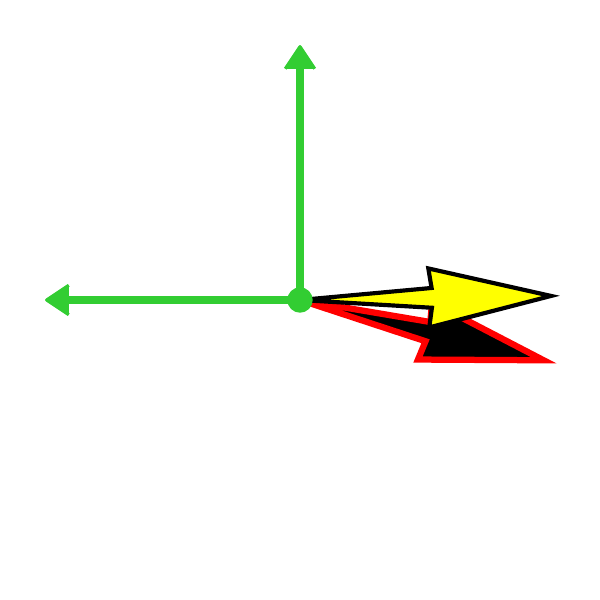}
&
\labelpicA{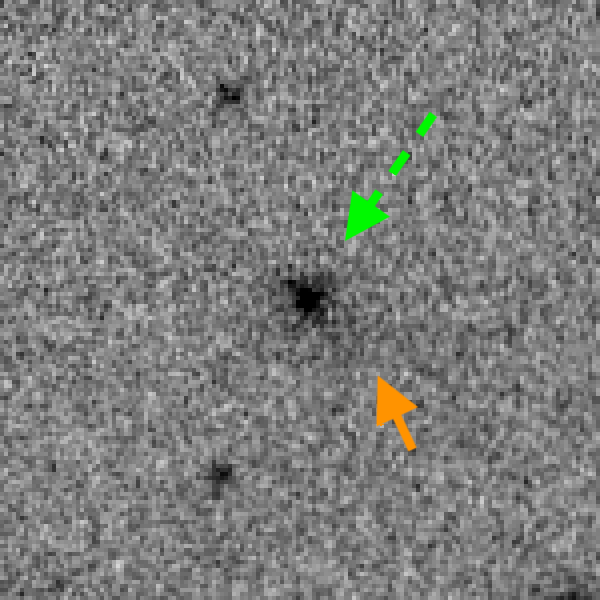}{(m)}{261P}{0.19}{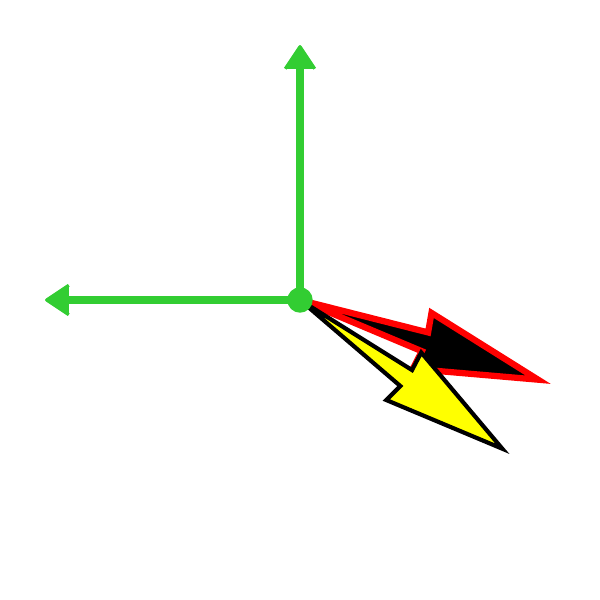}
&
\labelpicA{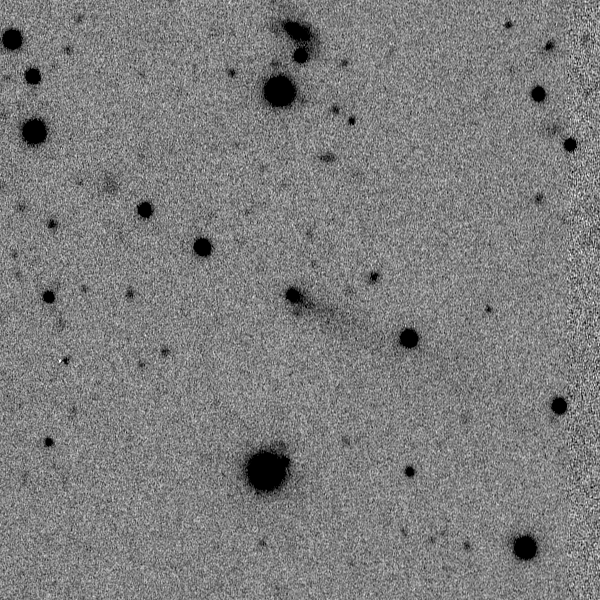}{(n)}{299P}{0.19}{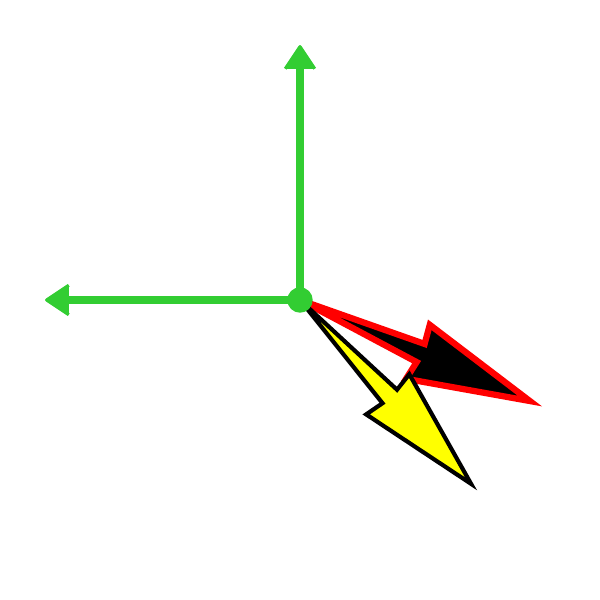}
&
\labelpicA{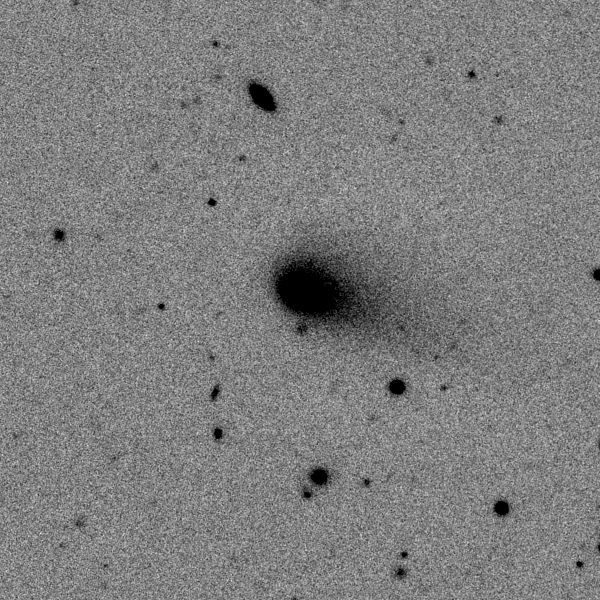}{(o)}{302P}{0.19}{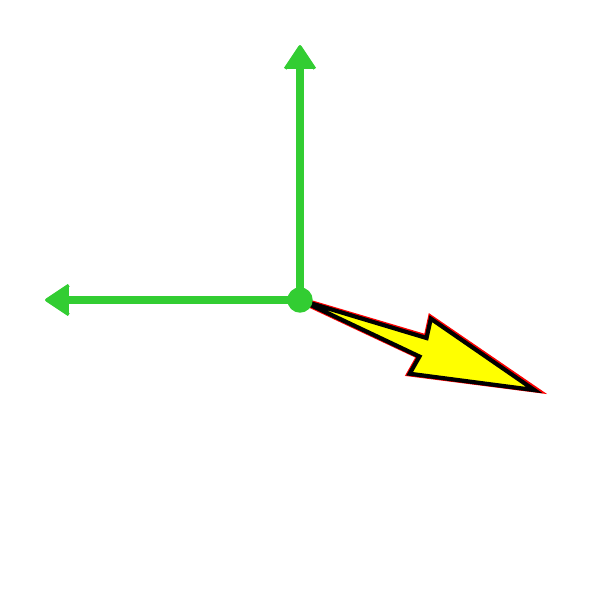} \\[0.0144\linewidth]
\labelpicA{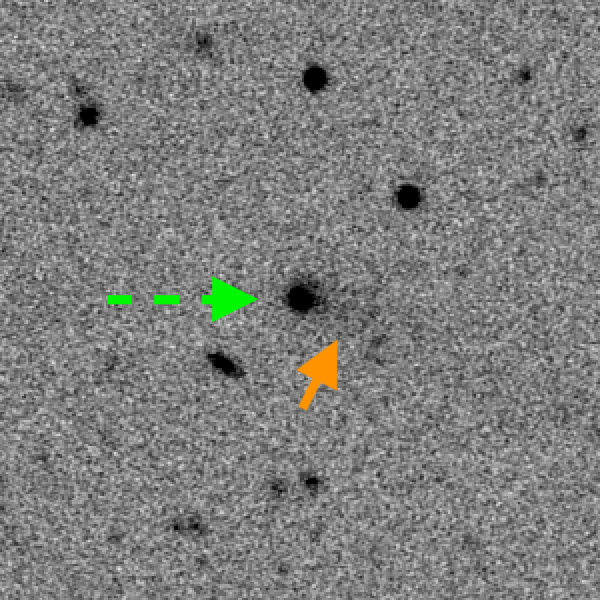}{(p)}{306P}{0.19}{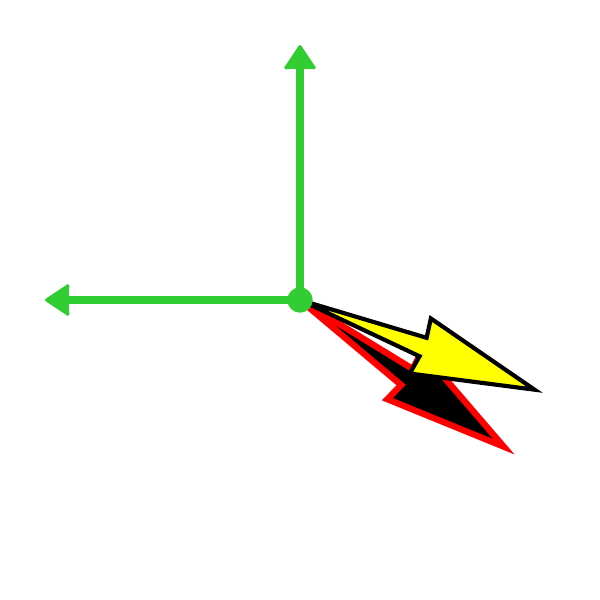}
&
\labelpicA{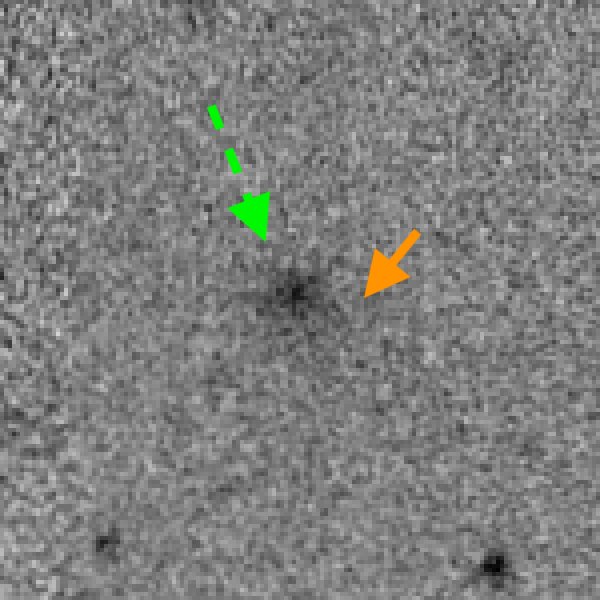}{(q)}{351P}{0.19}{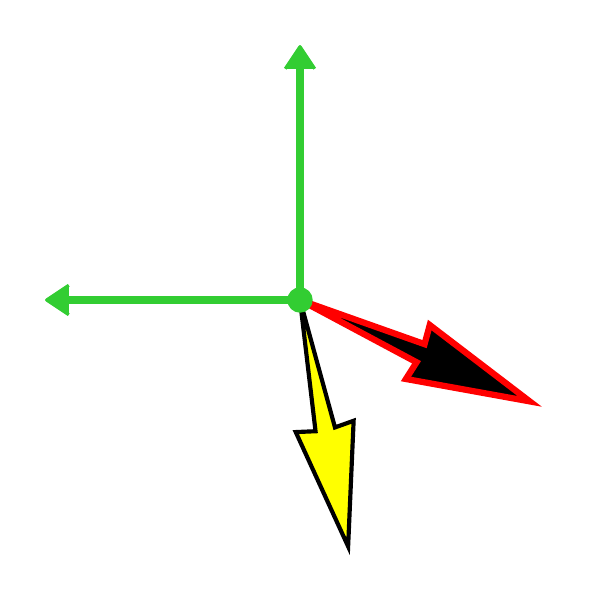}
&
\labelpicA{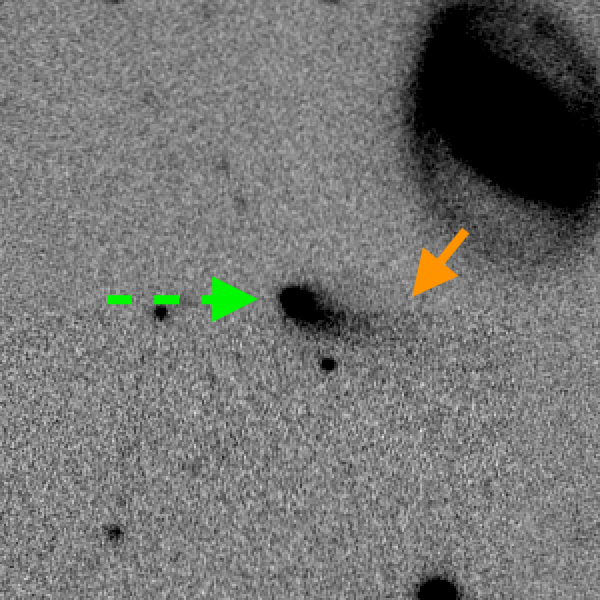}{(r)}{441P}{0.19}{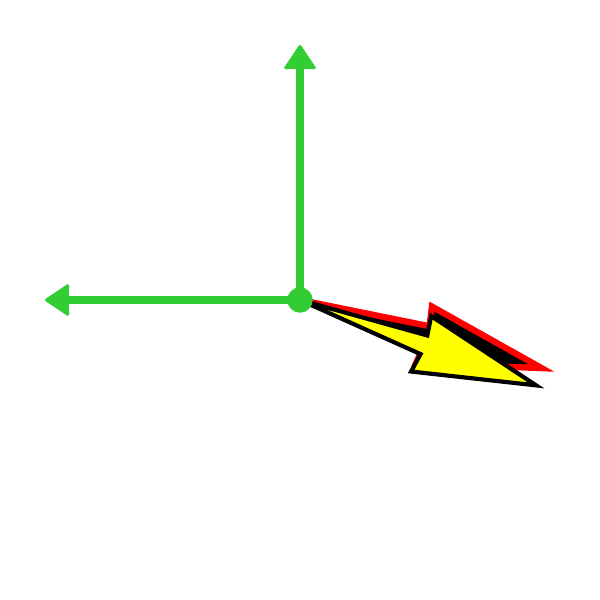}
&
\labelpicA{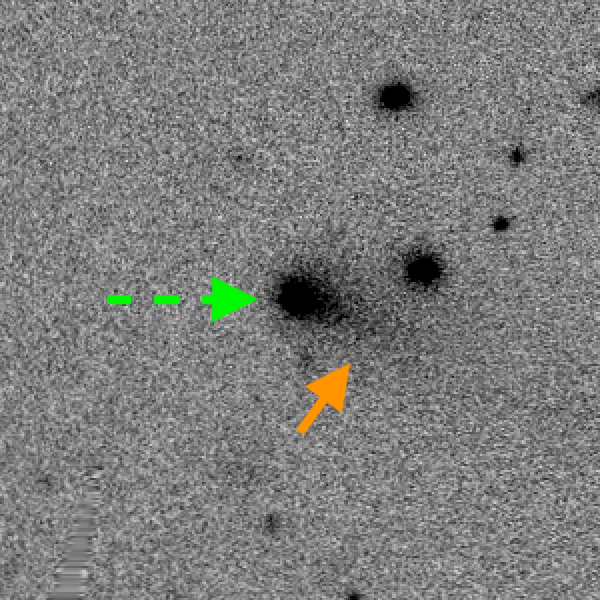}{(s)}{486P}{0.19}{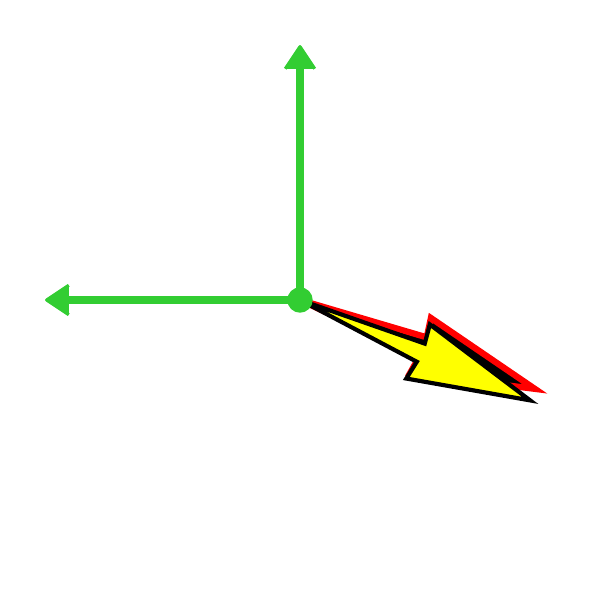}
&
\labelpicA{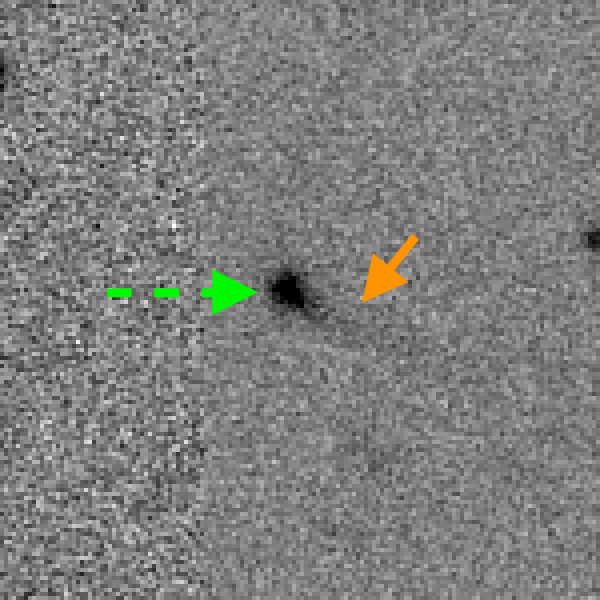}{(t)}{510P}{0.19}{} \\[0.0144\linewidth]
\labelpicA{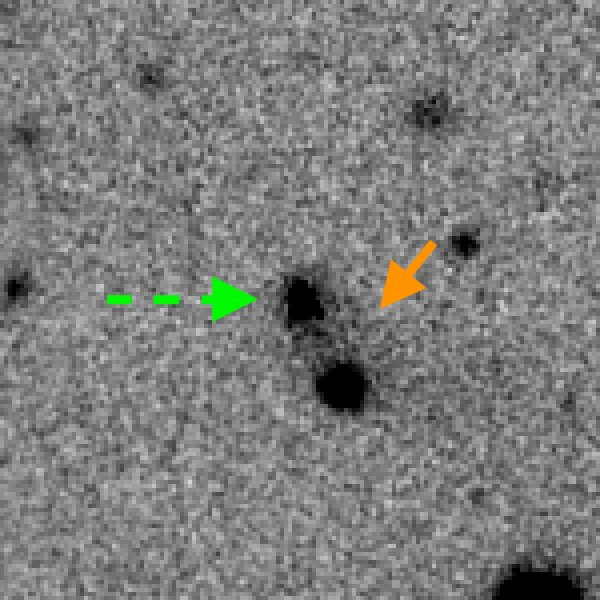}{(u)}{{\scriptsize P/2016 P5}}{0.19}{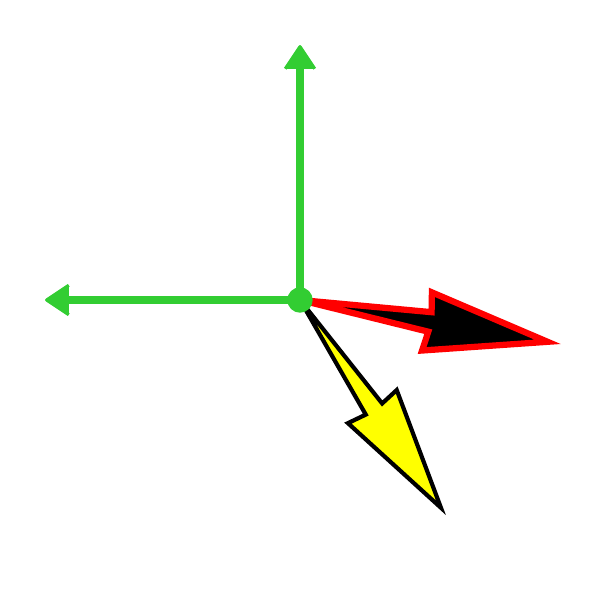}
&
\labelpicA{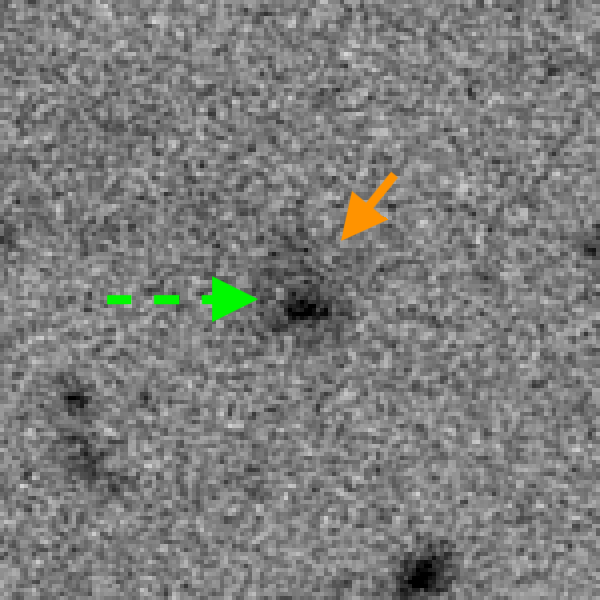}{(v)}{{\scriptsize C/2021 Q6}}{0.19}{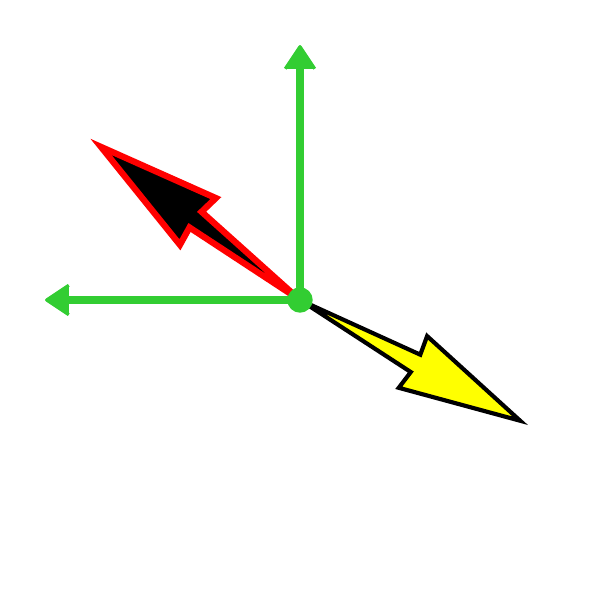}
&
\labelpicA{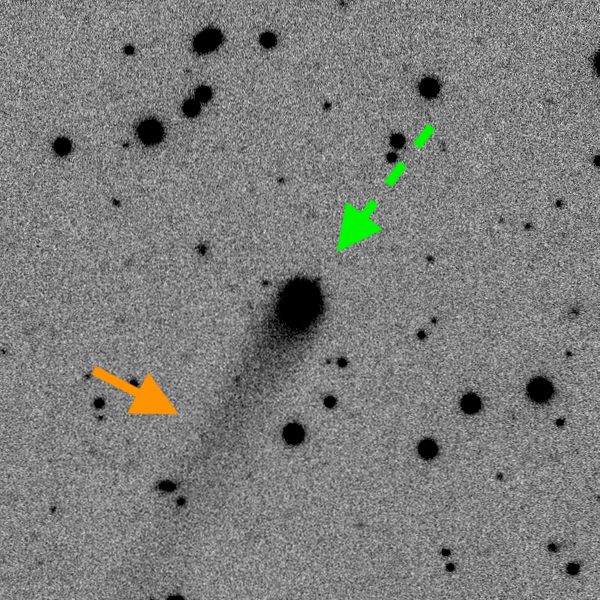}{(w)}{{\scriptsize C/2024 G6}}{0.19}{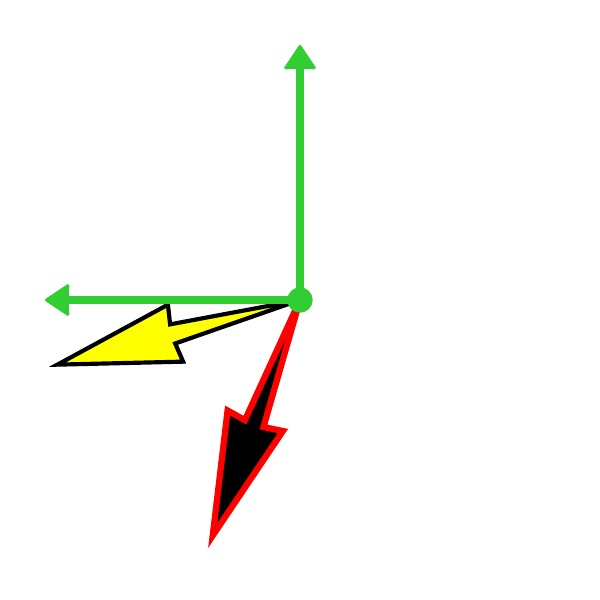}
&
\labelpicA{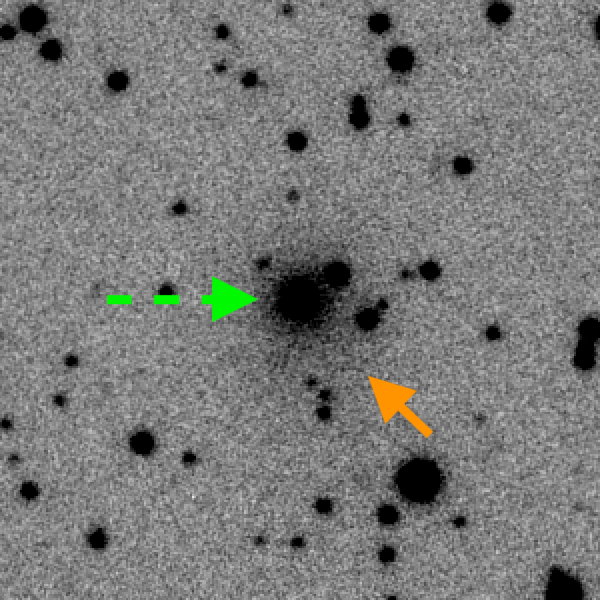}{(x)}{{\scriptsize C/2024 J3}}{0.19}{}
&
\labelpicA{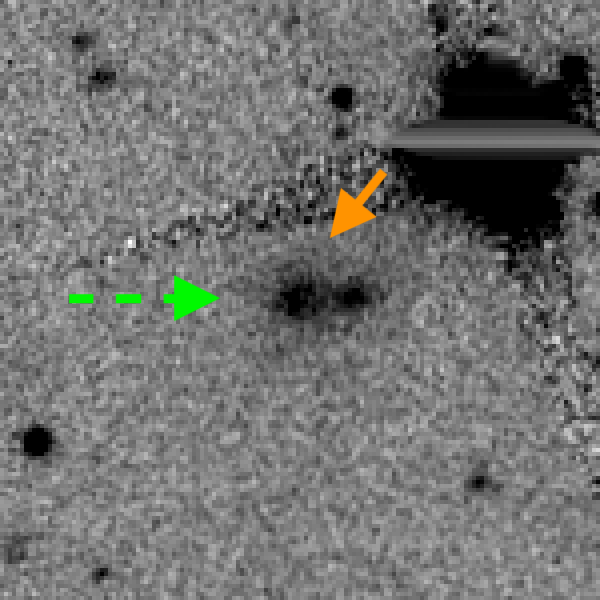}{(y)}{{\scriptsize C/2025 M1}}{0.19}{}
\end{tabular*}
\caption{Ephemeris-centered $600\times600$ pixel ($120\arcsec\times120\arcsec$) \acf{EDP2} cutouts of activity score $\ge5$ objects.
Yellow arrows show the anti-solar direction, black arrows with red outlines show the negative heliocentric-velocity direction, green arrows show north (up) and east (left). Dashed green markers indicate the object; solid orange markers highlight visible cometary activity. The source-visit dates, \ac{TAI} times, and \added{\ac{LSST}} bands \added{for each object} are: (a) 2008 QZ$_{44}$, 2025-07-20 05:08:10, $g$; (b) 2025 NC$_{6}$, 2025-07-20 06:16:00, $g$; (c) 3I/ATLAS, 2025-07-20 04:15:08, $u$; (d) 12P/Pons-Brooks, 2025-07-29 03:06:48, $r$; (e) 13P/Olbers, 2025-07-12 23:51:44, $g$; (f) 65P/Gunn, 2025-07-04 06:07:41, $r$; (g) 77P/Longmore, 2025-07-10 07:51:42, $i$; (h) 78P/Gehrels, 2025-07-20 05:11:29, $g$; (i) 99P/Kowal 1, 2025-07-20 08:47:37, $g$; (j) 131P/Mueller 2, 2025-07-20 06:20:32, $g$; (k) 188P/LINEAR-Mueller, 2025-07-22 05:04:11, $g$; (l) 243P/NEAT, 2025-07-20 08:20:02, $g$; (m) 261P/Larson, 2025-07-20 02:44:25, $u$; (n) 299P, 2025-07-24 05:45:12, $g$; (o) 302P/Lemmon-PANSTARRS, 2025-07-23 08:13:17, $r$; (p) 306P/LINEAR, 2025-07-23 08:19:01, $r$; (q) 351P/Wiegert-PANSTARRS, 2025-07-20 02:49:08, $u$; (r) 441P/PANSTARRS, 2025-07-20 05:19:40, $g$; (s) 486P/Leonard, 2025-07-08 08:30:13, $z$; (t) 510P/Boattini, 2025-07-19 06:03:41, $g$; (u) P/2016 P5 (COIAS), 2025-07-18 02:24:02, $g$; (v) C/2021 Q6 (PANSTARRS), 2025-07-24 05:41:42, $g$; (w) C/2024 G6 (ATLAS), 2025-07-20 03:11:32, $g$; (x) C/2024 J3 (ATLAS), 2025-07-19 01:56:22, $u$; and (y) C/2025 M1 (PANSTARRS), 2025-07-14 03:09:02, $r$. Panels (e), (h), (n), (o), and (w) are shown at full size; the others are centered crops ($2\times$ or $4\times$ zoom). Visit and detector values are selection provenance and do not identify the coadd itself. 
\label{fig:comet-gallery}}
\end{figure*}

With the returned dataset's \texttt{DataLink} we query the \ac{SODA} cutout service\footnote{\url{https://dp2.lsst.io/tutorials/api/api-102-3.html}}. The coordinates we queried were the predicted positions of \cometL{3I} at specific times, so deep coadded images from other visits/times should not show the object. The object was, however, plainly visible in multiple images (Figure \ref{fig:comet-gallery}), indicating the presence of shallow regions of \ac{EDP2} deep coadds. \ac{SODA} can return the image pixels + \ac{WCS} + primary \ac{HDU} in \texttt{cutout-sync} mode, which is the fastest mode and sufficient to produce cutout images (see Section \ref{subsec:cutouts}) suitable for human examination.

\subsection{Orbit Catalogs and Sorcha Searches (RSP)}
\label{subsec:orbitcatalogs} \label{sorchasearchs}

Here we describe our process for identifying and downloading the necessary input catalogs, running simulations to indicate where small solar system bodies may appear in the dataset, and applying filters to increase the odds of finding active and trailed sources while optimizing performance. We also mention \texttt{Ponder}\footnote{\url{https://github.com/lincc-frameworks/ponder}}, our orchestrator for the \texttt{Sorcha} survey simulator. \texttt{Ponder} is currently under development and will be fully described in a later publication. \added{\texttt{Sorcha} generates ephemerides with the \texttt{ASSIST} extension \citep{2023PSJ.....4...69H,hanno_rein_2023_7778017} for the \texttt{REBOUND} $N$-body code \citep{reinREBOUNDOpensourceMultipurpose2012}, which uses the \texttt{IAS15} integrator \citep{reboundias15} to integrate objects in the field of the Sun, Moon, planets, and $16$ massive asteroids.}

\subsubsection{Input Datasets}\label{subsubsec:inputdatasets}\label{sec:maglims}

We used \texttt{Ponder} to produce a Butler-derived ``visit database'' (similar to the previously distributed database\footnote{\url{https://survey-strategy.lsst.io/progress/prelsst/prelsst_dp2.html}}) in the format needed for \texttt{Sorcha}; columns include pointing information such as date and time, boresight RA and Dec, and broadband filter. 
We downloaded the {\tt MPCORB} (minor planet orbit catalog) and {\tt AllCometEls} (comet catalog) files on UT 2026 August 27 from the \ac{MPC} \textit{Data} webpage\footnote{\url{https://www.minorplanetcenter.net/data}}.

Using the visit database as input, we used \texttt{Ponder} to run two \texttt{Sorcha} survey simulations (one each for minor planets and comets). We imposed a limiting apparent magnitude of 26 across all bands as a starting point \added{that balances the compute cost of running the survey simulator against the expected single-visit image depth}; \added{nominal} single-visit image depth is shallower \added{($u\simeq23.8$, $g\simeq24.5$, $r\simeq24.0$, $i\simeq23.4$, $z\simeq22.7$, $y\simeq22.0$)}  \citep{jones_smtn-002_2021,bianco_pstn-054_2022,schwambRecommendationsVeraRubin2025}. We did not consider the dataset viable for recovering exceptionally faint sources via, for example, shift-and-stack tools. We disabled signal-to-noise, saturation, fading, linking, vignetting, trailing-loss, and randomization cuts, and applied no phase function or cometary activity model. We do not pre-filter to account for chip boundaries or the gaps between them in our circular footprint. Thus, the output is a catalog of predicted locations, not a claim of activity, detection, or detectability. 

The resulting catalog contained each object's observation time, filter (band), predicted RA and Dec, visit ID, and detector. After the full comet run, we carried out population-level simulations for \acp{NEO} by limiting \texttt{MPCORB} to objects with perihelion distance $q=a(1-e) < 1.3~\mathrm{au}$. Later, we filtered to a subset of known active objects that are not included in the \ac{MPC} \texttt{AllCometEls} file (e.g., \comet{133P}, 2008 QZ$_{44}$), plus 2025 NX$_{200}$, 2025 MT$_{356}$ (Section \ref{subsec:rediscovery}), and 2025 NC$_6$ (Section \ref{subsec:2025N6}) as active in a later screening of the general \texttt{MPCORB} dataset. Table \ref{tab:comet-results} shows a sample of the resulting output; the full tabular data is included with the online edition of this manuscript.

\subsection{Image Products and Source-visit Provenance}\label{subsec:provenance}\label{subsec:image_proucts}

Our catalog of ephemerides produced by
\texttt{Sorcha} represents predicted positions for small solar system bodies at different times, which correspond to unique visits (camera exposures) by Rubin's LSSTCam. The matching image products available to data-rights holders are \texttt{lsst.deep\_coadd} images, rather than per-visit products (e.g., \texttt{lsst.visit\_image}, \texttt{lsst.preliminary\_visit\_image}). Our catalog visit, detector, position, and \ac{TAI} (the native time system in Rubin data) epoch are thus predicted target observations, not unique identifiers or timestamps that necessarily correlate to an \texttt{lsst.deep\_coadd}.

\begin{figure}
    \centering
    \includegraphics[width=1\linewidth]{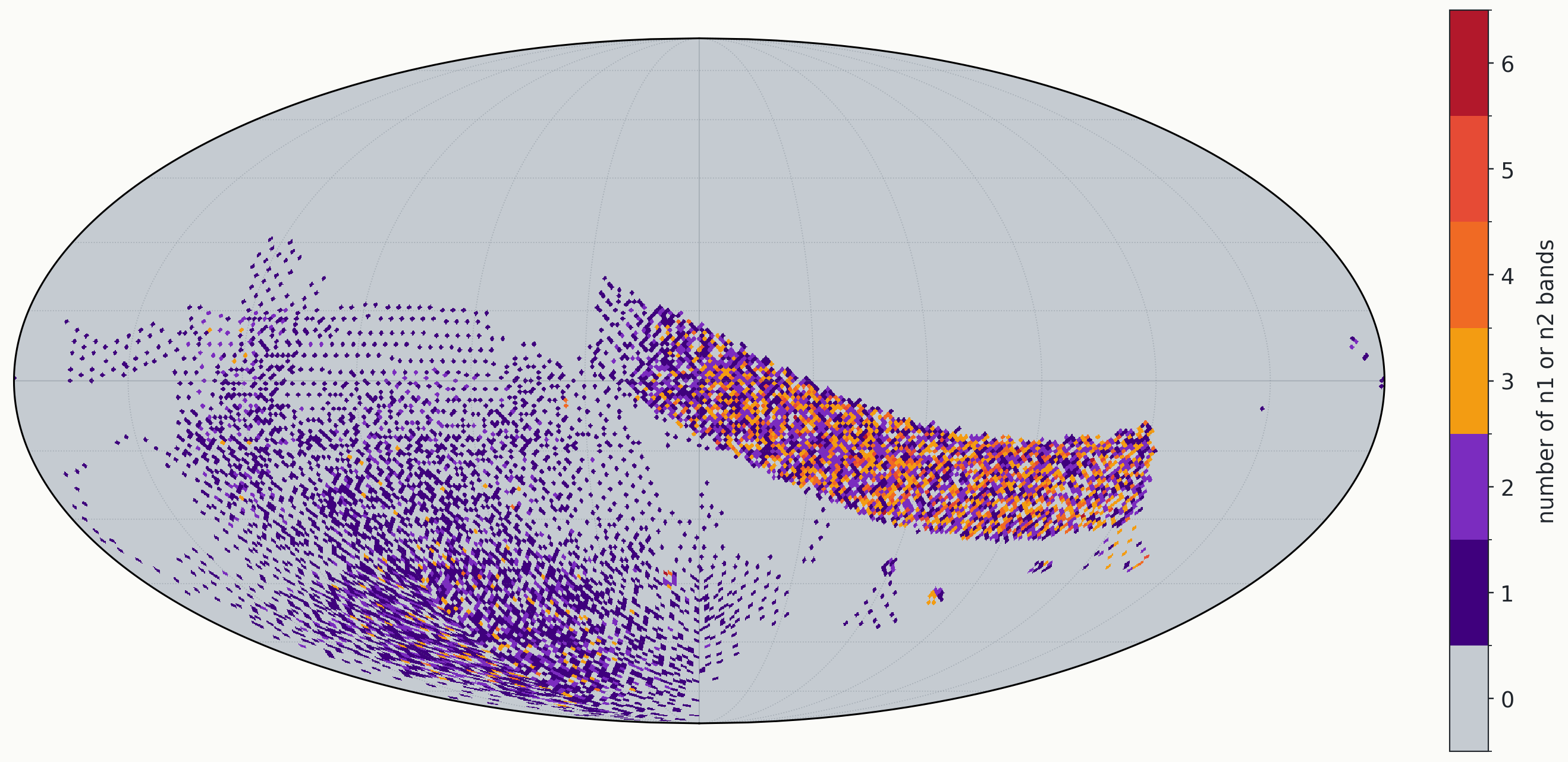}
    \caption{
    Sky coverage with shallow coverage of the Rubin \ac{EDP2} dataset. 
    Each bin on the sky represents the number of bands with shallow (\texttt{n1} or \texttt{n2}) area, ranging from 0 (no coverage) to 6 (all six of the $ugrizy$ bands were present with shallow data).
    }
    \label{fig:skycoverage}
\end{figure}

To reduce service queries to the cutout service, we apply a sequence of increasingly specific filters before retrieving image pixels. First, we sample the local consolidated exposure-time map, \texttt{\detokenize{deepCoadd_exposure_time_consolidated_map_sum}}, using the Butler and direct \texttt{HealSparse} \citep{rykoffHealSparseSparseRepresentation2026,verac.rubinobservatoryteamVeraRubinObservatory2026} access\added{; \texttt{HealSparse} provides a sparse representation of maps in the Hierarchical Equal Area isoLatitude Pixelization (HEALPix)\footnote{\url{http://healpix.sf.net}} scheme \citep{2005ApJ...622..759G}}. We divide the accumulated exposure time by the nominal exposure time (38~s in $u$, 30~s in $grizy$) to estimate the number of contributing exposures. We reject only positions that are sufficiently deep (e.g., $\ge$ \texttt{n3}) and covered by exactly one coadd footprint. We retain unknown, borderline, and overlapping cases because the consolidated map may represent a deeper coadd than an overlapping coadd that qualifies as \texttt{n1} or \texttt{n2}. This initial screen typically reduced the number of candidates by 40\% to 60\%.

Second, after \ac{SIA} identifies candidate \texttt{\detokenize{lsst.deep_coadd}} products, we query a local \texttt{\detokenize{coadd_inputs.parquet}} export of the Butler \texttt{\detokenize{deep_coadd_input_summary}}. This provides a conservative, patch-level visit/membership screen; we reject a candidate only when the predicted visit is known to be absent from its tract, patch, and band. Candidates with missing metadata or failed look-ups proceed to exact provenance. This phase reduces the number of candidates by an additional 10\% to 20\%.

For surviving candidates, combined-exposure retrieval returns both the serialized \texttt{CellCoadd} and its image pixels. We inspect the \texttt{CellCoadd} \texttt{PROVENANCE/CONTRIBUTIONS} table to identify source visits contributing within 5\arcsec of the predicted position; \added{this represents the ideal case (when the object is where it is predicted to be), as we cannot know \textit{a priori} if the object is off-center (or by how much) for objects with positional uncertainties higher than this nominal radius}. We retain only candidates with a predicted visit and a local source+visit count of one or two. Retrieving provenance and pixels together avoids a separate provenance request for the same candidate, although additional image fragments may still be required when the requested cutout crosses patch boundaries. Roughly one in six queries resulted in a viable cutout image.

\begin{figure*}
    \centering
    \begin{tabular}{cc}
        \includegraphics[width=0.45\linewidth]{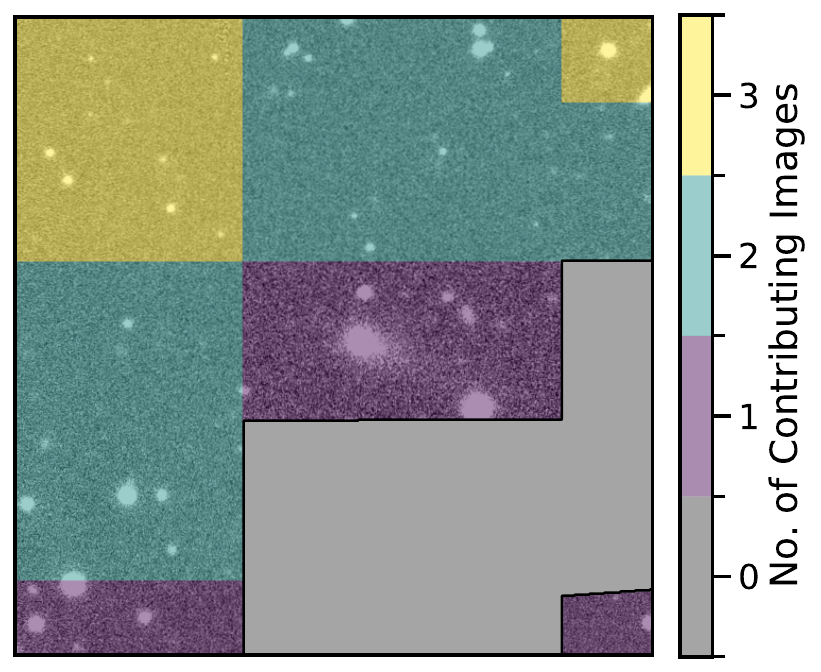} & \includegraphics[width=0.45\linewidth]{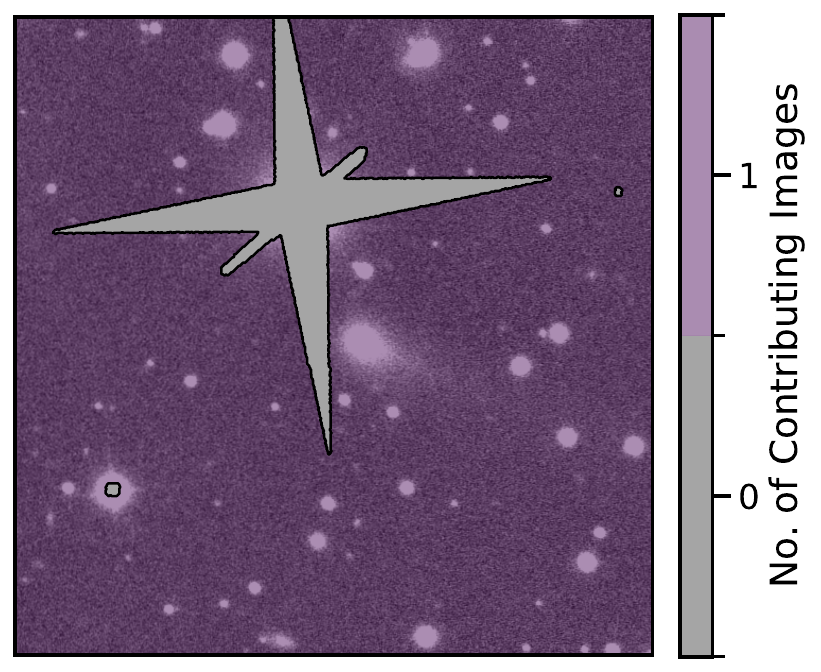} \\
    \end{tabular}
    \caption{78P/Gehrels (centered) ``shallow'' \texttt{deep\_coadd} cutout images from Rubin \ac{EDP2}. Both images are 600$\times$600~pixels (120\arcsec$\times$120\arcsec), reprojected with North up and East left.  In both panels, a tail is clearly visible, oriented at a position angle of $\sim240\degr$ East of North. The object occupies an \texttt{n1} region (purple, single-image depth), with adjacent \texttt{n0} regions (gray, no image data).
    \textbf{Left:} $z$-band, with 0--3 contributing visits. 
    \textbf{Right:} $g$-band, almost entirely \texttt{n1} with diffraction spikes from a nearby bright star resulting in \texttt{n0} area. 
    }
    \label{fig:mondrian}
    \label{fig:78P}
\end{figure*}

Once we determine that a shallow \texttt{lsst.deep\_coadd} is available for the (RA, Dec) position, we produce the cutout and measure provenance directly from the \texttt{CellCoadd} contribution tables within 5\arcsec~of each predicted position. We produced final cutouts only when the \ac{SODA} cutout service supplied local (within 5\arcsec) one- and two-source visit data, respectively, which we describe as \texttt{n1} and \texttt{n2} cutouts. Missing regions are described as \texttt{n0}. \texttt{n3} and higher layers may appear in the produced cutout, as shown in Figure \ref{fig:78P}, as long as the center region is \texttt{n1} or \texttt{n2}.

The two source visits of an \texttt{n2} cutout can be separated temporally by days to weeks (e.g., 2008~QZ$_{44}$'s UT 2025 June 21 cutout combines visits from UT 2025 June 21 and UT 2025 July 18). The epoch and visit we attach to each cutout are those of the visit whose predicted position coincides with the object; we verified this correspondence for every cutout in Table~\ref{tab:comet-results}, so the moving object is present in that visit only, and the other contributing visit to the \texttt{n2} supplies background (e.g., stars, galaxies, noise) but not a second image of the object. Thus, while morphology may be approximately preserved in \texttt{n1} and \texttt{n2} cutouts, photometry is not directly usable; in a weighted-mean coadd, an object's flux is diluted by roughly the weight of the visit that contains it, and any outlier rejection applied by the pipelines during coaddition can partially clip the moving source.
Quantitative photometry \citep[e.g., dust production rate $Af\rho$;][]{1984AJAHearnAfp} from these products would require a contributing-visit correction derived from the \texttt{CellCoadd} provenance, or should be restricted to \texttt{n1} cutouts; we do not attempt it here.

\subsection{Cutouts, Rendering, and Reproducibility}\label{subsec:cutouts}

We produce 600$\times$600-pixel (120\arcsec$\times$120\arcsec~at Rubin's 0.2\arcsec pixel$^{-1}$ scale) cutouts, matching the angular extent of cutouts used for our other citizen science projects \citep{chandlerActiveAsteroidsCitizen2024}. If multiple sets of pixels are returned, we mosaic the resulting pixels. LSSTCam observes at effectively arbitrary orientation with respect to the celestial sphere, so we use a bilinear interpolation to reproject (warp) the example tiles onto the centered $600\times600$-pixel \ac{ICRS} north-up, east-left-oriented grid at 0.2\arcsec pixel$^{-1}$ and reject outputs with uncovered pixels. 

We transfer the \ac{FITS} pixels 1:1 to a \ac{PNG} canvas. To enhance contrast in the \ac{PNG} image, we sample the center $\sim75\%$ of the image and perform iterative rejection (up to 5 iterations) to exclude values outside $\pm5\sigma$ \citep{chandlerActiveAsteroidsCitizen2024}. We save the \ac{PNG} canvas 1:1 (pixel-by-pixel, i.e., no additional resampling or other pixel alteration) after applying an exact vertical row reversal so that the display orientation follows the \ac{FITS} \ac{WCS} north up east left convention (this is a \ac{FITS} array vs. \ac{PNG} array correction). We set non-finite (NaN) pixels to transparent ($\alpha=0$). \added{The \ac{PNG} images typically preserve the appearance of cometary activity fairly well (or even improve it due to the contrast enhancement) when compared to the source \ac{FITS} images, but we note that the underlying deep coaddition processing already applied image processing (e.g., warping, smoothing) that diminishes faint/subtle activity features.}

During the cutout procedure, we preserve provenance including the object name, \ac{TAI} time, coordinates, visit, detector, filter, source-visit provenance (e.g., \texttt{n1}, \texttt{n2}), \ac{WCS}, and flux units in each \ac{FITS} file. We also query the JPL Horizons service for the (UTC-converted) epochs with the Rubin Observatory's site code, X05, in order to generate the anti-solar and anti-motion overlay arrows (Figure \ref{fig:comet-gallery}) in \ac{PDF}.

\subsection{Activity Identification}\label{subsec:activityfinding}
We visually examined the \ac{PNG} cutouts centered on the ephemeris-predicted positions of known small Solar System bodies to search for visible evidence of activity. We adopted the 0--9 rubric from \cite{chandlerChasingTailsActive2022a} with slight modifications. The subjective and progressive scale is as follows: 
0: We are unable to determine if the object is in the field (e.g., no image data, crowded field, image artifacts, or a single image for which we cannot confidently identify which near-center source is the object in question). 
1: We have likely identified the object, but it is too faint to determine conclusively. 
2: We have definitely identified the object, and it appears stellar (no activity indicators whatsoever). 
3: The identified object shows some non-stellar features, but we consider these unlikely to be true indicators of activity. 
4: One or more features are present that may be activity, and likely warrant further investigation, but the feature is still ambiguous. Scores $\le4$ are considered ``inactive'' in this work. 
5: There is activity that is unambiguous, though it is a marginal detection; morphological analysis is likely impossible. Scores $\ge5$ are treated as ``active'' for the purposes of this work. 
6: Activity is present and clearly visible, but is likely insufficient to study the morphology in a single image. 
7 -- 9: These describe increasingly apparent activity, peaking at the most prominent and bright activity features.

\subsection{Transient Search}\label{sec:transientmethods}

\added{We cross-matched the 2025 transients reported to the \ac{TNS}, the \ac{ZTF} Bright Transient Survey, and the Rochester supernova list against our \ac{EDP2} coverage map. We kept objects that were bright during the \ac{EDP2} visit window and also fell on a shallow (\texttt{n1} or \texttt{n2}) region. The window spans the \ac{DP2} pointing database from 2025 April 25 to 2026 January 7 (29,756 visits; 16,790 contribute to \texttt{deep\_coadd} products). We searched for transients discovered between 2025 January 1 and 2026 February 15 and found 4,415 transients in shallow coverage-map bins (a map built from the 28,698 \ac{DP2} visits listed in the release documentation); at the time of our search (2026 October 24), 227 had already been photometrically and spectroscopically classified and verified as ``bright'' (public photometry brighter than 21.5~mag between 2025 April 1 and 2026 January 31, a slightly wider timespan window than the pointing-database).

We measured exact provenance (Section \ref{subsec:provenance}) for 47 of the objects we considered to be the best demonstrations for this manuscript (Appendix~\ref{app:transients}). For each candidate, we read the \texttt{CellCoadd} provenance table for every \texttt{deep\_coadd} patch that contained the position, which, in turn, gave the visits and detectors contributing to the cell at those coordinates.
Because \ac{EDP2} does not include single-visit images, we queried the individual visit source catalogs for the nearest (<1\arcsec) source of the \ac{TNS} position. From these results, we retrieved the \ac{PSF} magnitude when we found a source; otherwise, we used the depth (i.e., the detector 5-sigma magnitude limit). These are direct-image \ac{PSF} magnitudes with no host subtraction, so they include any host light inside the \ac{PSF} footprint. Thus, for a transient on a bright host or in a nucleus, they overestimate the transient's brightness (we flag them in Appendix~\ref{app:transients}). Our highlighted examples have hosts that are either somewhat separated ($\gtrsim$1\arcsec) or $\gtrsim$2~mag fainter than the supernova, and their single-visit magnitudes agree with the public difference-imaging light curves to $\le$0.2~mag (Section~\ref{sec:supernovae}).}

\added{Unlike the comet and \ac{NEO} search, where a coadd must hold just one or two visits (otherwise the relative flux diminishes or is gone entirely), supernovae are stationary and stay bright for weeks to months, so we relaxed the visit requirements; an $n$-visit coadd holds a weighted mean of the flux over its contributing epochs, with weights set by each contributor's noise (Section \ref{subsec:provenance}). A transient at full (or nearly constant) brightness then appears at full flux whenever every contributing visit falls within an interval of nearly constant brightness, and at a fraction $k$/$n$ when only $k$ of them do, but the weights must also be considered. 
Here, ``nearly constant'' means the data fall within an interval when the flux changed slowly, such as during the few days around a supernova's peak brightness, the weeks to months after peak when most supernovae fade,
or the plateau of a Type II supernova. The coadd flux is a weighted time average, so it misrepresents lightcurves when images straddle the rapid rise
but it can remain representative long after the peak, even when the inputs span weeks (e.g., SN~2025mmt; Section~\ref{sec:supernovae}). We used the provenance table to test each band, so we deemed bands with $n>2$ acceptable when all of their epochs fell within one of these intervals.
}

\added{To show what objects were at each position before the \ac{EDP2} timespan, we produced comparison cutouts in a similar fashion to Section \ref{subec:3I}. Here we utilized only Legacy Survey DR10 data \citep{deyOverviewDESILegacy2019}, reprojecting to match the \ac{EDP2} cutout, because the longer timespan of transient events relative to our solar system cases made Rubin-only deep coadds unusable. We also stacked comparison images to increase depth when identifying potential transient hosts. We displayed these deeper stacks alongside the shallow \ac{EDP2} coadd, plus an alternate-band \ac{EDP2} coadd if one was available. The Legacy Survey imaging predates all of these transients, and its depth is comparable to (or greater than) Rubin's single-visit depth (Section \ref{sec:maglims}), so a source visible in Rubin and absent in the Legacy Survey comparison image should be new. For cases where the Legacy Survey DR10 lacked a band at a desired position, we substituted a Pan-STARRS1 stack in that band \citep{chambersPanSTARRS1Surveys2016}.}

\section{Results} \label{sec:results}

\subsection{Recovery and Activity Yield} \label{subsec:recovery_and_activity}\label{subsec:results_catalog}

In this work, we targeted active objects and \acp{NEO} with trail lengths $\ge5$~pix (on-sky motion $\gtrsim0.8^\circ$/day for the 30~s $grizy$ exposures, $\gtrsim0.6^\circ$/day for 38~s $u$-band exposures), though we produced all available \ac{NEO} cutouts. As a reminder, we pre-screened to filter out images where the object was in an area that was too deep ($n>2$). We then kept only cutouts that contained visits corresponding to our \texttt{Sorcha} catalog (Section \ref{sec:methods}). 

Of the 14\,030 potential comet detections, 7\,020 were pre-screened as too deep, \added{6,339 had no \texttt{deep\_coadd} at the desired position or the desired visit was not present}, and 631 (221 \texttt{n1}, 410 \texttt{n2}) cutouts were produced; \added{the remaining 40 cutouts had no usable image data or the center was not \texttt{n1}/\texttt{n2}}. Of the 284 known active objects not in the \texttt{AllCometEls} file, we produced 26 cutouts (7 \texttt{n1}, 19 \texttt{n2}) and did not yet have pre-screening to apply at the time. Including the two \cometL{3I} images showing activity, we identified activity in 59 of the $\sim$700 cutouts, or $\sim$9\%; Figure \ref{fig:comet-gallery} shows a sample of the active images.

We broke down the \ac{NEO} detections by orbital class. Of 12 possible Aten cutouts, 5 were too deep, and none produced cutouts. Of the 1\,705 possible Apollo cutouts, 839 were too deep, and we produced 61 cutouts (23 \texttt{n1}, 38 \texttt{n2}). Of the 6\,488 Amor-class possible cutouts, 3\,123 were too deep; we successfully made 275 cutouts (81 \texttt{n1} and 194 \texttt{n2} images). We identified 8 images (spanning 7 unique objects) with trail lengths over 50 pixels, and another 21 \added{(spanning 15 unique objects)} with lengths between 25 and 50 pixels.

\begin{deluxetable*}{lcccccrcc}
\tabletypesize{\scriptsize}
\tablecaption{Scored EDP2 Cometary Activity Observations.\label{tab:comet-results}}
\tablehead{
\colhead{Object name} & \colhead{2025 Datetime} & \colhead{Band} &
\colhead{Visit} & \colhead{Coadded visit} & \colhead{$r_H$} & \colhead{$f$} & \colhead{$\alpha$} & \colhead{Score} \\
\colhead{} & \colhead{(UTC)} & \colhead{} & \colhead{} & \colhead{} & \colhead{(au)} & \colhead{(deg)} & \colhead{(deg)} & \colhead{}
}
\startdata
2008 QZ$_{44}$ & Jun-21 06:38:59 & $r$ & 2025062000499 & 2025071700433 & 2.523 & 320.4 & 18.2 & 8 \\
2008 QZ$_{44}$ & Jul-07 05:47:44 & $y$ & 2025070600472 & 2025070900407 & 2.483 & 325.1 & 13.9 & 5 \\
2008 QZ$_{44}$ & Jul-20 05:07:33 & $g$ & 2025071900417 & 2025072300334 & 2.454 & 328.9 & 9.3 & 8 \\
2008 QZ$_{44}$ & Jul-24 05:45:15 & $g$ & 2025072300334 & 2025071900417 & 2.446 & 330.1 & 7.8 & 8 \\
2025 NC$_{6}$ & Jul-20 06:15:23 & $g$ & 2025071900502 & \nodata & 2.291 & 35.0 & 20.6 & 7 \\
3I/ATLAS & Jul-13 03:58:55 & $z$ & 2025071200555 & \nodata & 4.110 & 282.8 & 6.2 & 5 \\
3I/ATLAS & Jul-20 04:14:31 & $u$ & 2025071900384 & 2025072100254 & 3.879 & 284.1 & 9.0 & 5 \\
12P/Pons-Brooks & Jul-29 02:39:00 & $r$ & 2025072800112 & 2025072800142 & 5.643 & 139.8 & 9.6 & 6 \\
12P/Pons-Brooks & Jul-29 03:06:11 & $r$ & 2025072800142 & 2025072800112 & 5.644 & 139.8 & 9.6 & 6 \\
13P/Olbers & Jul-12 23:51:07 & $g$ & 2025071200280 & \nodata & 4.624 & 123.2 & 5.4 & 9 \\
13P/Olbers & Jul-23 00:27:06 & $u$ & 2025072200021 & 2025071900104 & 4.713 & 123.9 & 7.3 & 5 \\
65P/Gunn & Jul-04 06:07:04 & $r$ & 2025070300470 & 2025062100424 & 2.928 & 3.9 & 3.3 & 6 \\
65P/Gunn & Jul-19 01:13:41 & $g$ & 2025071800106 & 2025071900129 & 2.931 & 7.2 & 7.7 & 5 \\
65P/Gunn & Jul-21 01:20:20 & $u$ & 2025072000087 & \nodata & 2.931 & 7.6 & 8.4 & 6 \\
77P/Longmore & Jul-10 07:51:05 & $i$ & 2025070900669 & \nodata & 4.495 & 146.7 & 13.0 & 6 \\
77P/Longmore & Jul-24 07:51:32 & $i$ & 2025072300482 & 2025070900669 & 4.493 & 148.0 & 12.3 & 6 \\
78P/Gehrels & Jul-07 03:28:27 & $z$ & 2025070600368 & \nodata & 3.248 & 257.9 & 7.6 & 7 \\
78P/Gehrels & Jul-19 04:09:09 & $u$ & 2025071800231 & \nodata & 3.193 & 259.9 & 4.2 & 6 \\
78P/Gehrels & Jul-20 05:10:52 & $g$ & 2025071900422 & 2025072200190 & 3.188 & 260.1 & 4.0 & 9 \\
78P/Gehrels & Jul-23 02:47:30 & $g$ & 2025072200190 & \nodata & 3.175 & 260.5 & 3.4 & 8 \\
99P/Kowal & Jul-20 08:47:01 & $g$ & 2025071900655 & 2025071700211 & 6.133 & 104.8 & 2.1 & 7 \\
128P/Shoemaker-Holt & Jul-22 08:03:28 & $r$ & 2025072100576 & \nodata & 3.590 & 291.6 & 15.7 & 5 \\
131P/Mueller 2 & Jul-20 06:19:55 & $g$ & 2025071900509 & \nodata & 2.746 & 301.3 & 17.8 & 5 \\
188P/LINEAR-Mueller & Jun-21 07:01:21 & $i$ & 2025062000528 & 2025070100595 & 3.205 & 287.5 & 16.0 & 6 \\
188P/LINEAR-Mueller & Jul-22 05:03:34 & $g$ & 2025072100354 & 2025071900399 & 3.096 & 293.4 & 9.7 & 7 \\
243P/NEAT & Jul-20 08:19:26 & $g$ & 2025071900613 & \nodata & 2.820 & 300.1 & 8.0 & 6 \\
243P/NEAT & Jul-22 04:57:48 & $u$ & 2025072100349 & \nodata & 2.815 & 300.6 & 7.4 & 5 \\
261P/Larson & Jul-20 02:43:49 & $u$ & 2025071900278 & 2025072100256 & 2.355 & 300.9 & 1.6 & 7 \\
261P/Larson & Jul-22 03:19:45 & $u$ & 2025072100256 & 2025071900278 & 2.347 & 301.5 & 0.9 & 5 \\
299P/Catalina-PANSTARRS & Jul-19 04:02:13 & $u$ & 2025071800223 & \nodata & 3.808 & 77.4 & 6.0 & 6 \\
299P/Catalina-PANSTARRS & Jul-24 05:44:35 & $g$ & 2025072300333 & 2025071900417 & 3.820 & 78.1 & 4.8 & 8 \\
302P/Lemmon-PANSTARRS & Jul-08 08:01:30 & $y$ & 2025070700611 & 2025071100845 & 3.334 & 22.0 & 17.7 & 7 \\
302P/Lemmon-PANSTARRS & Jul-12 08:24:53 & $y$ & 2025071100845 & \nodata & 3.337 & 22.7 & 17.5 & 8 \\
302P/Lemmon-PANSTARRS & Jul-23 08:12:40 & $r$ & 2025072200537 & \nodata & 3.345 & 24.6 & 16.7 & 9 \\
303P/NEAT & Jul-18 00:13:03 & $g$ & 2025071700063 & \nodata & 2.963 & 300.4 & 5.9 & 5 \\
303P/NEAT & Jul-18 00:47:37 & $r$ & 2025071700110 & 2025071300126 & 2.963 & 300.4 & 5.9 & 5 \\
303P/NEAT & Jul-19 01:48:22 & $u$ & 2025071800136 & \nodata & 2.959 & 300.6 & 6.2 & 5 \\
306P/LINEAR & Jul-11 08:02:02 & $z$ & 2025071000634 & 2025070900622 & 1.297 & 341.8 & 51.5 & 5 \\
306P/LINEAR & Jul-23 08:18:24 & $r$ & 2025072200545 & 2025072300495 & 1.278 & 352.0 & 52.5 & 6 \\
351P/Wiegert-PANSTARRS & Jul-20 02:48:31 & $u$ & 2025071900284 & 2025072100249 & 3.190 & 23.2 & 1.7 & 6 \\
441P/PANSTARRS & Jul-20 05:19:03 & $g$ & 2025071900434 & 2025072300340 & 3.335 & 350.9 & 6.8 & 8 \\
486P/Leonard & Jul-08 08:29:36 & $z$ & 2025070700650 & \nodata & 2.398 & 30.5 & 24.8 & 8 \\
486P/Leonard & Jul-10 07:21:03 & $z$ & 2025070900627 & 2025070700650 & 2.401 & 31.1 & 24.7 & 7 \\
486P/Leonard & Jul-22 08:05:26 & $r$ & 2025072100579 & \nodata & 2.424 & 34.7 & 23.5 & 9 \\
491P/Spacewatch-PANSTARRS & Jul-20 02:50:08 & $u$ & 2025071900286 & 2025072000149 & 3.974 & 46.8 & 0.8 & 6 \\
491P/Spacewatch-PANSTARRS & Jul-21 02:14:53 & $u$ & 2025072000149 & \nodata & 3.975 & 46.9 & 0.8 & 6 \\
510P/Boattini & Jul-19 06:03:04 & $g$ & 2025071800340 & 2025072300366 & 5.128 & 318.8 & 6.8 & 7 \\
513P/Broughton & Nov-29 03:07:46 & $i$ & 2025112800164 & \nodata & 3.461 & 33.8 & 10.9 & 7 \\
P/2016 P5 (COIAS) & Jul-18 02:23:25 & $g$ & 2025071700225 & 2025072200191 & 4.651 & 82.7 & 2.5 & 7 \\
P/2016 P5 (COIAS) & Jul-22 03:11:33 & $u$ & 2025072100250 & 2025071800233 & 4.653 & 83.1 & 2.0 & 5 \\
P/2016 P5 (COIAS) & Jul-23 02:48:10 & $g$ & 2025072200191 & \nodata & 4.653 & 83.2 & 1.9 & 8 \\
C/2021 Q6 (PANSTARRS) & Jul-24 05:41:05 & $g$ & 2025072300328 & 2025071900407 & 9.154 & 25.5 & 3.0 & 7 \\
C/2023 H3 (PANSTARRS) & Jul-20 02:50:08 & $u$ & 2025071900286 & \nodata & 6.027 & 49.2 & 0.4 & 5 \\
C/2023 V4 (Camarasa-Duszanowicz) & Jun-21 03:36:45 & $i$ & 2025062000256 & \nodata & 4.964 & 123.2 & 6.3 & 5 \\
C/2024 G6 (ATLAS) & Jul-20 03:10:55 & $g$ & 2025071900309 & \nodata & 6.594 & 341.9 & 7.4 & 9 \\
C/2024 J3 (ATLAS) & Jul-19 01:55:45 & $u$ & 2025071800144 & 2025072000100 & 5.677 & 291.2 & 2.2 & 8 \\
C/2025 L2 (MAPS) & Jul-05 04:57:19 & $y$ & 2025070400460 & \nodata & 3.301 & 315.7 & 12.8 & 7 \\
C/2025 M1 (PANSTARRS) & Jul-14 03:08:25 & $r$ & 2025071300337 & 2025062400139 & 13.225 & 309.6 & 3.3 & 5 \\
C/2025 M1 (PANSTARRS) & Jul-20 03:54:43 & $u$ & 2025071900360 & 2025072100163 & 13.208 & 309.8 & 3.6 & 5 \\
\enddata
\tablecomments{$f$ is the true anomaly angle and $\alpha$ the phase angle. \added{Visit is the source visit at the center of the object's predicted position/datetime/band; coadded visit is the second source visit contributing to the central 5\arcsec\ of the coadd if present, otherwise \nodata indicates that only one visit contributed.} Scores are visual activity scores \citep{chandlerActiveAsteroidsCitizen2024} assigned by C.O.C. during cutout review; we list only rows with scores $\ge5$. \added{J.M. independently confirmed comets as active.} For display, we round times to the nearest second and angles to the nearest 0.1$^\circ$.}
\end{deluxetable*}

Table~\ref{tab:comet-results} lists every \ac{EDP2} cutout for which we scored cometary activity at $\ge5$, together with the source visits, band, true anomaly angle $f$, and phase angle $\alpha$ at the predicted epoch; Figure~\ref{fig:comet-gallery} shows a selection of images from that dataset. We used the \cometL{3I} comparison images (Section \ref{subec:3I}) to validate the cutout system; example images appear in Figure \ref{fig:3i-gallery}.

The sample in Table~\ref{tab:comet-results} is conditioned on (1) the input orbit and magnitude catalogs along with the adopted magnitude cutoff (Section \ref{subsubsec:inputdatasets}), and (2) shallow-coadd availability (Section~\ref{sec:methods}). Thus, the sample is a census of what the \ac{EDP2} shallow regions happened to contain at the predicted epochs rather than of the active population. Moreover, the $\sim$9\% activity fraction above refers to cutouts of objects already known to be active, and the panels of Figure~\ref{fig:comet-gallery} were chosen to show clear coma and tail morphology rather than the full range of appearances.

\subsection{Sample Statistics and Observing Geometry}\label{subsec:stats}

The activity sample comprises 59 cutouts spanning 32 objects (Table~\ref{tab:comet-results}): 29 designated comets, the interstellar comet \cometL{3I}, and two asteroid-designated objects (2008~QZ$_{44}$ and 2025 NC$_6$). At the time of observation, 16 objects were inbound and 16 were outbound. Heliocentric distances range from $r_h=1.28$~au (\comet{306P}, nine days before perihelion) to $r_h=13.2$~au for C/2025~M1 (PANSTARRS), with five detections beyond $r_h=6$~au (C/2023~H3 (PANSTARRS) at $r_h=6.03$~au, \comet{99P} at $r_h=6.13$~au, C/2024~G6 (ATLAS) at 6.59~au, C/2021~Q6 (PANSTARRS) at $r_h=9.15$~au, and C/2025~M1 at $r_h=13.2$~au). All but one epoch fall between 2025 June 21 and July 29; \comet{513P} was imaged on 2025 November 29. 26 cutouts are \texttt{n1} and 33 are \texttt{n2} products. By band, the cutouts are 16~$u$, 19~$g$, 10~$r$, 5~$i$, 5~$z$, and 4~$y$.

Many of our detections occurred at very small phase angles ($\alpha<3\degr$; e.g., C/2023~H3 at $\alpha=0.4\degr$, \comet{491P} at $\alpha=0.8\degr$, \comet{261P} at $\alpha=0.9\degr$, P/2016~P5 at $\alpha=1.9\degr$, and \comet{99P} at $\alpha=2.1\degr$; Table~\ref{tab:comet-results}). Near opposition, the anti-solar vector points nearly along the line of sight, so dust tails are strongly foreshortened and the activity we detect there may appear more in the form of coma rather than as an extended tail \citep{finsonTheoryDustComets1968,fulleMotionCometaryDust2004}; at the same time, the dust benefits from the opposition brightness surge \citep{meechObservationsCometHalley1987,schleicherNarrowbandPhotometryComet1998,kolokolovaPhysicalPropertiesCometary2004,bertiniScatteringPhaseFunction2017}, which favors coma detection. 16 of the 59 activity cutouts are in the $u$ band, where cometary images are rare in the literature (and \ac{LSST} will markedly change this landscape; \citealt{jonesSolarSystemScience2009,solontoi_detecting_2010,schwambLargeSynopticSurvey2018,kelleyCommunityChallengesEra2020}); single 38~s $u$-band exposures showing resolved comae are themselves a demonstration of the depth available, and the $u$ band admits the CN (0--0) emission band near 388~nm \citep{ahearnEnsemblePropertiesComets1995,farnhamHBNarrowbandComet2000,ivezicLSSTScienceDrivers2019}, so a gas contribution to these detections cannot be excluded without follow-up spectroscopy or narrowband imaging.

\subsection{Activity Relative to the Observational Record}\label{subsec:activityComparison}

\begin{deluxetable*}{lccccrrr}
\tabletypesize{\scriptsize}
\tablecaption{Active Object Information and Same-apparition Minor Planet Center Comparison.\label{tab:arc-extension}}
\tablehead{
\colhead{Object} & \colhead{$q$ (au)} & \colhead{$T_p$} & \colhead{Rubin epochs} & \colhead{MPC arc (non-X05)} & \colhead{$N_{\rm obs}$} & \colhead{X05 nights} & \colhead{$\Delta$ (d)}
}
\startdata
2008 QZ44 & 2.350 & 2025-10-25 & 2025-06-21--07-24 & 2025-05-23 to 2025-12-18 & 119 & 8 & inside \\
2025 NC6 & 2.155 & 2025-04-15 & 2025-07-20 & 2025-07-09 to 2025-09-15 & 33 & 13 & inside \\
3I/ATLAS & 1.356 & 2025-10-29 & 2025-07-13--07-20 & 2025-05-08 to 2026-04-14 & 8421 & 4 & inside \\
12P & 0.781 & 2024-04-21 & 2025-07-29 & 2020-06-10 to 2025-06-07 & 8199 & 0 & $+$51.6 \\
13P & 1.173 & 2024-06-30 & 2025-07-12--07-23 & 2023-08-13 to 2025-07-31 & 2456 & 0 & inside \\
65P & 2.914 & 2025-06-16 & 2025-07-04--07-21 & 2021-12-05 to 2026-09-10 & 1341 & 0 & inside \\
77P & 2.333 & 2023-03-27 & 2025-07-10--07-24 & 2021-11-03 to 2025-10-19 & 854 & 1 & inside \\
78P & 2.014 & 2026-06-25 & 2025-07-07--07-23 & 2025-04-25 to 2026-09-11 & 818 & 3 & inside \\
99P & 4.725 & 2022-04-12 & 2025-07-20 & 2019-10-08 to 2024-07-25 & 1177 & 11 & inside \\
128P & 3.056 & 2026-07-17 & 2025-07-22 & 2025-08-26 to 2026-01-31 & 147 & 0 & $-$35.1 \\
131P & 2.416 & 2026-02-15 & 2025-07-20 & 2025-07-26 to 2026-01-08 & 167 & 5 & inside \\
188P & 2.555 & 2026-04-13 & 2025-06-21--07-22 & 2025-06-20 to 2026-08-24 & 100 & 5 & inside \\
243P & 2.455 & 2026-02-25 & 2025-07-20--07-22 & 2025-06-30 to 2026-08-24 & 74 & 2 & inside \\
261P & 2.049 & 2025-12-27 & 2025-07-20--07-22 & 2025-05-25 to 2026-02-13 & 316 & 8 & inside \\
299P & 3.151 & 2024-04-30 & 2025-07-19--07-24 & 2022-02-05 to 2026-08-21 & 601 & 8 & inside \\
302P & 3.289 & 2025-03-09 & 2025-07-08--07-23 & 2023-03-27 to 2026-01-30 & 820 & 0 & inside \\
303P & 2.484 & 2026-02-19 & 2025-07-18--07-19 & 2025-07-19 to 2025-08-23 & 16 & 0 & $-$1.1 \\
306P & 1.247 & 2025-08-01 & 2025-07-11--07-23 & 2025-06-29 to 2025-11-27 & 66 & 0 & inside \\
351P & 3.131 & 2025-03-26 & 2025-07-20 & 2025-05-22 to 2026-08-11 & 264 & 7 & inside \\
441P & 3.326 & 2025-09-09 & 2025-07-20 & 2022-01-25 to 2025-11-10 & 132 & 12 & inside \\
486P & 2.308 & 2025-04-03 & 2025-07-08--07-22 & 2024-04-18 to 2026-01-13 & 538 & 0 & inside \\
491P & 3.715 & 2024-09-06 & 2025-07-20--07-21 & 2024-04-17 to 2026-09-10 & 413 & 8 & inside \\
510P & 4.898 & 2026-09-09 & 2025-07-19 & 2025-06-20 to 2026-08-09 & 51 & 1 & inside \\
513P & 3.276 & 2025-06-13 & 2025-11-29 & 2024-07-01 to 2026-01-11 & 195 & 0 & inside \\
P/2016 P5 & 4.433 & 2023-05-29 & 2025-07-18--07-23 & 2022-03-23 to 2025-09-28 & 182 & 7 & inside \\
C/2021 Q6 & 8.709 & 2024-03-25 & 2025-07-24 & 2021-01-02 to 2025-09-16 & 389 & 0 & inside \\
C/2023 H3 & 5.232 & 2024-02-18 & 2025-07-20 & 2023-03-16 to 2025-07-22 & 69 & 4 & inside \\
C/2023 V4 & 1.122 & 2024-05-30 & 2025-06-21 & 2023-11-05 to 2025-05-23 & 1153 & 0 & $+$28.8 \\
C/2024 G6 & 6.430 & 2026-02-20 & 2025-07-20 & 2024-04-10 to 2026-09-02 & 1604 & 0 & inside \\
C/2024 J3 & 3.865 & 2026-11-25 & 2025-07-19 & 2024-05-06 to 2026-09-07 & 3057 & 1 & inside \\
C/2025 L2 & 2.834 & 2025-12-20 & 2025-07-05 & 2025-06-02 to 2026-06-02 & 384 & 1 & inside \\
C/2025 M1 & 10.828 & 2029-08-03 & 2025-07-14--07-20 & 2021-05-18 to 2026-07-14 & 80 & 0 & inside \\
\enddata
\tablecomments{$T_p$ is the perihelion date of the current passage (JPL Horizons integration; C/2025 M1, \comet{77P} and 2025 NC$_6$ use JPL SBDB elements). ``Rubin epochs'' are the UTC dates of the activity score~$\ge5$ \ac{EDP2} cutouts. ``MPC arc (non-X05)'' is the first and last observation of the apparition (previous aphelion to the next, or the whole passage for long-period objects) reported to the MPC by any station other than Rubin (X05), with $N_{\rm obs}$ the number of such observations, retrieved from the \ac{MPC} observations API on UT 2026 September 11. ``X05 nights'' counts nights of the apparition on which Rubin astrometry is present in the MPC. $\Delta$ is the number of days by which the Rubin activity epochs fall outside every \ac{MPC} observation of the apparition, Rubin's own X05 astrometry included: negative values precede the first observation, positive values follow the last, and ``inside'' means the epochs lie within the arc. Relative to non-Rubin observers alone, two further objects fall outside the arc: \comet{99P} ($+$359.9 d), \comet{131P} ($-$6.6 d). For P/2016~P5 the aphelion-bounded arc includes 2022--2023 survey positions; relative to the 2025 return alone Rubin precedes the first non-Rubin position (F52, UT 2025 July 29) by 11.3~d, and likewise \comet{77P} by 18.1~d (UT 2025 July 28). Because Rubin's \texttt{deep\_coadd} epochs are predicted source-visit times rather than measured astrometry, these are epoch comparisons, not new astrometric arcs.}
\end{deluxetable*}

Table~\ref{tab:arc-extension} places each activity epoch we identified against the \ac{MPC} astrometric arc of the same apparition, and against the dated coma and tail reports we could locate in the \accite{COBS}{zakrajsekIntroducingCometObservation2018a,zakrajsekCometObservationDatabase2025}\footnote{\url{https://cobs.si/}}, the \ac{CBET} and \ac{MPEC} publications, and observer archives (retrieved 2026 September 11--12). We classify these into \added{six} outcomes:

\textbf{New Activity Epochs}: For 2008~QZ$_{44}$, 2025 NC$_6$, \comet{306P}, and \comet{243P}, other observers tracked the object astrometrically through our epochs, but no readily available dated coma or tail report exists for the return (this does not necessarily mean they were not active or seen to be active, only that the activity was not reported); thus, the Rubin images are the only activity evidence to the best of our knowledge. For \comet{306P}, outside images from UT 2025 August 3 (J.~Maikner), August 31 (M.~J\"ager, G.~Rhemann and E.~Prosperi), and September 18 (R.~Fichtl) exist, but to our knowledge none shows unambiguous activity.\footnote{\url{http://www.aerith.net/comet/catalog/0306P/2025-pictures.html}}

\textbf{Early Activity Extends Observations Arc}: These activity detections extend both the activity arc and the observational (astrometric) arc \added{earlier} for the objects. For \comet{128P} and \comet{303P}, the Rubin activity images predate every observation of the 2025 return reported to the \ac{MPC} (by 35~d and 1.1~d), and for \comet{131P} they predate every non-Rubin observation (by 6.6~d), with \added{the first \ac{MPC} observations being} Rubin's \added{astrometry from July 10}.

\textbf{Activity Extension}: Our activity reports precede the earliest \added{evidence of} activity (to the best of our knowledge) for the return. \added{For \comet{510P} the first outside coma image/report follows our report by 10} days, for \comet{188P} by 34 days\footnote{H.~Sato measured $\sim$7\arcsec
coma on UT 2025 July 25 image (T.~Ikemura) \url{http://www.aerith.net/comet/catalog/0188P/2026-pictures.html}}, and for P/2016~P5 by 40~days\footnote{Coma reported to \ac{COBS} by J.~Maikner for UT 2025 August 27; his August 26 report lists none. A UT 2025 August 13 image (H.~Sato) does not resolve the morphology,  \url{https://britastro.org/cometobs/2016p5/2016p5_20250813_hsato.html}.}.

\textbf{Later Activity Extends Observation Arc}: These detections extend both the activity record and the observation (astrometric) arc. For C/2023~V4 (29~d), \comet{12P} (52~d), and \comet{99P} (360~d relative to non-Rubin observers, whose record ends in 2024 July) the images follow the last reported observation of the object. In each case, they are also the latest dated activity evidence, extending the documented activity of \comet{12P} to $r_h=5.64$~au, C/2023~V4 to $r_h=4.96$~au, and \comet{99P} to $r_h=6.13$~au, the last some 1~au beyond the \ac{ATLAS} photometry of \citet{gillanActivityJupiterFamilyComets2025}. 

\textbf{Later Activity Extension}:
For these objects, our Rubin epochs lie within the astrometric arc of the apparition, but follow every outside coma or tail report we could locate, so they extend the activity record alone. The last such report precedes our images by 21~months for \comet{77P} (\ac{COBS} coma of UT 2023 October 5) and by 26~months for C/2023~H3
(UT 2023 May 28, CBET 5267). For C/2021~Q6 we found no morphology report after the discovery circular (UT 2021 September 3, CBET 5032),
although amateur images through 2022 November exist that we did not assess at the \ac{PSF} level, so that interval is an upper bound rather than a measured duration. C/2025~M1 is the marginal case, following the \ac{CFHT}\added{-detected} coma and tail of UT 2025 June 24 by only 20~d; its value lies in the depth of the detection at $r_h=13.2$~au rather than in the interval, since archival imaging already places the onset of its activity in 2023 April. 

\textbf{Interspersed Activity Observations}: The remaining objects (\comet{13P}, \comet{65P}, \comet{78P}, \comet{261P}, \comet{299P}\footnote{UT 2025 July 31 image (M.~J\"ager, G.~Rhemann, E.~Prosperi), \url{http://www.aerith.net/comet/catalog/0299P/2024-pictures.html}.},  \comet{302P}, \comet{351P}, \comet{441P}, \comet{486P}, \comet{491P}, \comet{513P}, \cometL{3I}, C/2024~G6, C/2024~J3, and C/2025~L2) lie inside a record in which others reported activity both before and after our epochs. These serve as consistency checks on the morphology we recover from single- and two-visit coadds, rather than as new results. However, these data are still worth scrutiny because, for example, Rubin may provide the highest resolution or deepest imaging available, as with \cometL{3I} \citep{chandlerNSFDOEVeraRubin2026}.

\subsection{2025 NC6}\label{subsec:2025N6}
We screened a selection of asteroids from the general \texttt{MPCORB} run designed to produce cutouts for the \textit{Rubin Comet Catchers} project. In addition to the re-discoveries described in Section \ref{subsec:rediscovery}, we identified one object, \added{2025 NC$_6$}, as clearly showing a diffuse tail (Figure \ref{fig:comet-gallery}) in the coincident anti-solar and anti-motion directions as projected on the sky; the object was 96 days post-perihelion (true anomaly angle of $f=35\degr$) at 2.29~au. The orbit is consistent with that of a \ac{JFC}, with a Tisserand parameter with respect to Jupiter of $T_\mathrm{J}= 2.80$ (where $2<T_\mathrm{J}<3$ nominally indicates cometary orbits; \citealt{levisonCometTaxonomy1996}), though further dynamical integration would be needed to rule out membership of other classes such as the quasi-Hildas \citep{tothQuasiHildaSubgroupEcliptic2006,gil-huttonCometCandidatesQuasiHilda2016}. The MPC record includes Rubin observations (X05) and discovery observations obtained with the \ac{KPNO} Bok 90-inch telescope (V00). As of 2026 September 16, no comet designation or outside activity report could be found, and precovery observations in the \ac{MPC} extended the arc of the object back to UT 2015 March 29. Our \ac{SBSAR}\footnote{\url{https://sbsar.net}} campaign extended the arc to UT 2007 May 10; a search for activity could thus reveal two additional epochs, but is outside the scope of this work.

\subsection{2008 QZ44}\label{subsec:2008QZ44}

\citet{chandlerCometaryActivity2008QZ442023} reported an anti-solar tail in \ac{CFHT} MegaPrime images from 2008 November 20 ($r_h=2.43$~au) and a faint tail in \ac{DECam} images of 2017 November 12--13 ($r_h=2.90$~au), both after perihelion. The orbit of 2008~QZ$_{44}$ is that of a \ac{JFC} ($T_J=2.82$, $q=2.35$~au, $P=8.6$~yr). Our four \ac{EDP2} epochs (2025 June 21 to July 24, $r_h=2.45$--2.52~au) lie 93--126~d before the 2025 October 25 perihelion; this, in conjunction with Zhang et al. (submitted) would mark the \added{object's} first \added{reported} pre-perihelion activity, to the best of our knowledge, and its third consecutive active apparition (\added{following activity in the} 2008 and 2017 \added{apparitions}). We also observed 2008~QZ$_{44}$ active in 2025 with the \ac{ARC} 3.5~m telescope at the \ac{APO}; we will report this independent confirmation in a parallel Note (Zhang et al., in preparation) and cross-reference it here.

\subsection{Independent Rediscovery of Activity}\label{subsec:rediscovery}

We ``rediscovered'' \comet{243P} and \comet{510P} as active objects when we found visible evidence of activity for 2025~MT$_{356}$ and 2025~NX$_{200}$. We later learned both objects had not yet been reconciled into their comet counterparts. 2025~MT$_{356}$, with six X05 positions (2025 June 21, July 18, July 20), at $a=3.826$~au and $e=0.360$, matches \comet{243P}. 2025~NX$_{200}$ also had six X05 positions (2025 July 2, 21, 22), and with $a=6.47$~au, $q=4.88$~au, matches 510P. In both cases, we visually confirmed the comets matched their respective object's cutouts (Figure \ref{fig:comet-gallery}).

For \comet{510P}, our July 19 activity image precedes Oribe's stellar recovery images of July 24 and the first published cometary appearance, the Pan-STARRS2 6\arcsec\ tail of July 29 (\ac{CBET} 5590); for \comet{243P}, no observer has reported a coma or tail this return. The \ac{MPC} has already linked a third Rubin designation, 2025~OZ$_{695}$, to active Centaur P/2005~S2 (Skiff) as described in \ac{MPEC} 2026-N37, so these are independent re-detections of known comets rather than discoveries. Nevertheless, the two objects (\comet{243P} and \comet{510P}) serendipitously served as both testbeds and proofs of concept for the methods outlined in this manuscript.

\subsection{Trailed NEOs}\label{subsec:neoResults}

\begin{figure}[ht!]
\centering
\begin{tabular}{cc}
     \labelpicA{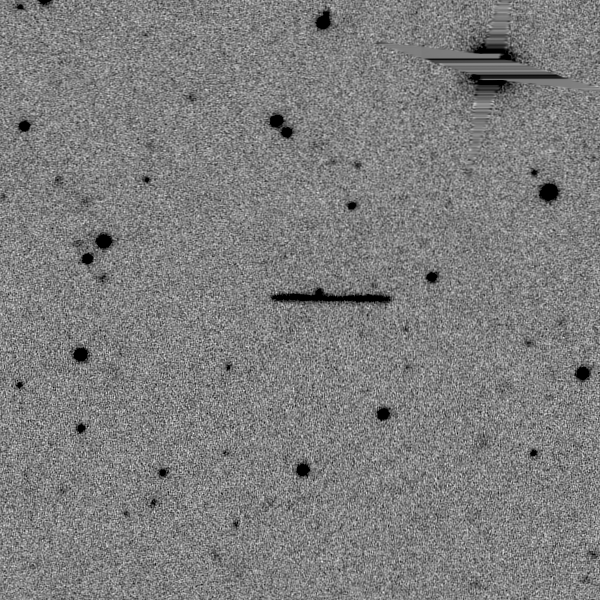}{(a)}{118 $\Delta_\mathrm{pix}$}{0.46}{}\hfill
 & \labelpicA{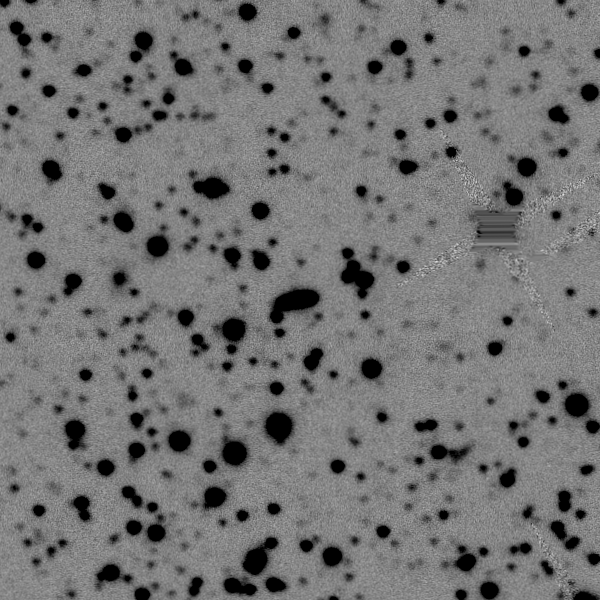}{(b)}{33 $\Delta_\mathrm{pix}$}{0.46}{}
 \\
\end{tabular}
\caption{Ephemeris-centered $600\times600$ pixel ($120\arcsec\times120\arcsec$) \acf{EDP2} cutouts, with north up and east left, showing two \added{examples of} highly trailed \acfp{NEO}. 
The labels give the rounded predicted trail length in pixels ($\Delta_\mathrm{pix}$). 
The \added{objects,} source-visit dates, \ac{TAI} times, \ac{LSST} bands, and trail lengths are: 
(a) 2025 OP$_1$, 2025-07-23 02:39:11, $g$, $\Delta_\mathrm{pix}$=118~pix;
(b) 1999 MN, 2025-07-01 09:20:16, $i$, $\Delta_\mathrm{pix}=33$~pix. 
\label{fig:trail-gallery}}
\end{figure}

The same ephemeris-to-cutout workflow applied to the $q<1.3$~au subset of \texttt{MPCORB} yielded trailed \ac{NEO} images with predicted motion of $5$ or more pixels over a 30~s $grizy$ or 38~s $u$-band exposure. Figure~\ref{fig:trail-gallery} shows two examples with predicted \added{trails} of 118 and 33 pixels. We restrict \added{the scope of this} Letter to demonstrating that trails are recoverable from the shallow coadds; \ac{NEO} analysis will appear in a separate paper (Chow et al., in preparation).

\begin{figure*}
    \centering
    \setlength{\tabcolsep}{1.5pt}
    \newcommand{\galw}{0.153\textwidth}\newcommand{\rowlabel}[1]{\rotatebox{90}{\makebox[\galw]{#1}}}
    \footnotesize
    \begin{tabular}{c@{\hspace{3pt}}cccccc}
        & \normalsize 2025nbh & \normalsize 2025qaz & \normalsize 2025mmt & \normalsize 2025acrq & \normalsize 2025abgi & \normalsize 2025kdz \\
        & 60\arcsec$\times$60\arcsec & 30\arcsec$\times$30\arcsec & 30\arcsec$\times$30\arcsec & 30\arcsec$\times$30\arcsec & 60\arcsec$\times$60\arcsec & 60\arcsec$\times$60\arcsec \\[4pt]
        & $grz$ stack & $giz$ stack & $griz$ stack & $g$ & $g$ & $gri$ RGB \\
        & 2016--2018 & 2019--2021 & 2016--2021 & 2017--2018 & 2013--2017 & 2016--2021 \\
        \rowlabel{LSDR10} &
        \includegraphics[width=\galw]{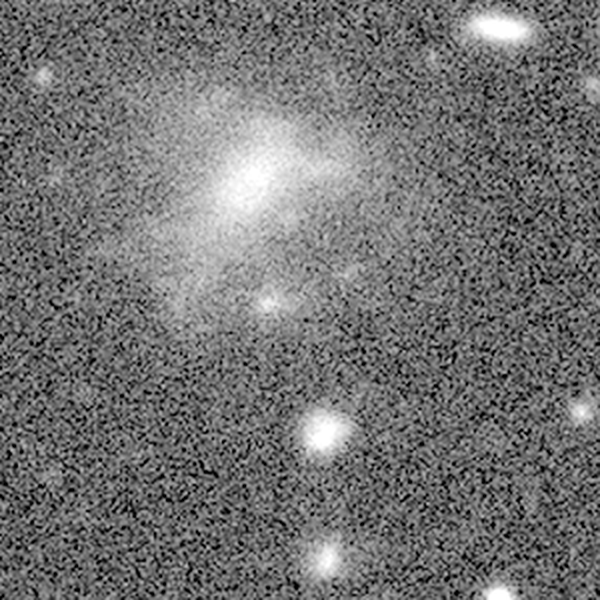} &
        \includegraphics[width=\galw]{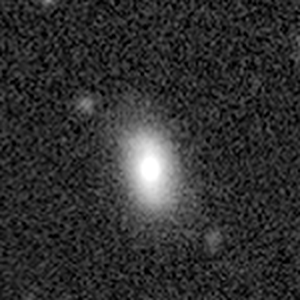} &
        \includegraphics[width=\galw]{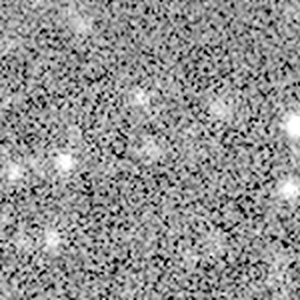} &
        \includegraphics[width=\galw]{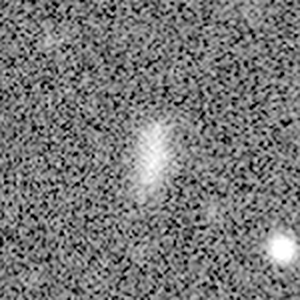} &
        \includegraphics[width=\galw]{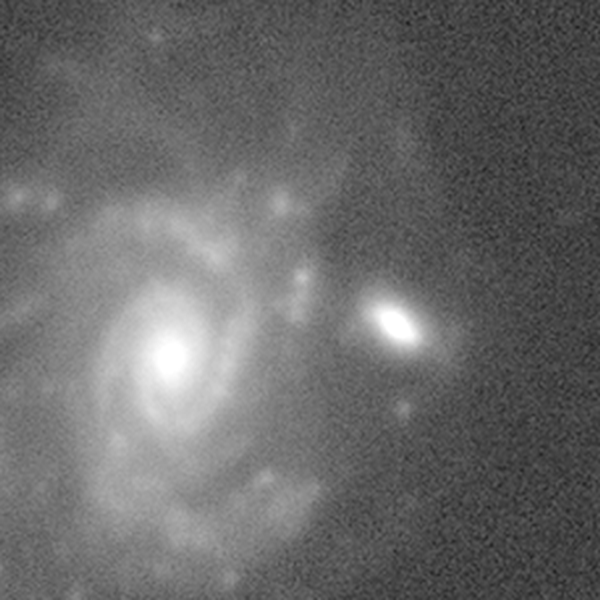} &
        \includegraphics[width=\galw]{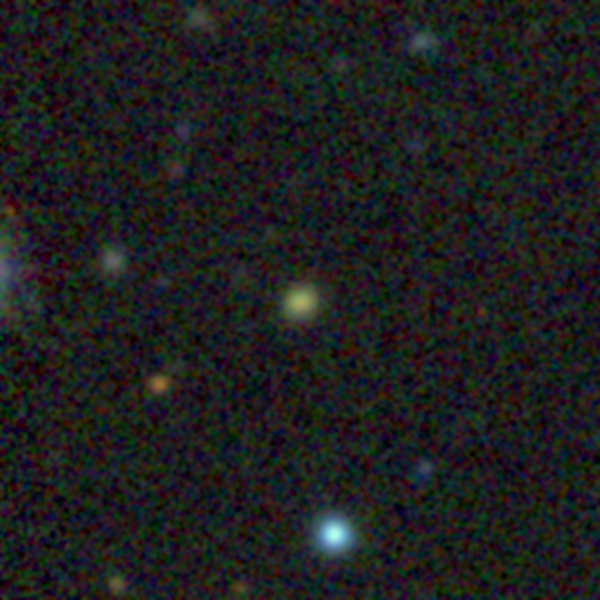} \\
        \rowlabel{Rubin EDP2} &
        \includegraphics[width=\galw]{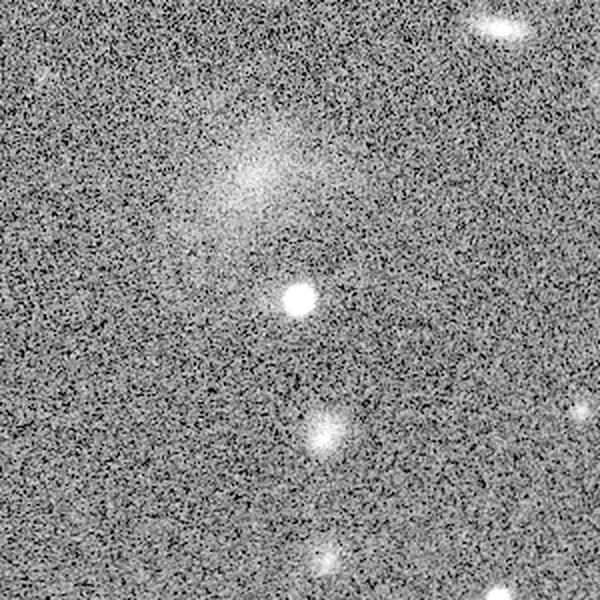} &
        \includegraphics[width=\galw]{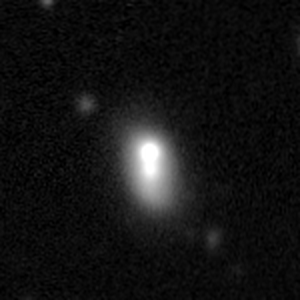} &
        \includegraphics[width=\galw]{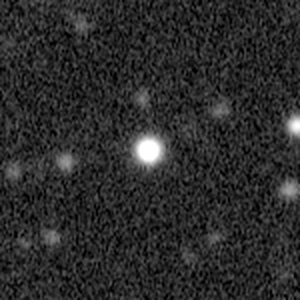} &
        \includegraphics[width=\galw]{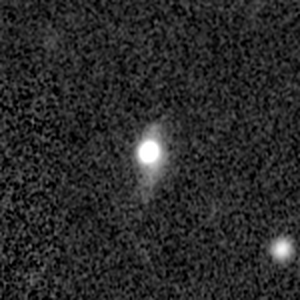} &
        \includegraphics[width=\galw]{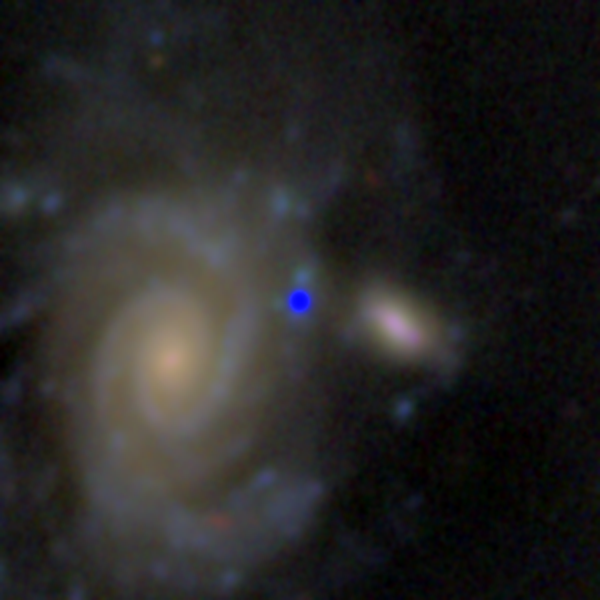} &
        \includegraphics[width=\galw]{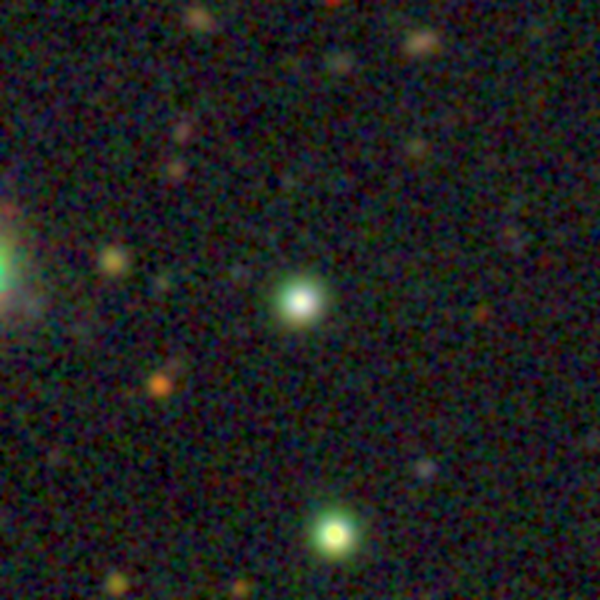} \\
        & $ri$ stack & $ugrizy$ stack & $grizy$ stack & \texttt{n2} $g$ & $gri$ RGB & $gri$ RGB \\
        & Jul 22--24 & Jul 1--Aug 11 & Jul 1--Sep 14 & Jul 22, Nov 9 & Jul 1--Nov 9 & Jun 21--Sep 22 \\
    \end{tabular}
    \caption{
    Example transients in \ac{EDP2} coadds (bottom row; 2025 UTC dates) and pre-2025 Legacy Survey DR10 (LSDR10) comparison images (top row). Labels give the cutout size, bands, and observation dates for each panel. ``Stack'' panels combine all listed bands to maximize host detectability; ``RGB'' panels are color composites. The 2025acrq \ac{EDP2} panel is an \texttt{n2} coadd of one pre-explosion visit and one taken four days after discovery. SN~2025nbh lies in the outskirts of the ultra-diffuse galaxy \acs{AGC}~114905, the diffuse source to its northeast; SN~2025abgi is on an outer arm of PGC~073035 (Section~\ref{sec:supernovae}). The transient is centered in each panel. North is up, east is left. Each panel has its own display stretch.
    }
    \label{fig:transients}\label{fig:supernovae}
\end{figure*}

\subsection{Supernovae}\label{sec:supernovae}
\added{Of the 47 transients for which we measured provenance, 36 fell into at least one \texttt{n1} or \texttt{n2} region in an \ac{EDP2} band; the remaining 11 either had no \texttt{deep\_coadd} coverage or only appeared in deeper (\texttt{n}$>2$) cells. Most shallow \ac{EDP2} cells were built from 2025 June--July visits; the shallow cells predate (by 15 to 124 days) the discovery of objects discovered after August, and most of these pre-discovery visits have no catalog source at the position, giving 5$\sigma$ limits of 21.5--24.8 mag (across all six bands). The few sources that are present can include host light. For example, the $g\approx23.3$ source at the position of the superluminous SN~2025ukr, 23--24 days before discovery, matches its compact host in Legacy Survey DR10 ($g=23.17$, $\sim$0.5\arcsec\ away), leaving no detectable supernova flux to $g\approx24.5$
some 44 days before its $g=18.2$ peak. That is about 4~mag deeper than the \ac{ZTF} limits a week later. Most later visits containing those transients fell in cells with $\ge3$ input visits, and those earlier visits diluted the transient signal (Appendix~\ref{app:transients}). The Rubin \texttt{n1} magnitudes for all objects we highlight here agree with publicly listed light curves to $\le$0.2~mag.}

\added{Here we showcase seven transient events recovered from \ac{EDP2} data, with corresponding images for six of them in Figure \ref{fig:transients}. Table~\ref{tab:transients} lists the remaining objects with measured provenance. Noteworthy events include \ac{TDE} 2025qhf, which has an \texttt{n1} $u$-band cell ($u = 18.73$) and an \texttt{n2} $g$ cell (combining $g = 18.58$ and $18.70$). However, it is nuclear (i.e., the direct-image magnitude includes host light), and its pre-2025 host dominates every image, so we did not include it in the gallery.}

\added{\textbf{SN 2025nbh} (Type II, $z = 0.0165$, \citealt{dasZTFTransientClassification2025}; \ac{ATLAS} discovery 2025 June 7, \citealt{tonryATLASTransientDiscovery2025}) falls in an \texttt{n1} $r$-band coadd taken UT 2025 July 22, 45 days post-discovery. The single-visit catalog measures $r = 18.80 \pm 0.01$, matching the $r = 18.77$ reported by \ac{ZTF} that same night. The $i$-band coadd is also \texttt{n1}, from two nights later. The Legacy Survey DR10 image appears empty at the same coordinates. The galaxy 13\arcsec{} northeast is \ac{AGC} 114905, a gas-rich ultra-diffuse galaxy whose rotation curve appears to require little or no dark matter \citep{mancerapinaNoNeedDark2022, mancerapinaExploringNatureDark2024, afruniLongLifeUltra2025}. The supernova lies about 4.7 kpc from the galaxy's center, within an extended feature (visible in Figure \ref{fig:supernovae}), and the \accite{SNID}{blondinDeterminingTypeRedshift2007} redshift of $z=0.0165$ agrees with the galaxy's redshift of $z=0.0181$--$0.0183$ via \ac{NED} and \ac{SIMBAD}, to within \ac{SNID} nominal precision for a Type II. We are not aware of any previous supernova reports for this galaxy.}

\added{\textbf{SN 2025qaz} (Type Ia, $z = 0.10$, \citealt{quirola-vasquezEPESSTOTransientClassification2025}; \ac{ZTF} discovery 2025 June 28, \citealt{sollermanZTFTransientDiscovery2025}) falls in an \texttt{n1} $g$-band coadd taken UT 2025 July 20 (three days after the \ac{ZTF} peak) with $g = 18.50$, 0.2 mag brighter than \ac{ZTF} measurements from the same night. The source is flagged as extended because the \ac{PSF} flux includes light from the host (clearly visible in the background in Figure \ref{fig:supernovae}), a galaxy 2\arcsec away that the Legacy Survey DR10 panel shows without the supernova present. The $u$-band coadd combines two visits four days apart. During that time, the supernova faded from magnitude 19.33 to 19.80.}

\added{\textbf{SN 2025mmt} (Type II, $z = 0.11$, \citealt{huberSCATTransientClassification2025}; Pan-STARRS discovery 2025 May 27, \citealt{chambersPanSTARRSTransientDiscovery2025}) demonstrates a deeper (\texttt{n2}) stack; the $g$-band coadd combines visits from 2025 July 20 and August 29. Both show a point source within $\sim$0.13\arcsec of the position ($g = 20.32$ and $20.59$ respectively; \ac{ZTF} measured $g=20.44$ on July 24). The $r$-band coadd combines seven visits from July 1 to September 14; all detected the supernova as it declined in brightness from $r = 19.85$ to $20.50$. A faint source is visible in the Legacy Survey DR10 panel, possibly the host dwarf at $g = 24.5$ in the Legacy Survey DR10 catalog, $\sim$0.16\arcsec from the position.}

\added{\textbf{SN 2025acrq} (Type Ia, $z = 0.088$; \ac{ZTF} discovered 2025 November 5, \citealt{sollermanZTFTransientDiscovery2025a}; classified by \ac{SEDM}, \citealt{fremlingZTFTransientClassification2025}) is a before-and-after case made entirely from Rubin data. Its $g$-band coadd cell was built from two visits: one pre-explosion (UT 2025 July 22 06:49, 106 days before discovery), and one post-explosion (UT 2025 November 9 02:41, four days after discovery and while it was still brightening toward its November 27 \ac{ZTF} peak). We found no detection (5$\sigma$ limit of 24.5 mag) within 1\arcsec of its position in the July visit's source catalog, but the November source catalog revealed a source $\sim$0.19\arcsec from the position; it was measured at $g = 18.72 \pm 0.01$, within 0.04 mag of \ac{ZTF} and \ac{ATLAS} photometry the same night. The \texttt{n1} $r$-band coadd (2025 July 20) shows only the host with a centroid $\sim$1\arcsec away at $g = 20.9$ in pre-2025 Legacy Survey imaging. Because only one of the two $g$ input visits overlapped temporally with the supernova, the coadd should retain roughly half its single-visit flux; the coadd object table had no measurement at that position, so this is our estimate from the provenance only. The \ac{NEEDLE} classifier \citep{aamerSLSNTDECandidates2025} also flagged the object as a superluminous-supernova candidate.}

\added{\textbf{SN 2025abgi} (Type II, $z = 0.023$; \ac{ATLAS} discovery 2025 October 21, \citealt{tonryATLASTransientDiscovery2025a}; classified by \ac{SEDM}, \citealt{sollermanZTFTransientClassification2025}) exploded on an outer arm of PGC~073035 (MCG~$-$01$-$01$-$008) a face-on spiral at $z = 0.0236$, the unambiguous-host exemplified in Figure~\ref{fig:transients}. The galaxy is well resolved in both the pre-2025 Legacy Survey image and the Rubin color composite, and the supernova appears as a blue point on the arm only in the Rubin composite. The $g$-band coadd cell combines one pre-explosion visit (2025 July 22) and two back-to-back exposures (2025 November 9 02:00 UTC) 19 days post-discovery. The latter's source catalogs measure $g = 18.27$ and $18.28 \pm 0.01$ at $\sim$0.09\arcsec from the position. Because this $g$-band cell has three inputs (\texttt{n3}) rather than one or two, it allows a direct measurement of the dilution caused by coaddition.
The coadd object table gives $g = 18.86$ at the position, 0.58 mag fainter than the single November visits, compared to the 0.44 mag expected if two of three equal-weight inputs contained the source; the difference thus reveals the per-visit weights in the coadd. The $r$-band cell, built from four visits between July 1 and August 17 (all pre-explosion), shows the same arm without the supernova. The \ac{TNS} entry also lists a Rubin alert-pipeline internal name, from a later (2026) Rubin detection.}

\added{\textbf{SN 2025kdz} (superluminous Type II, $z = 0.233$; \ac{ATLAS} discovery 2025 May 7, \citealt{tonryATLASTransientDiscovery2025b}; classified with Lick/Kast, \citealt{chornockZTFTransientClassification2025}) is (1) the most distant object in our set, and (2) a case in which the flux was not diluted, in contrast to SN 2025acrq. It stayed between $z=18.4$ and $g=18.9$ mag for months, so both $g$-band visits (2025 July 19 and 24, 73 and 78 days after discovery, respectively) and both $z$-band visits (2025 July 5 and 10, 59 and 64 days after discovery) fell on the plateau. The single-visit source catalogs had $g = 18.87$ and $18.90$, and $z = 18.41$ and $18.42$, each at $\sim$0.25\arcsec from the position; the coadd object table shows $g = 18.885 \pm 0.002$ and $z = 18.412 \pm 0.003$, equal to the flux mean of the inputs to $\sim$0.003 mag in both bands. In other words, when every input has the transient at the same brightness, as on a plateau or within a few days of peak, the coadd nominally recovers the full flux. The host is a compact source 0.23\arcsec away at $g = 21.2$ in Legacy Survey DR10 (Figure \ref{fig:transients}), $\sim$2.3 mag fainter than the supernova.
}

\added{These four cases span the dilution relation of Section~\ref{sec:transientmethods}, where $k$ of $n$ input visits contain the transient. When all inputs hold the transient ($k = n$) at nearly constant brightness, the coadd recovers the full flux; the SN~2025kdz coadd magnitudes equal the mean flux of its inputs to $\sim$0.003~mag, and SN~2025mmt's $g$-band coadd is roughly representative despite its two inputs spanning 40 days of slow brightness decline. When only some inputs hold the transient, the flux is diluted, for example SN~2025abgi ($k/n = 2/3$) measures 0.58~mag fainter in the coadd than in its November single visit data, and SN~2025acrq ($k/n = 1/2$) should retain roughly half of its single-visit flux (i.e., 0.75~mag fainter), but the coadd object table has no measurement at that position, so we derive this measurement via provenance only.} 

\added{\textbf{SN 2025qqj} (Type Ia, $z = 0.063$; \ac{ZTF} discovered 2025 July 6, \citealt{wiseZTFTransientDiscovery2025}; classified with the Liverpool Telescope, \citealt{wiseZTFTransientClassification2025}) illustrates Rubin's potential for contributing photometry and image data in bands absent from the discovery and other surveys. The \ac{TNS} public photometry we used is from \ac{ZTF} ($g$, $r$), \ac{ATLAS} (orange), Pan-STARRS ($w$), \accite{GOTO}{steeghsGravitationalwaveOpticalTransient2022} $L$-band, and \accite{LAST}{ofekLargeArraySurvey2023} clear filter; no $u$-band or ultraviolet measurement existed for this supernova at the time we searched (2026 October 24). Rubin's \texttt{n2} $u$-band (UTC 2025 July 19 05:03, July 22 04:39) visits are 13 and 16 days post-discovery and 8 and 5 days pre-\ac{ZTF} $g$-band peak (July 27). Their single-visit source catalogs recorded $u = 18.43 \pm 0.01$ and $18.49 \pm 0.01$ at 0.12\arcsec from the supernova position, so it was already fading in $u$ while still brightening in $g$, as expected for a Type Ia (Ia $u$-band maxima precede $B$- and $g$-band maxima by a few days;  \citealt{hsiaoKCorrectionsSpectralTemplates2007}). The remaining \ac{EDP2} coadds at this sky location are deeper ($g$=\texttt{n3}, $r$=\texttt{n7}, $i=$\texttt{n5}, $z$=\texttt{n6}, $y$=\texttt{n4}), but the supernova is detected in all bands. The Rubin $g$ magnitude (17.97) on July 20 matches the public light curve. The host galaxy is $\sim$0.9\arcsec away at $g = 19.0$ in pre-2025 Legacy Survey imaging; the first $u$-band visit is flagged as extended in the source table, and the second as point-like, so some host light may enter the $u$ \ac{PSF} flux. We were not able to find any public $u$-band measurements to check against, which is exactly why we highlight this object: for transients discovered by wide-field surveys, Rubin's $u$, $z$, and $y$ coverage may often be the only photometry available for those bands, and here it falls within the week of peak brightness.}

\section{Discussion and Summary}\label{sec:summary}

We identified shallow layers of single-image (\texttt{n1}) or two-image (\texttt{n2}) depth in the \acf{EDP2} dataset released by the NSF-DOE Vera C. Rubin Observatory (Figure \ref{fig:skycoverage}), an important facet of this dataset with broad applications to astronomy beyond the examples in this Letter. We demonstrated a method for identifying known small Solar System bodies in these shallow data, especially objects that can be definitively identified in single-visit depth data -- comets and \acp{NEO} -- because of their respective ``tails and trails.'' \added{We further demonstrate the utility of our approach to transient science, especially supernovae.}  Accompanying this Letter are (1) a catalog of objects we identified as displaying visible evidence of activity in the form of a tail or coma, and (2) images of these objects in \ac{PNG} format. We have made a Jupyter notebook tutorial available online \citep{chandlerNotebooksTailsTrails2026} for use with the \ac{RSP}.

Running the \texttt{Sorcha} survey simulator with our \texttt{Ponder} orchestrator for all known comets yielded $14\,030$ predicted comet positions within \ac{EDP2}. Applying our shallow-coadd pre-screen and source-visit provenance cuts left $\sim700$ cutouts (Section~\ref{subsec:recovery_and_activity}), of which we scored $59$ cutouts spanning $32$ objects as showing a coma or tail (Table~\ref{tab:comet-results}). Figure \ref{fig:comet-gallery} highlights the best examples. The detections range from $r_h=1.28$~au (\comet{306P}, nine days pre-perihelion) to $r_h=13.2$~au (C/2025~M1) and include five comets beyond $r_h=6$~au. Notably, we made 16 detections in the $u$ band, \added{in which observations are rare}.

We discovered that minor planet 2025 NC$_6$ is active (Section \ref{subsec:2025N6}). We also independently rediscovered activity in minor planets 2025~MT$_{356}$ and 2025~NX$_{200}$ (Section \ref{subsec:rediscovery}), which have been identified (but not yet linked) as comets \comet{243P} and \comet{510P}, respectively. All three detections demonstrate the utility of our approach.
The same ephemeris-to-cutout workflow applied to the \ac{NEO} population produced 336 cutouts, 29 of them with predicted trails longer than 25 pixels (Section~\ref{subsec:neoResults}); a separate work on these objects (Chow et al., in preparation) is forthcoming.

Set against the \ac{MPC} record (Table~\ref{tab:arc-extension}), the Rubin images of 2008~QZ$_{44}$, 2025~NC$_6$, \comet{306P}, and \comet{243P} are the only evidence of activity for their respective returns that we could locate. For \comet{188P}, \comet{510P} and P/2016~P5, they are the earliest activity evidence of the return, preceding the first outside tail/coma report or image by 10--40~d. 2008~QZ$_{44}$ is now active for a third consecutive apparition and, for the first time in the reported record, before perihelion (Section~\ref{subsec:2008QZ44}). The activity images of \comet{128P} and \comet{303P} precede every reported observation of their 2025 returns; those of \comet{131P} precede every non-Rubin observation. The \comet{12P} and C/2023 V4 activity images postdate the latest observations in the \ac{MPC} record by 51.6 and 28.8 days, respectively. The \comet{99P} activity image postdates the latest non-Rubin observation by 359.9 days (at a heliocentric distance of $r_h=6.13$~au), but falls within the MPC arc when Rubin's astrometry is included. The Rubin images provide the latest evidence for the apparitions of \comet{77P} and C/2023 H3, following previously dated activity evidence from 2023 October and 2023 May, respectively.

The catalog and cutouts support activity-onset and activity-persistence comparisons of the kind made here, target selection for follow-up, and archival precovery searches keyed to the quoted epochs. However, photometry and quantities derived from it, such as $Af\rho$, are not directly usable from the \texttt{n2} products because a moving object's flux is diluted by the coaddition weights and may be clipped by outlier rejection (Section \ref{subsec:provenance}). Quantitative measurements therefore require either the contributing-visit correction described above, or the corresponding single-visit images; uncorrected measurements from the coadds should at most be regarded as approximate.

We note several caveats to the method described in this paper. The predicted object positions come from an activity-free brightness model with a 26th-magnitude cut and no detectability model, so the $\sim$9\% activity fraction describes only what is contained in the shallow regions rather than the active fraction of any population. Background confusion is severe in \texttt{n1} visits and worse when an \texttt{n2} cutout combines two epochs of field stars and galaxies; the gallery panels were chosen for clear morphology and are not a census of appearances. Moreover, ephemeris uncertainty should be negligible for the numbered comets analyzed here (though they can deviate; see Figure \ref{fig:comet-gallery}), but matters for short-arc objects. 

Our comparison with the outside record is also limited by the sources searched (the \ac{MPC}, \ac{COBS}, \acp{CBET}, \acp{MPEC}, and observer archives) and by the fact that a magnitude or astrometric position is not evidence of activity. The earliest and latest activity dates quoted here are relative to that record and can move as archival images are examined. Finally, near opposition the anti-solar direction lies close to the line of sight, so several of our activity detections appear as comae rather than tails (Section~\ref{subsec:stats}). Each epoch quoted in this Letter is the time of the single visit containing the corresponding object; the \texttt{deep\_coadd} from which the cutout was drawn carries no timestamp of its own, and the other contributing visit of an \texttt{n2} product \added{nominally} adds only background (Section~\ref{sec:methods}).

Many of the objects described in this Letter warrant follow-up study, with three in particular meriting immediate observation. \comet{510P} passed perihelion on 2026 September 9 at $q=4.898$~au, and is presently near opposition; our 2025 July 19 image is the earliest activity from its return, and no morphology had been reported during the 2026 approach \added{(we observed it ourselves, as described below).} \comet{99P} is at $r_h=6.7$~au and near opposition; our detection extend its documented activity beyond the $r_h=6.13$~au \ac{EDP2} image, already the most distant on record for this comet. No one has reported observations of \comet{303P} since 2025 August 23, and only 16 positions over five nights have been reported for its 2025--2026 return. More generally, the 2024 apparitions of \comet{12P}, \comet{13P}, and C/2023~V4 and the 2025--2026 returns of \comet{128P}, \comet{131P}, and \comet{243P} each have a gap of weeks to months between our images and the nearest outside morphology report, so archival single-visit images and current follow-up can both convert these epoch comparisons into activity durations. 

To show that the methodology we present in this work (alongside Table \ref{tab:arc-extension}) serves as a valuable observing tool, we carried out observations on UT 2026 September 12 with the \accite{ARCTIC}{huehnerhoffAstrophysicalResearchConsortium2016} on \ac{APO}'s 3.5~m \ac{ARC} telescope in $r$-band. We observed \comet{99P}, \comet{299P}, \comet{302P}, \comet{303P}, \comet{351P}, \comet{486P}, \comet{491P}, \comet{510P}, and P/2016~P5. We found that all appeared active with a tail and/or coma, with \comet{99P} activity visible in stacked images, thereby extending \comet{99P} activity to 6.7 au. Notably, \comet{510P} was three days past perihelion, and astrometry of \comet{303P} had not been reported since 2025 August 23, so these are its first positions since perihelion. We will report astrometry and photometry separately.

The full \ac{DP2} release will also allow submission of the X05 astrometry from the detection nights. The predicted-position catalog can be regenerated for later data previews with the same notebook, and the methods described here naturally extend to any moving or transient source recognizable in a single exposure. The larger \textit{Rubin Comet Catchers} volunteer classification of the candidate image sets, and the population-level activity census it enables, will be reported separately (Chandler et al., in preparation), as will the full \ac{NEO} analysis (Chow et al., in preparation); the trail gallery here is primarily meant to show that trails are retrievable in \texttt{n1} images and survive the coaddition in \texttt{n2} images, and are recoverable with the same access pattern as tails.

\added{The same provenance-driven access pattern we describe applies to stationary transients as well. Of the 47 publicly listed 2025 supernovae and other transients (as of 2026 October 24) that we measured coadd provenance, 36 fell in \texttt{n1} or \texttt{n2} \ac{EDP2} cells, usually in bands or at depths not reached by the discovering surveys. Rubin single-visit magnitudes for the supernovae we highlight (Section \ref{sec:supernovae}) agree with the publicly available light curves to $\sim$0.2~mag, and shallow cells that predate later discoveries yield 5$\sigma$ pre-discovery limits of 21.5 to 24.8~mag. We showcased seven events (six in Figure \ref{fig:transients}) that illustrated recovery and the dilution relationship (Section~\ref{sec:transientmethods}), from no dilution at all (2025kdz) to one of two inputs (2025acrq).
}

\added{Among them, we identify SN~2025nbh as a Type II supernova in the outskirts of \ac{AGC}~114905 (Section \ref{sec:supernovae}), the gas-rich, ultra-diffuse galaxy at the center of the debate over whether or not such galaxies can exist without dark matter; we find no prior report of a supernova in this galaxy. A Type II event indicates a young stellar population and thus recent star formation in a system usually described by its gas and kinematics. With a light curve and spectra, it also offers an independent distance via the expanding-photosphere or standardized-candle methods \citep{hamuyTypeIISupernovae2002,dejaegerMeasurementHubbleConstant2020} that would test the 76~Mpc Hubble-flow distance informing the galaxy's mass budget. That opportunity existed in Rubin images taken 2025 July, but data-rights holders could only act on it once \ac{EDP2} was released a year later. 

The same provenance tables we describe in this work also expose a pitfall of the coaddition process; shallow coadds built while a transient was bright will contain that transient. Thus, templates assembled from early survey visits will carry the imprint of every supernova and outburst that happened during their construction.
}

Rubin deep coadds will contain regions with varying numbers of contributing visits throughout the \ac{LSST}. Although we \added{primarily} exploit the shallowest regions here, the methods we describe can more generally identify regions with low, high, or specific numbers of contributing visits, enabling the selection of coadded imaging according to the depth requirements of a given science case.

\begin{acknowledgments}

Many thanks to Arthur and Jeanie Chandler for their ongoing support.
C.O.C., M.W., W.B., J.K., and A.J.C. are supported by Schmidt Sciences. 
C.O.C. gratefully acknowledges support from the NASA CSSFP (grant No. 80NSSC26K0380).
I.C. acknowledges the support of the Natural Sciences and Engineering Research Council of Canada (NSERC) and is partially funded by an NSERC Postgraduate Scholarship-Doctoral. 
C.O.C., I.C., J.A.K., J.M., J.R.A.D., and P.S.F. acknowledge support from the DiRAC Institute in the Department of Astronomy at the University of Washington. The DiRAC Institute is supported through generous gifts from the Charles and Lisa Simonyi Fund for Arts and Sciences, Janet and Lloyd Frink, and the Washington Research Foundation. 
A.F.G. acknowledges support from the European Union's Horizon Europe Research and Innovation Programme under Grant Agreement No. 101131928 (ACME). We thank M. Schwamb (Queen's University Belfast) for her feedback and for organizing the Sprint that brought the \texttt{Ponder} project together. \added{We thank Mario Juri\'{c} for his self-described role of ``dream crusher.''}

\added{We thank Breakthrough Listen, the Astrophysics Research Centre at Queen’s University Belfast (QUB), and the QUB School of Mathematics and Physics for their generous support of the LSST Solar System First Data Sprint Belfast (February 2026). Breakthrough Listen is managed by the Breakthrough Initiatives, sponsored by the Breakthrough Prize Foundation (http://www.breakthroughinitiatives.org). We thank LSST-DA, Breakthrough Listen, the UW eScience Institute, and the DiRAC institute for their support of the SSSC First Data Sprint Seattle (November 2026).}

This material is based upon work supported in part by the National Science Foundation through Cooperative Agreements AST-1258333 and AST-2241526 and Cooperative Support Agreements AST-1202910 and 2211468 managed by the Association of Universities for Research in Astronomy (AURA), and the Department of Energy under Contract No. DE-AC02-76SF00515 with the SLAC National Accelerator Laboratory managed by Stanford University. Additional Rubin Observatory funding comes from private donations, grants to universities, and in-kind support from LSST-DA Institutional Members. This publication is based in part on proprietary Rubin Observatory Legacy Survey of Space and Time (LSST) data, and was prepared in accordance with the Rubin Observatory data rights and access policies. All authors of this publication meet the requirements for co-authorship of proprietary LSST data. This research uses services or data provided by the Rubin Science Platform at NSF-DOE Vera C. Rubin Observatory, which is jointly funded by the U.S. National Science Foundation and the U.S. Department of Energy, Office of Science.

This research has made use of data and/or services provided by the International Astronomical Union's Minor Planet Center. This research has made use of the Comet Observation Database (COBS) hosted by \v{C}rni Vrh Observatory. This research has made use of NASA's Astrophysics Data System Bibliographic Services. This research has made use of the Science Explorer, funded by NASA under Cooperative Agreement 80NSSC21M00561. \added{Some of the results in this paper have been derived using the healpy and HEALPix packages. The SPICE Resource files used in this work are described in \citet{1996P&SS...44...65A,2018P&SS..150....9A}.}

Portions of the software used to query archives, produce cutouts, and assemble tables for this work were written with the assistance of large-language-model coding tools (Anthropic \texttt{Claude Code}, Opus 4.8, \added{Opus 5.5,} and Fable 5.1; OpenAI \texttt{Codex}, 5.6-sol, 5.6-luna, \added{6-luna,} and 6-Astra). The authors ran, checked, and interpreted all code and verified every quoted observation date and source against the primary record. Generative AI tools, including \texttt{ChatGPT} (OpenAI) and \texttt{Grammarly}, were used to assist with drafting, editing, refining, and organizing the manuscript text. The authors reviewed and approved all scientific content, interpretations, and conclusions and take full responsibility for the final manuscript.
\end{acknowledgments}

\begin{contribution}
C.O.C. led the study's conceptualization and methodology, software development, investigation, formal analysis, data curation, visualization, project administration, and preparation of the original manuscript. C.O.C. also produced the image cutouts and carried out the visual activity search.

J.M. co-led the visual activity search \added{(including inspection)} and cutout production, and developed methods for coadd depth analysis and visualization.

I.C. developed the NEO methodology and carried out the NEO analysis.

A.F.G. contributed to the conceptualization of the activity analysis and validation of cometary activity.

P.S.F. contributed to the conceptualization and methodology of the source-visit provenance optimization.

\added{M.L.G. contributed to the analysis and validation of transient science, and to the methodology and optimization of source-visit provenance.}

C.O.C. conceived the Ponder software framework. J.M. developed its initial implementation. M.W. leads Ponder's ongoing development, architecture, and maintenance, with additional software contributions from W.B. J.K. provides technical supervision of Ponder development, with A.C. providing overall project supervision.

J.R.A.D. secured funding to support the project and organized a sprint week that accelerated its completion. 

H.H.H., M.S.P.K., P.B., J.A.K., and I.C. contributed to review and editing of the manuscript.

\end{contribution}

\facilities{\acs{ARC}:3.5m (\acs{ARCTIC}), 
Rubin:Simonyi (LSSTCam)
}

\software{
        \texttt{Astropy}  \citep{2013A&A...558A..33A,2018AJ....156..123A,2022ApJ...935..167A},
        \texttt{acronym} \citep{weisenburgerAcronymAutomaticReduction2017}, 
        \texttt{ASSIST} \citep{2023PSJ.....4...69H,hanno_rein_2023_7778017}, 
        {\tt astrometry.net} \citep{langAstrometryNetBlindAstrometric2010}, 
        \texttt{astroquery} \citep{2019AJ....157...98G},
        JPL Small-Body Identification (\texttt{SB\_IDENT}) API, 
        Claude \texttt{Opus 4.8}, \texttt{Opus 5.5} and \texttt{Fable 5.1} by Anthropic, for software development and analysis, 
        \texttt{Codex} \texttt{5.6-sol}, \texttt{5.6-luna}, \texttt{6-luna}, and \texttt{6-Astra} for software development and analysis, 
        \texttt{Black} (\url{https://black.readthedocs.io/en/stable/faq.html}),
        \texttt{healpy} \citep{Zonca2019,2005ApJ...622..759G},
        \texttt{HealSparse} \citep{rykoffHealSparseSparseRepresentation2026}, 
          JPL Horizons \citep{giorginiHorizonsJPLsOnLine1996}, 
        Jupyter Notebooks \citep{soton403913}, 
        \texttt{LSST Science Pipelines and the Data Butler} \citep{juricLSSTDataManagement2017,boschOverviewLSSTImage2019,jennessVeraRubinObservatory2022}, 
        \texttt{Matplotlib} \citep{hunterMatplotlibPythonPlotting2014,Hunter:2007},
        \texttt{Numba} \citep{2015llvm.confE...1L},
        \texttt{NumPy} \citep{harrisArrayProgrammingNumPy2020}, 
        {\tt pandas} \citep{rebackPandasdevPandasPandas2022,mckinney-proc-scipy-2010}, 
        \texttt{Ponder},
        \texttt{Pooch} \citep{uieda2020}, 
        \texttt{PyTables} \citep{pytables}, 
        \texttt{pyVO} \citep{grahamPyVOPythonAccess2014}, 
        {\tt REBOUND} \citep{reinREBOUNDOpensourceMultipurpose2012,reboundias15}, 
        \texttt{reproject} \citep{2022ApJ...935..167A}, 
        \texttt{rubin\_sim} \citep{angeli_end--end_2014, 2018Icar..303..181J,peter_yoachim_2022_7087823},
        \texttt{Rubin Science Platform} (RSP) \citep{juric_lse-319_2019}, 
        \texttt{SAOImageDS9} \citep{joyeNewFeaturesSAOImage2006}, 
        \texttt{sbpy} \citep{2019JOSS....4.1426M},
        \texttt{SciPy} \citep{2020SciPy-NMeth},
        \texttt{SkyBot} \citep{berthierSkyBoTNewVO2006}, 
        \texttt{Sorcha} \citep{merrittSorchaSolarSystem2025a,2025AJ....170..100M,holmanSorchaOptimizedSolar2025}, 
        \texttt{SpiceyPy} \citep{2020JOSS....5.2050A}, 
        \texttt{sqlite} (\url{https://www.sqlite.org/index.html}), 
        \texttt{sqlite3} (\url{https://docs.python.org/3/library/sqlite3.html}), 
        \texttt{tqdm} \citep{casper_da_costa_luis_2023_8233425}, 
        \acs{CADC} Solar System Object Information Search \citep{gwynSSOSMovingObjectImage2012} 
          }

\appendix

\section{Sky Coverage Breakdown}\label{sec:skycoverage}

\begin{figure*}
    \centering
    \setlength{\tabcolsep}{3pt}
    \begin{tabular}{@{}c@{\hspace{6pt}}c@{}}
        \textbf{$u$ band} & \textbf{$g$ band} \\
        \includegraphics[width=0.48\textwidth]{\detokenize{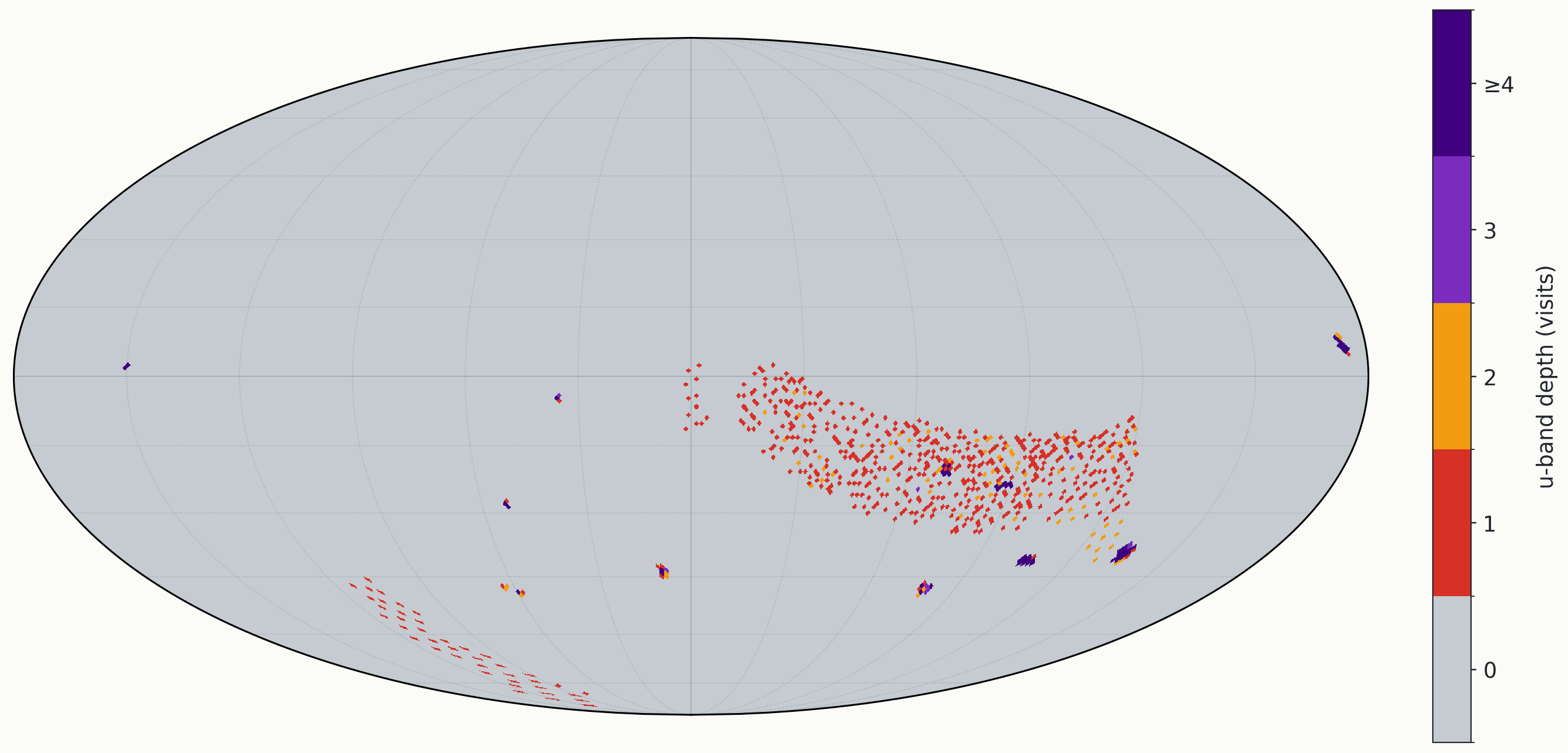}} &
        \includegraphics[width=0.48\textwidth]{\detokenize{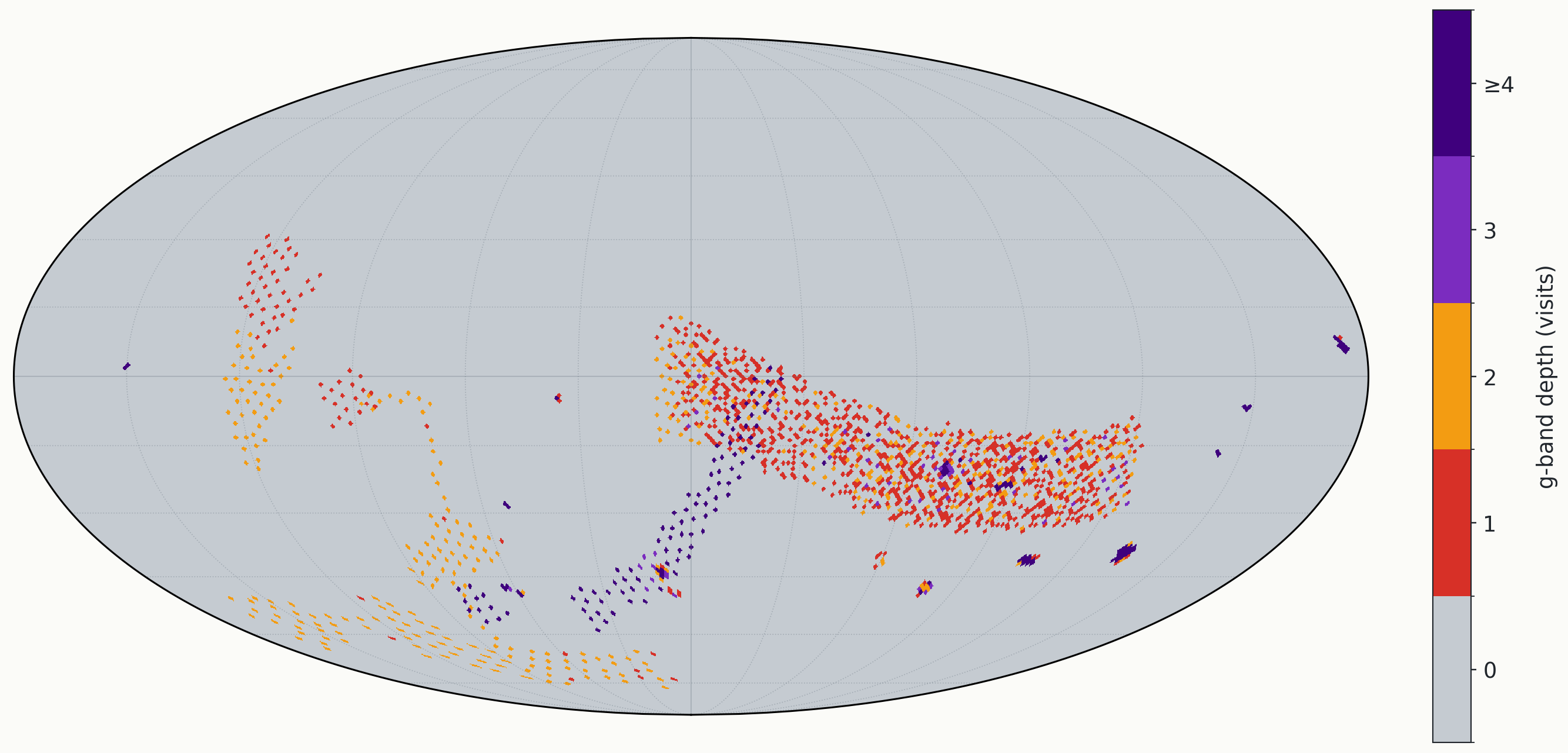}} \\
        \textbf{$r$ band} & \textbf{$i$ band} \\
        \includegraphics[width=0.48\textwidth]{\detokenize{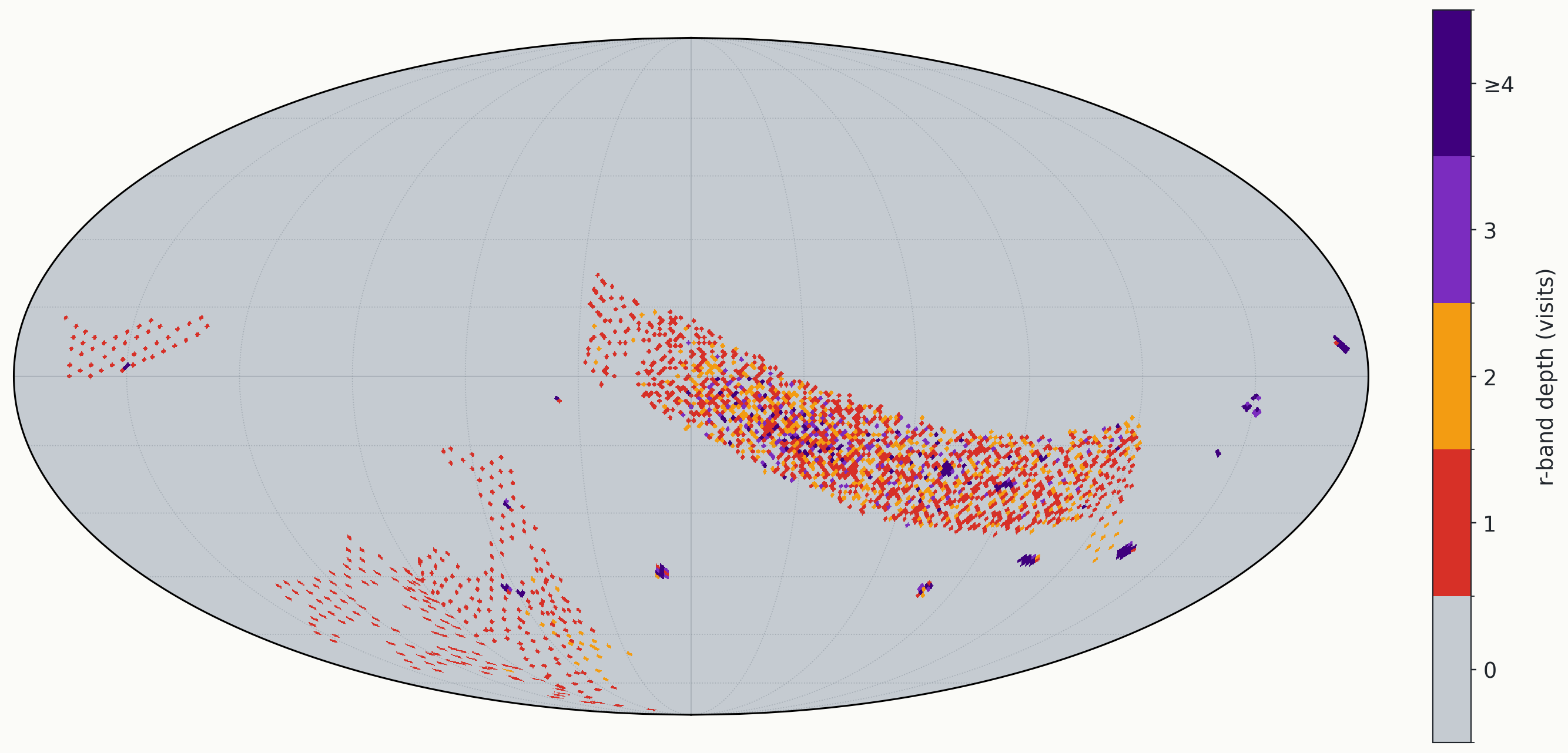}} &
        \includegraphics[width=0.48\textwidth]{\detokenize{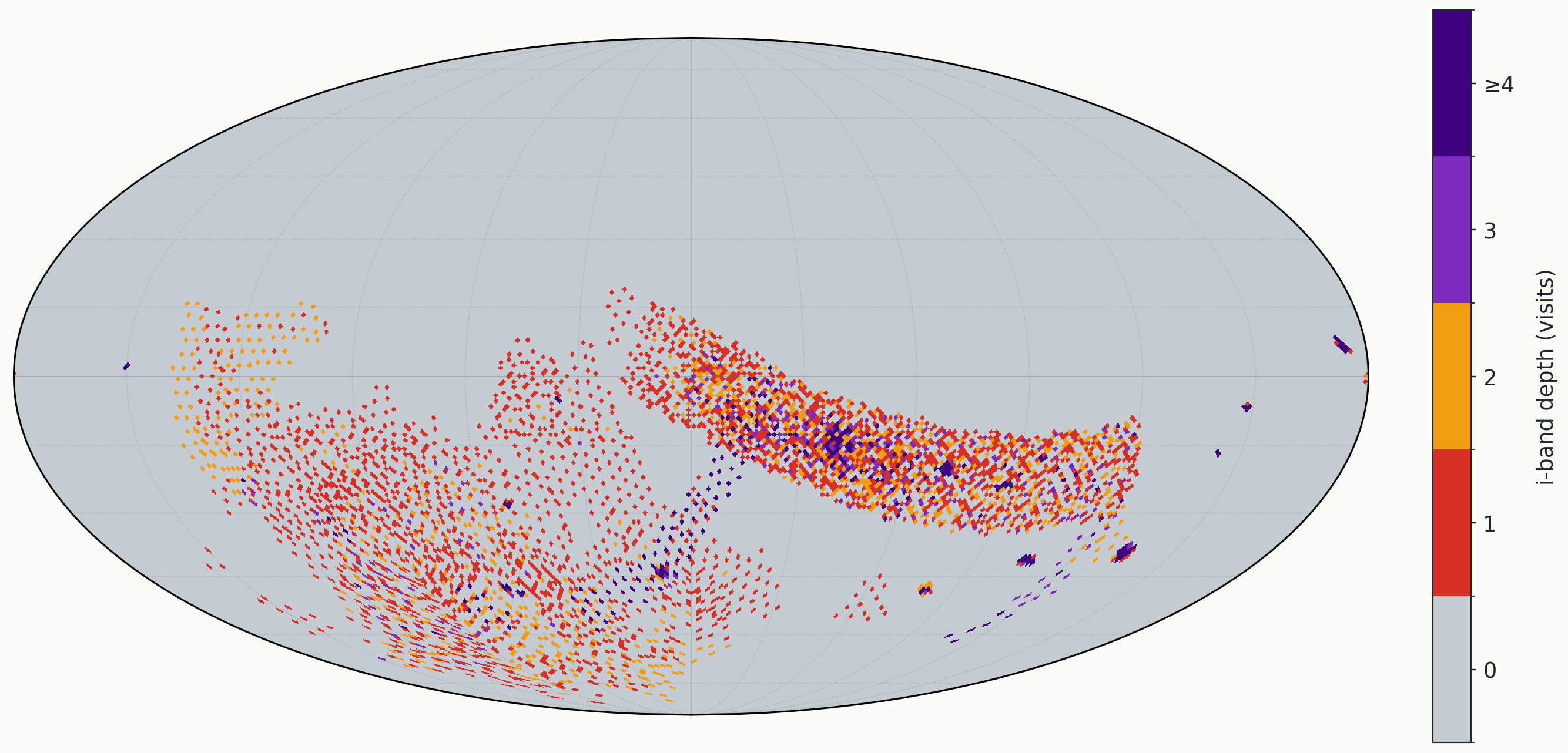}} \\
        \textbf{$z$ band} & \textbf{$y$ band} \\
        \includegraphics[width=0.48\textwidth]{\detokenize{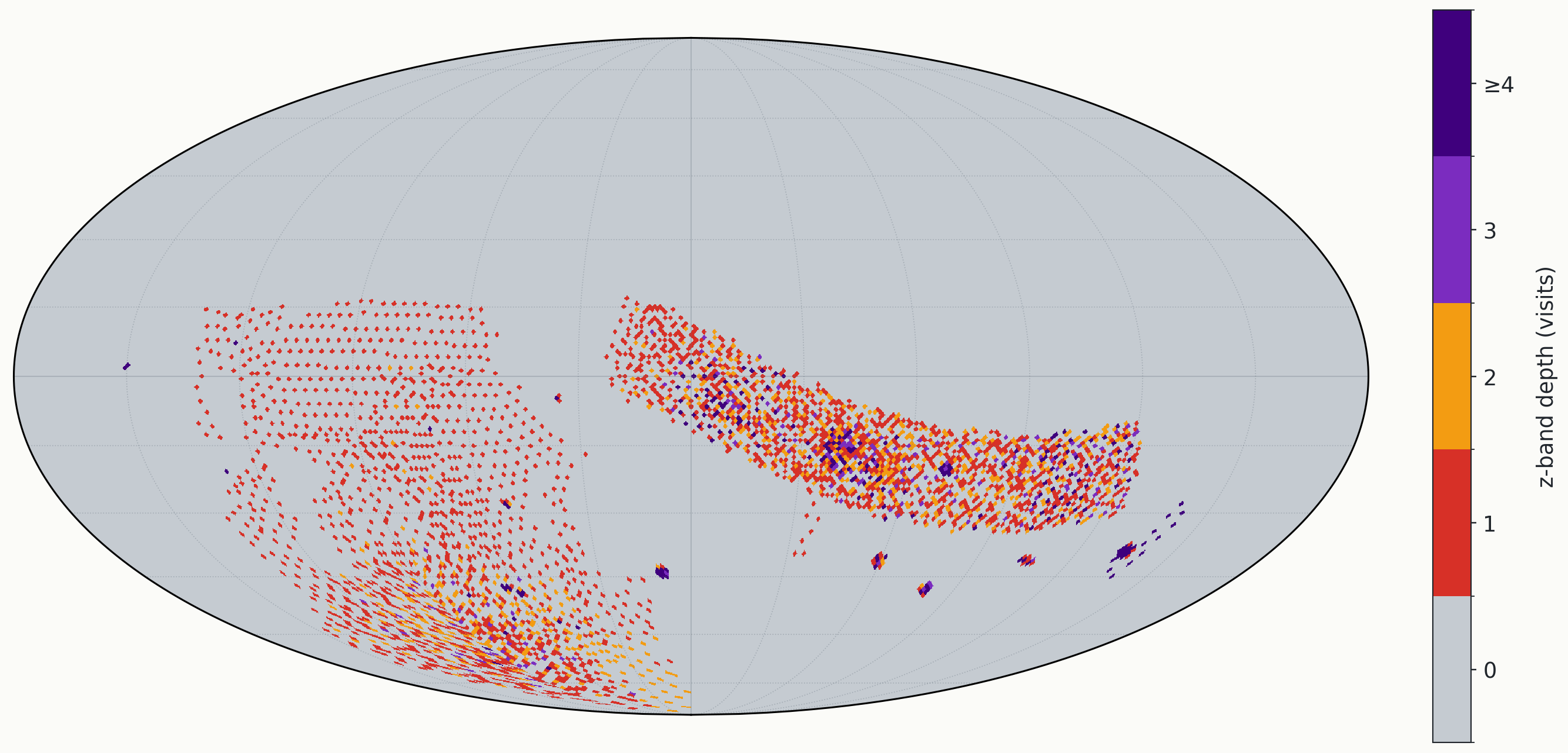}} &
        \includegraphics[width=0.48\textwidth]{\detokenize{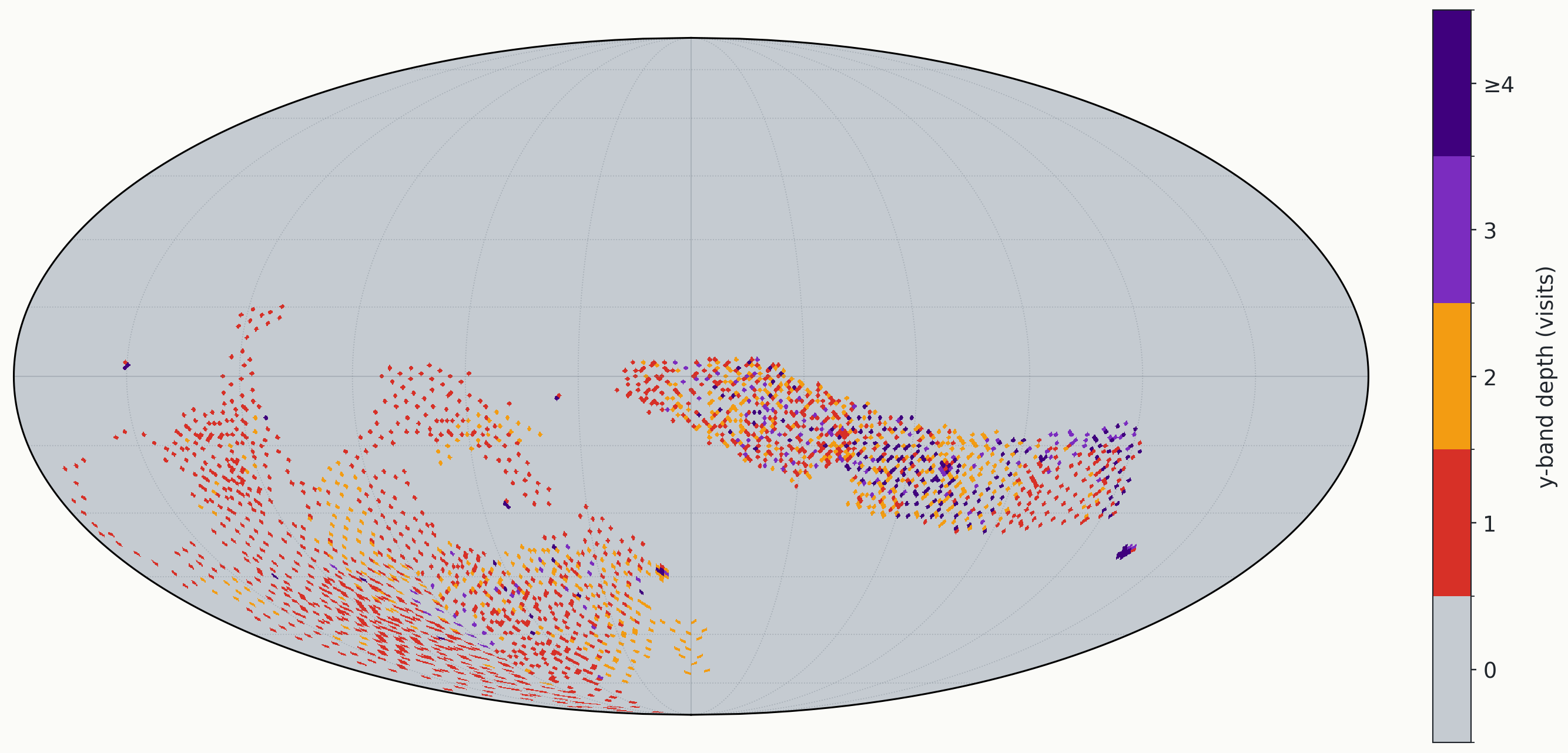}}
    \end{tabular}
    \caption{Sky coverage of the Rubin \ac{EDP2} dataset by band. The six panels show the supplied sky-coverage maps for the $u$, $g$, $r$, $i$, $z$, and $y$ bands.}
    \label{fig:skycoverage-by-band}
\end{figure*}

Figure \ref{fig:skycoverage-by-band} shows \ac{EDP2} sky coverage depth (number of contributing images to the \texttt{deep\_coadd}) broken down by band ($ugrizy$).

\section{Transients in Shallow Regions}\label{app:transients}

\added{Table~\ref{tab:transients} lists 41 of the 47 public transients for which we measured \texttt{deep\_coadd} provenance at the reported position (Section~\ref{sec:transientmethods}); the other six have no \texttt{deep\_coadd} coverage and are named in the table's notes. We provide (by band) the number of visits contributing to the coadd cell that holds the object's position.
For cells with one or two inputs (i.e., \texttt{n1} and \texttt{n2}) we list each contributing visit's 2025 UTC date and its offset in days from the public discovery date, plus the \ac{PSF} magnitude of the nearest source within 1\arcsec in that visit's source (single-visit) catalog when a visit was present, otherwise the detector 5$\sigma$ limit. We indicate ``blend'' for visits where the transient's position fell inside a large blend whose deblending was skipped, so the catalog carries no measurement. Classifications, redshifts, and discovery dates are from the \ac{TNS}. We provide the full per-visit tables, coadd object photometry, and comparison cutouts alongside the online edition.}

\begin{table*}[t]
\centering
\scriptsize
\setlength{\tabcolsep}{2.5pt}
\renewcommand{\arraystretch}{1.0}
\newcommand{\visitcell}[1]{\parbox[t]{9.9cm}{\strut #1\strut}}
\caption{Public transients with measured \texttt{deep\_coadd} provenance in \ac{EDP2}.\label{tab:transients}}
\begin{tabular}{@{}llllll@{}}
\hline\hline
Object & Class & $z$ & Disc.\ (2025) & $n_{ugrizy}$ & \visitcell{\texttt{n1}/\texttt{n2} visits: 2025 date (days from discovery) Rubin \ac{PSF} mag, 5$\sigma$ limit ($>$), or ``blend''} \\
\hline
2025imr & SLSN-I & 0.09 & 04-27 & 2/6/5/7/8/6 & \visitcell{$u$ 07-20 (+84) 21.65; 07-20 (+84) 21.58} \\
2025kdz & SLSN-II & 0.233 & 05-07 & 0/2/7/9/2/5 & \visitcell{$g$ 07-19 (+73) 18.87; 07-24 (+78) 18.90; $z$ 07-05 (+59) 18.41; 07-10 (+64) 18.42} \\
2025lwt & SN Ia & 0.09 & 05-23 & 1/3/11/11/5/3 & \visitcell{$u$ 07-22 (+60) 23.35} \\
2025lws & SN Ia & 0.07 & 05-24 & 0/4/5/6/7/9 & \visitcell{--} \\
2025lxp & SN Ia & 0.045 & 05-25 & 2/2/11/8/6/3 & \visitcell{$u$ 07-19 (+55) 21.08; 07-22 (+58) 21.24; $g$ 07-20 (+56) 19.30; 08-17 (+84) 19.14} \\
2025mmt & SN II & 0.11 & 05-27 & --/2/7/5/9/6 & \visitcell{$g$ 07-20 (+54) 20.32; 08-29 (+94) 20.59} \\
2025mma & SN Ia & 0.076 & 05-29 & 1/3/3/6/12/8 & \visitcell{$u$ 07-19 (+51) $>$23.9} \\
2025nbh & SN II & 0.017 & 06-07 & --/--/1/1/--/-- & \visitcell{$r$ 07-22 (+45) 18.80; $i$ 07-24 (+47) 18.88} \\
2025nno & SN II & 0.019 & 06-14 & --/0/3/3/4/-- & \visitcell{--} \\
2025nou & SN Ia & 0.041 & 06-15 & 1/2/5/6/4/5 & \visitcell{$u$ 07-22 (+37) 19.45; $g$ 07-20 (+35) 17.67; 07-24 (+39) 17.86} \\
2025nxc & SN Ic & 0.039 & 06-17 & --/4/5/10/7/2 & \visitcell{$y$ 07-08 (+21) blend; 07-10 (+23) blend} \\
2025nxh & SN Ia & 0.081 & 06-17 & 1/1/3/4/4/7 & \visitcell{$u$ 07-19 (+32) 20.50; $g$ 07-20 (+33) blend} \\
2025ois$^{a}$ & SN Ia & 0.04 & 06-20 & --/1/2/2/3/-- & \visitcell{$g$ 11-09 (+142) 18.30; $r$ 07-20 (+30) 16.68; 07-22 (+32) blend; $i$ 07-04 (+14) 16.29; 07-10 (+20) 16.60} \\
2025qhf & TDE & 0.093 & 06-26 & 1/2/6/10/7/3 & \visitcell{$u$ 07-19 (+23) 18.73; $g$ 07-20 (+24) 18.58; 08-17 (+52) 18.70} \\
2025qaz & SN Ia & 0.1 & 06-28 & 2/1/3/7/7/5 & \visitcell{$u$ 07-19 (+21) 19.33; 07-23 (+25) 19.80; $g$ 07-20 (+22) 18.50} \\
2025qqj & SN Ia & 0.063 & 07-06 & 2/3/7/5/6/4 & \visitcell{$u$ 07-19 (+13) 18.43; 07-22 (+16) 18.49} \\
2025saa & SN Ia & 0.077 & 07-06 & 0/4/9/7/4/-- & \visitcell{--} \\
2025ukr & SLSN-I & 0.22 & 08-16 & 1/2/5/13/3/3 & \visitcell{$u$ 07-19 (-28) $>$23.8; $g$ 07-23 (-24) 23.38; 07-24 (-23) 23.27} \\
2025ugi & SN Ic & 0.027 & 08-17 & --/0/3/0/4/2 & \visitcell{$y$ 07-10 (-38) blend; 07-14 (-34) blend} \\
2025umq & SN Ia & 0.159 & 08-17 & --/--/1/0/2/-- & \visitcell{$r$ 07-22 (-26) blend; $z$ 07-08 (-40) $>$22.8; 07-11 (-37) $>$22.6} \\
2025vnx & SN Ia & 0.04 & 08-23 & 2/3/5/4/4/10 & \visitcell{$u$ 07-20 (-34) blend; 07-21 (-33) 23.46} \\
2025vrt & SN Ia & 0.07 & 08-25 & --/6/3/2/1/-- & \visitcell{$i$ 07-04 (-52) $>$23.4; 07-10 (-46) $>$23.1; $z$ 07-10 (-46) $>$22.6} \\
2025voy & SN Ia & 0.081 & 08-27 & 1/2/4/6/6/12 & \visitcell{$u$ 07-21 (-37) $>$23.8; $g$ 07-04 (-54) blend; 07-20 (-38) $>$23.9} \\
2025wib & SN Ia & 0.063 & 08-27 & --/1/2/4/5/-- & \visitcell{$g$ 07-20 (-38) blend; $r$ 07-18 (-40) $>$23.6; 07-19 (-39) blend} \\
2025wla & SN Ia & 0.1 & 08-28 & 1/3/3/3/7/1 & \visitcell{$u$ 07-20 (-39) $>$23.6; $y$ 07-10 (-49) $>$21.5} \\
2025wpj & SN & 0.044 & 09-01 & 2/2/6/9/8/7 & \visitcell{$u$ 07-19 (-44) blend; 07-23 (-40) blend; $g$ 07-20 (-43) blend; 08-17 (-15) blend} \\
2025xby & SN Ia & 0.058 & 09-08 & 2/2/2/5/4/3 & \visitcell{$u$ 07-19 (-51) $>$23.8; 07-22 (-48) $>$23.9; $g$ 07-20 (-50) $>$24.7; 07-24 (-46) $>$24.3; $r$ 07-18 (-52) $>$24.1; 07-22 (-48) $>$24.1} \\
2025xej & SN Ia & 0.078 & 09-09 & 2/4/5/8/6/11 & \visitcell{$u$ 07-19 (-52) 23.42; 07-22 (-49) $>$23.8} \\
2025xqv & SN Ic & 0.031 & 09-09 & 1/3/5/21/17/11 & \visitcell{$u$ 07-19 (-52) $>$23.6} \\
2025xsq & SN Ia & 0.169 & 09-11 & 1/6/3/12/4/8 & \visitcell{$u$ 07-20 (-53) $>$23.8} \\
2025xma & SN Ia & 0.113 & 09-13 & 1/2/5/3/6/4 & \visitcell{$u$ 07-21 (-54) 22.56; $g$ 07-04 (-71) blend; 07-14 (-61) blend} \\
2025xln & SN Ia & 0.082 & 09-14 & 1/1/2/4/2/0 & \visitcell{$u$ 07-19 (-57) $>$23.9; $g$ 07-20 (-56) blend; $r$ 06-21 (-85) blend; 07-23 (-53) blend; $z$ 07-10 (-66) blend; 07-22 (-54) blend} \\
2025yom & SN Ia & 0.073 & 09-25 & 1/2/4/4/5/3 & \visitcell{$u$ 07-20 (-67) $>$23.8; $g$ 07-19 (-68) $>$24.7; 07-22 (-65) $>$24.8} \\
2025ytq & SN II & 0.035 & 09-27 & --/5/8/7/6/6 & \visitcell{--} \\
2025aadp & SN Ia-91T-like & 0.06 & 10-12 & --/2/5/5/10/7 & \visitcell{$g$ 07-20 (-84) blend; 08-29 (-44) blend} \\
2025abbq & SN II & 0.023 & 10-19 & 1/2/3/4/2/3 & \visitcell{$u$ 07-22 (-89) blend; $g$ 07-20 (-91) blend; 07-24 (-87) blend; $z$ 06-21 (-120) blend; 07-10 (-101) blend} \\
2025abgi & SN II & 0.023 & 10-21 & 1/3/4/8/3/2 & \visitcell{$u$ 07-20 (-93) blend; $y$ 07-08 (-105) blend; 07-09 (-104) blend} \\
2025acrq & SN Ia & 0.088 & 11-05 & --/2/1/2/4/-- & \visitcell{$g$ 07-22 (-106) $>$24.5; 11-09 (+4) 18.72; $r$ 07-20 (-108) 22.22; $i$ 07-04 (-124) 22.30; 07-10 (-118) $>$23.0} \\
2025aedb & SN IIP & 0.022 & 11-17 & --/2/7/4/3/5 & \visitcell{$g$ 07-19 (-121) $>$24.7; 08-29 (-80) $>$24.7} \\
2025afth & SN II & 0.026 & 12-03 & 0/3/4/6/4/3 & \visitcell{--} \\
2025ahnj$^{a}$ & SN Ibn & 0.067 & 12-17 & --/2/3/3/5/3 & \visitcell{$g$ 11-09 (-38) 21.64; 11-09 (-38) 21.55} \\
\hline
\end{tabular}
\tablecomments{Provenance and photometry measured on the \acf{RSP}, 2026 October 22--24. $n_{ugrizy}$: visits contributing to the coadd cell at the reported position, in order $u/g/r/i/z/y$ (``--'' no \texttt{deep\_coadd} coverage; 0, cell with no inputs). Last column, for cells with one or two inputs: \ac{PSF} magnitude of the nearest source within 1\arcsec in the visit's source catalog, otherwise the detector 5$\sigma$ limit when no source matched, or ``blend'' where the position fell inside a large blend whose deblending was skipped. Objects with no shallow cell (--) have three or more visits in every covered band. Six further screened objects (2025aarn, 2025aaxd, 2025acsy, 2025ajvx, 2025vdq, 2025yii) have no \texttt{deep\_coadd} coverage at their positions and are omitted. Classifications, redshifts, and discovery dates are from the \ac{TNS}. $^{a}$Measurements of the host nucleus, not the transient: for 2025ois the Rubin source is about 2~mag brighter than the \ac{ZTF} light curve, and for 2025ahnj the two $g$ visits predate discovery by 38 days.}
\end{table*}

\bibliography{paper}{}
\bibliographystyle{aasjournalv7}

\input{Acronyms.tex}

\end{document}

%% file: Acronyms.tex

\begin{acronym}[JSONP]\itemsep0pt 
\acro{2MASS}{Two Micron All Sky Survey}
\acro{AA}{active asteroid}
\acro{ACO}{asteroid on a cometary orbit}
\acro{AGC}{Arecibo General Catalog}
\acro{AI}{artificial intelligence}
\acro{API}{Application Programming Interface}
\acro{APT}{Aperture Photometry Tool}
\acro{ARC}{Astrophysical Research Consortium}
\acro{ARCTIC}{Astrophysical Research Consortium Telescope Imaging Camera}
\acro{AOS}{Active Optics System}
\acro{APO}{Apache Point Observatory}
\acro{ARO}{Atmospheric Research Observatory}
\acro{AstOrb}{Asteroid Orbital Elements Database}
\acro{ASU}{Arizona Statue University}
\acro{ATLAS}{Asteroid Terrestrial-impact Last Alert System}
\acro{AURA}{Association of Universities for Research in Astronomy}
\acro{BASS}{Beijing-Arizona Sky Survey}
\acro{BLT}{Barry Lutz Telescope}
\acro{CADC}{Canadian Astronomy Data Centre}
\acro{CASU}{Cambridge Astronomy Survey Unit}
\acro{CATCH}{Comet Asteroid Telescopic Catalog Hub}
\acro{CBAT}{Central Bureau for Astronomical Telegrams}
\acro{CBET}{Central Bureau for Electronic Telegrams}
\acro{CCD}{charge-coupled device}
\acro{CEA}{Commissariat a l'Energes Atomique}
\acro{CFHT}{Canada-France-Hawaii Telescope}
\acro{CFITSIO}{C Flexible Image Transport System Input Output}
\acro{CNEOS}{Center for Near Earth Object Studies} 
\acro{CNRS}{Centre National de la Recherche Scientifique}
\acro{COBS}{Comet Observations Database} 
\acro{CPU}{Central Processing Unit}
\acro{CTIO}{Cerro Tololo Inter-American Observatory}
\acro{DAPNIA}{Département d'Astrophysique, de physique des Particules, de physique Nucléaire et de l'Instrumentation Associée}
\acro{DART}{Double Asteroid Redirection Test}
\acro{DDF}{Deep Drilling Field}
\acro{DDT}{Director's Discretionary Time}
\acro{DECaLS}{Dark Energy Camera Legacy Survey}
\acro{DECam}{Dark Energy Camera}
\acro{DES}{Dark Energy Survey}
\acro{DESI}{Dark Energy Spectroscopic Instrument}
\acro{DCR}{differential chromatic refraction}
\acro{DCT}{Discovery Channel Telescope}
\acro{DOE}{Department of Energy}
\acro{DP2}{Data Preview 2}
\acro{DR}{Data Release}
\acro{DS9}{Deep Space Nine}
\acro{EDP2}{Early Data Preview 2}
\acro{ESA}{European Space Agency}
\acro{ESO}{European Southern Observatory}
\acro{ETC}{exposure time calculator}
\acro{ETH}{Eidgenössische Technische Hochschule}
\acro{FAQ}{frequently asked questions}
\acro{FITS}{Flexible Image Transport System}
\acro{FOV}{field of view}
\acro{FTN}{Faulkes Telescope North}
\acro{GEODSS}{Ground-Based Electro-Optical Deep Space Surveillance}
\acro{GIF}{Graphic Interchange Format}
\acro{GMOS}{Gemini Multi-Object Spectrograph}
\acro{GOTO}{Gravitational-wave Optical Transient Observer}
\acro{GPU}{Graphics Processing Unit}
\acro{GRFP}{Graduate Research Fellowship Program}
\acro{GRSS}{Gauss-Radau Small-body Simulator }
\acro{HARVEST}{Hunting for Activity in Repositories with Vetting-Enhanced Search Techniques}
\acro{HDU}{Header Data Unit}
\acro{HSC}{Hyper Suprime-Cam}
\acro{HST}{Hubble Space Telescope}
\acro{IAU}{International Astronomical Union}
\acro{ICRS}{International Celestial Reference System}
\acro{IMACS}{Inamori-Magellan Areal Camera and Spectrograph}
\acro{IMB}{inner Main-belt}
\acro{IMCCE}{Institut de Mécanique Céleste et de Calcul des Éphémérides}
\acro{INAF}{Istituto Nazionale di Astrofisica}
\acro{INT}{Isaac Newton Telescopes}
\acro{IP}{Internet Protocol}
\acro{IRSA}{Infrared Science Archive}
\acro{IRTF}{Infrared Telescope Facility}
\acro{ISR}{Instrumental Signature Removal}
\acro{ISO}{Interstellar Object}
\acro{ISO2}{International Organization for Standardization} 
\acro{ITC}{integration time calculator}
\acro{JAXA}{Japan Aerospace Exploration Agency}
\acro{JD}{Julian Date}
\acro{JFC}{Jupiter-family comet}
\acro{JPL}{Jet Propulsion Laboratory}
\acro{KBO}{Kuiper Belt object}
\acro{KOA}{Keck Observatory Archive}
\acro{KPNO}{Kitt Peak National Observatory}
\acro{LAST}{Large Array Survey Telescope}
\acro{LBC}{Large Binocular Camera}
\acro{LCO}{Las Cumbres Observatory}
\acro{LBCB}{Large Binocular Camera Blue}
\acro{LBCR}{Large Binocular Camera Red}
\acro{LBT}{Large Binocular Telescope}
\acro{LDT}{Lowell Discovery Telescope}
\acro{LINCC}{LSST Interdisciplinary Network for Collaboration and Computing}
\acro{LINEAR}{Lincoln Near-Earth Asteroid Research}
\acro{LMI}{Large Monolithic Imager}
\acro{LONEOS}{Lowell Observatory Near-Earth-Object Search}
\acro{LSST}{Legacy Survey of Space and Time}
\acro{MBC}{Main-belt comet}
\acro{MGIO}{Mount Graham International Observatory}
\acro{MITHNEOS}{MIT-Hawaii Near-Earth Object Spectroscopic Survey}
\acro{ML}{machine learning}
\acro{MMB}{middle Main-belt}
\acro{MMR}{mean-motion resonance}
\acro{MOST}{Moving Object Search Tool}
\acro{MPEC}{Minor Planet Electronic Circular} 
\acro{MzLS}{Mayall z-band Legacy Survey}
\acro{MPC}{Minor Planet Center}
\acro{MUSE}{Multi Unit Spectroscopic Explorer}
\acro{MuSCAT3}{Multicolor Simultaneous Camera for studying Atmospheres of Transiting exoplanets}
\acro{NAU}{Northern Arizona University}
\acro{NEA}{near-Earth asteroid}
\acro{NEAT}{Near-Earth Asteroid Tracking}
\acro{NEATM}{Near Earth Asteroid Thermal Model}
\acro{NED}{NASA/IPAC Extragalactic Database}
\acro{NEEDLE}{NEural Engine for Discovering Luminous Events}
\acro{NEO}{near-Earth object}
\acro{NEOCP}{Near-Earth Object Confirmation Page}
\acro{NEOWISE}{Near-Earth Object Wide-field Infrared Survey Explorer}
\acro{NGPS}{Next Generation Palomar Spectrograph}
\acro{NIHTS}{Near-Infrared High-Throughput Spectrograph}
\acro{NOAO}{National Optical Astronomy Observatory}
\acro{NOIRLab}{National Optical and Infrared Laboratory}
\acro{NRC}{National Research Council}
\acro{OMB}{outer Main-belt}
\acro{OSIRIS-REx}{Origins, Spectral Interpretation, Resource Identification, Security, Regolith Explorer}
\acro{NSF}{National Science Foundation}
\acro{PA}{position angle}
\acro{PANSTARRS}{Panoramic Survey Telescope and Rapid Response System}
\acro{Pan-STARRS1}{Panoramic Survey Telescope and Rapid Response System}
\acro{PANSTARRS1}{Panoramic Survey Telescope and Rapid Response System}
\acro{PDF}{Portable Document Format}
\acro{PI}{Principal Investigator}
\acro{PNG}{Portable Network Graphics}
\acro{PSI}{Planetary Science Institute}
\acro{PSF}{point spread function}
\acro{PTF}{Palomar Transient Factory}
\acro{PVI}{Preliminary Visit Image}
\acro{QH}{Quasi-Hilda}
\acro{QHA}{Quasi-Hilda Asteroid}
\acro{QHC}{Quasi-Hilda Comet}
\acro{QHO}{Quasi-Hilda Object}
\acro{RA}{Right Ascension}
\acro{REU}{Research Experiences for Undergraduates}
\acro{RMS}{root-mean-square}
\acro{RNAAS}{Research Notes of the American Astronomical Society}
\acro{RSP}{Rubin Science Platform}
\acro{SAFARI}{Searching Asteroids For Activity Revealing Indicators}
\acro{SBSAR}{Small-Body Search and Rescue}
\acro{SDSS}{Sloan Digital Sky Survey}
\acro{SMOKA}{Subaru Mitaka Okayama Kiso Archive}
\acro{SAO}{Smithsonian Astrophysical Observatory}
\acro{SBDB}{Small Body Database}
\acro{SDSS DR-9}{Sloan Digital Sky Survey Data Release Nine}
\acro{SFD}{Size-Frequency Distribution}
\acro{SIA}{Simple Image Access}
\acro{SLAC}{Stanford Linear Accelerator Center}
\acro{SEDM}{Spectral Energy Distribution Machine}
\acro{SIMBAD}{Set of Identifications, Measurements and Bibliography for Astronomical Data}
\acro{SNID}{Supernova Identification}
\acro{SN}{supernova}
\acro{SNR}{signal to noise ratio}
\acro{SOAR}{Southern Astrophysical Research Telescope}
\acro{SODA}{Server-side Operations for Data Access}
\acro{SSOIS}{Solar System Object Information Search}
\acro{SSP}{Solar System Processing}
\acro{SQL}{Structured Query Language}
\acro{SUP}{Suprime Cam}
\acro{SV}{Science Validation}
\acro{SwRI}{Southwestern Research Institute}
\acro{TAI}{Temps Atomique International}
\acro{TAP}{Telescope Access Program}
\acro{TBTs}{Test Bed Telescopes}
\acro{TDE}{tidal disruption event}
\acro{TNO}{Trans-Neptunian object}
\acro{TNS}{Transient Name Server}
\acro{ToO}{Target of Opportunity}
\acro{TRAPPIST}{Transiting Planets and Planetesimals Small Telescope}
\acro{UA}{University of Arizona}
\acro{UT}{Universal Time}
\acro{UCSC}{University of California Santa Cruz}
\acro{UCSF}{University of California San Francisco}
\acro{VATT}{Vatican Advanced Technology Telescope}
\acro{VIA}{Virtual Institute of Astrophysics}
\acro{VIRCam}{VISTA InfraRed Camera}
\acro{VISTA}{Visible and Infrared Survey Telescope for Astronomy}
\acro{VLT}{Very Large Telescope}
\acro{VST}{Very Large Telescope (VLT) Survey Telescope}
\acro{WEB}{Water Enriched Block}
\acro{WFC}{Wide Field Camera}
\acro{WIRCam}{Wide-field Infrared Camera}
\acro{WISE}{Wide-field Infrared Survey Explorer}
\acro{WCS}{World Coordinate System}
\acro{YORP}{Yarkovsky--O'Keefe--Radzievskii--Paddack}
\acro{ZTF}{Zwicky Transient Facility}
\end{acronym}